\documentclass[notitlepage,letterpaper,natbib,aps,prd,twocolumn,amsmath,amsfonts,nofootinbib,preprintnumbers,superscriptaddress,secnumarabic]{revtex4-1}
\pdfoutput=1
\usepackage{amssymb,amsmath,latexsym,mathrsfs}
\usepackage{url}
\usepackage{enumitem}
\usepackage{graphicx}
\usepackage[usenames,dvipsnames]{color}
\usepackage[breaklinks,colorlinks,urlcolor=blue,citecolor=blue,linkcolor=magenta]{hyperref}
 \usepackage{multirow}

\begin{document}

\title{Cosmological Constraints on Invisible Neutrino Decays Revisited}
\preprint{KCL-2019-57}
\author{Miguel Escudero}
\email{miguel.escudero@kcl.ac.uk}
\affiliation{Department of Physics, King's College London, Strand, London WC2R 2LS, UK}
\author{Malcolm Fairbairn}
\email{malcolm.fairbairn@kcl.ac.uk}
\affiliation{Department of Physics, King's College London, Strand, London WC2R 2LS, UK}
\begin{abstract}
Invisible neutrino decay modes are difficult to target at laboratory experiments, and current bounds on such decays from solar neutrino and neutrino oscillation experiments are somewhat weak. It has been known for some time that Cosmology can serve as a powerful probe of invisible neutrino decays. In this work, we show that in order for Big Bang Nucleosynthesis to be successful, the invisible neutrino decay lifetime is bounded to be $\tau_\nu > 10^{-3}\,\text{s}$ at 95\% CL. We revisit Cosmic Microwave Background constraints on invisible neutrino decays, and by using Planck2018 observations we find the following bound on the neutrino lifetime: $\tau_\nu > (1.3-0.3)\times 10^{9}\,\text{s}  \, \left({m_\nu}/{ 0.05\,\text{eV} }\right)^3$ at $95\%$ CL. We show that this bound is robust to modifications of the cosmological model, in particular that it is independent of the presence of dark radiation. We find that lifetimes relevant for Supernova observations ($\tau_\nu \sim 10^{5}\,\text{s}\,  \left({m_\nu}/{ 0.05\,\text{eV} }\right)^3$) are disfavoured at more than $5\,\sigma$ with respect to $\Lambda$CDM given the latest Planck CMB observations. Finally, we show that when including high-$\ell$ Planck polarization data, neutrino lifetimes $\tau_\nu = (2-16)\times 10^{9}\,\text{s} \,  \left({m_\nu}/{ 0.05\,\text{eV} }\right)^3$ are mildly preferred -- with a 1-2~$\sigma$ significance -- over neutrinos being stable. 
\end{abstract}

%%%%%%%%%%%%%%%%%%%%%%%%%%%%%%%%%%%%%%%%%%%%%%%%%%%%%%
\maketitle
%%%%%%%%%%%%%%%%%%%%%%%%%%%%%%%%%%%%%%%%%%%%%%%%%%%%%%

%%%%%%%%%%%%%%%%%%%%%%%%%%%%%%%%%%%%%%%%%%%%%%%%%%%%%%
\section{Introduction}
%%%%%%%%%%%%%%%%%%%%%%%%%%%%%%%%%%%%%%%%%%%%%%%%%%%%%%

At least two of the light active neutrinos are massive~\cite{Esteban:2018azc,deSalas:2017kay,Capozzi:2016rtj} and will therefore decay via suppressed loop interactions even without any new physics~\cite{Petcov:1976ff,HOSOTANI1981411,PhysRevD.25.766}. Given our knowledge of the Standard Model (SM) interactions, the active neutrino lifetimes are considerably larger than the age of the Universe $\tau_{\nu} > 10^{35}\, \text{yr}$, and therefore are too large to have any measurable implication for laboratory experiments, for astrophysics or for cosmology. However, many extensions of the SM do predict substantially shorter neutrino lifetimes, see for example~\cite{Chikashige:1980ui,Gelmini:1980re,Schechter:1981cv,Lee:1977tib,Berezhiani:1990wn,Berezhiani:1990jj,Joshipura:1992vn,Burgess:1992dt,Berezhiani:1992cd,PhysRevD.9.743,GEORGI1990196,Davidson:2005cs,Bell:2005kz,Lindner:2017uvt,Dvali:2016uhn}. 

The constraints on the neutrino lifetime are very much dependent upon the neutrino decay products. Radiative neutrino decays are strongly constrained by the non-observation of neutrino magnetic moments~\cite{Fujikawa:1980yx} in laboratory experiments $\tau_\nu\,$$\gtrsim$$10^{18} \, \text{yr}$~\cite{Beda:2013mta,Borexino:2017fbd}, by cosmic microwave background (CMB) spectral distortions $\tau_\nu \gtrsim 10^{12} \, \text{yr}$~\cite{Mirizzi:2007jd,Aalberts:2018obr}, by 21 cm cosmology~\cite{Chianese:2018luo}, and by astrophysical considerations $\tau_\nu \gtrsim 10^{20} \, \text{yr}$~\cite{PhysRevLett.64.2856,ARCEODIAZ20151,RAFFELT1999319}. In contrast, the constraints on invisible neutrino decays, namely those that do not involve photons in the final state, are considerably looser. This is a result of the difficulty in detecting the decay products from such a process and due to fact that light active neutrinos are usually highly boosted. 

Invisible neutrino decays are constrained by solar neutrino experiments~\cite{Beacom:2002cb,Picoreti:2015ika,SNO_nudec,Berryman:2014qha,Funcke:2019grs}. In normal ordering scenarios (NO), they lead to the limit $\tau_{\nu_2}/m_{\nu_2} \gtrsim 1.5\times 10^{-3}\,\text{s} \, \text{eV}^{-1}$~\cite{SNO_nudec}.  For inverted ordering (IO) the limits are $\tau_{\nu_1}/m_{\nu_1} \gtrsim 4\times 10^{-3}\,\text{s} \, \text{eV}^{-1}$ and $\tau_{\nu_2}/m_{\nu_2} \gtrsim 7\times 10^{-4}\,\text{s} \, \text{eV}^{-1}$~\cite{Berryman:2014qha}. Recently, Ref.~\cite{Funcke:2019grs} also reported constraints on $\tau_{\nu_3}/m_{\nu_3} > 2.2\times 10^{-5}\,\text{s}\,\text{eV}^{-1}$ by noting that electron neutrinos also mix with $\nu_3$ neutrinos. There are also constraints from atmospheric and long-baseline experiments~\cite{GonzalezGarcia:2008ru,Gomes:2014yua,Gago:2017zzy,Choubey:2018cfz} that lead to $\tau_{\nu_3}/m_{\nu_3} \gtrsim 3\times 10^{-10}\,\text{s} \, \text{eV}^{-1} $~\cite{GonzalezGarcia:2008ru}. 

In addition, the fact that the CMB spectrum is well fitted with free-streaming neutrino perturbations can be used~\cite{Hannestad:2004qu,Hannestad:2005ex,Basboll:2008fx,Archidiacono:2013dua,Bell:2005dr} to set strong constraints on the neutrino lifetime $\tau_\nu > 1.2 \times  10^{9}\, \text{s} \left(m_\nu/0.05\,\text{eV} \right)^{3}$~\cite{Archidiacono:2013dua}. 

\vspace{0.3cm}

\begin{figure*}[t]
\centering
\includegraphics[width=0.9\textwidth]{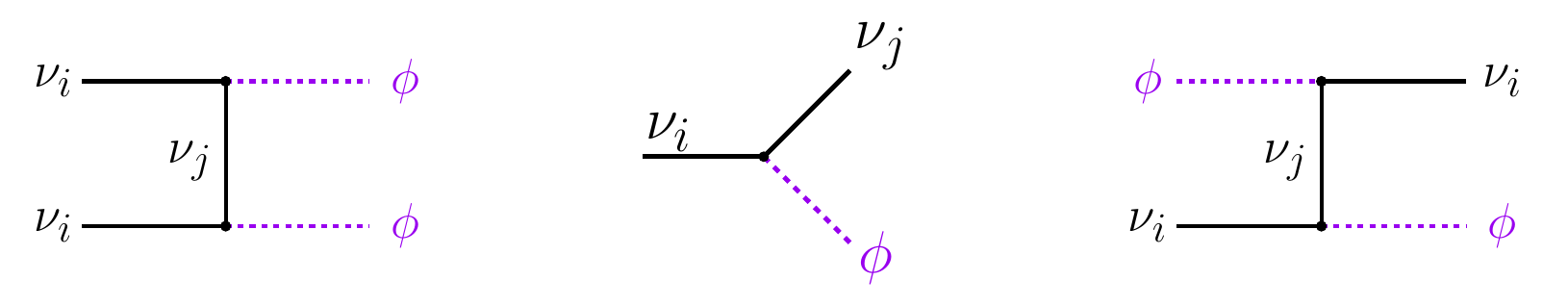} 
\vspace{-0.2cm}
\caption{Annihilation (left) and scattering (right) neutrino-$\phi$ diagrams as induced by the same interactions that trigger invisible neutrino decays (middle).  }\label{fig:decay-annihilation}
\end{figure*}

In this work, in light of these somewhat weak constraints on invisible neutrino decays, we study the impact of invisible neutrino decays upon Big Bang Nucleosynthesis (BBN), and also revisit the constraints on invisible neutrino decays derived from the CMB observations made by the Planck satellite.

In the first part of this paper, we exploit the fact that in order for neutrinos to decay invisibly, they should decay into massless or at least to very light species.  Because of this, the same interactions that trigger the decay may produce a thermal population of such light species prior to BBN and thereby augment the number of relativistic neutrino species in the early Universe, $N_{\rm eff}$. We show that, independently of the neutrino decay process and the neutrino type, neutrino lifetimes $\tau_\nu < 10^{-3}\,\text{s}$ are ruled out by the current measured primordial nuclei abundances. In this way we improve upon current constraints from accelerator and long-baseline experiments by 8 orders of magnitude, and by 2 orders of magnitude over current constraints from solar neutrino experiments. We note that similar phenomenology has been studied in the past when the $\tau$-neutrino was still allowed to be a mass eigenstate and considerably heavy ($m_{\nu_\tau} <23\,\text{MeV}$), see e.g.~\cite{Kolb:1991sn,Dolgov:1993uz,Kawasaki:1993gz,Dodelson:1994it,Dolgov:1996fp,Hannestad:1997ar}.

For the second part of the work, we calculate the effect of neutrino decays in the density perturbations of the neutrino fluid and use this to test the neutrino decay hypothesis against the 2018 temperature and polarization CMB power spectra as measured by the Planck satellite~\cite{Adam:2015rua,Ade:2015xua}. For previous CMB analysis see~\cite{Bashinsky:2003tk,Trotta:2004ty,Hannestad:2004qu,Hannestad:2005ex,Basboll:2008fx,Archidiacono:2013dua,Bell:2005dr}, particularly Ref.~\cite{Archidiacono:2013dua}. In this study, we are maximally conservative and perform analyses assuming various types of neutrino decay modes. We consider invisible neutrino modes in which an active neutrino decays into another active neutrino plus a massless scalar field and obtain a lower limit on the lifetime using the Planck 2018 data of $\tau_{\nu} > 1.3\times 10^{9}\,\text{s}  \, \left({m_\nu}/{ 0.05\,\text{eV} }\right)^3$ at 95\% CL. This bound is only 10\% more stringent than the previous limit obtained in Ref.~\cite{Archidiacono:2013dua} that used Planck 2013 data, but unlike Ref.~\cite{Archidiacono:2013dua} here we consider that only the two neutrinos that participate in the decay process are interacting. In addition, we explore the possible degeneracies between a finite neutrino lifetime and a variation in $N_{\rm eff}$, and show that contrary to previous expectations~\cite{Bell:2005dr,deSalas:2018kri,Denton:2018aml} even if only one neutrino species decays and a non-interacting $N_{\rm eff}$ is allowed to vary, neutrino lifetimes of $\tau_\nu < 0.9\times 10^{9}\,\text{s}  \, \left({m_\nu}/{ 0.05\,\text{eV} }\right)^3 $ are still excluded at 95\% CL by Planck CMB observations. Finally, we find that when including Planck 2018 high-$\ell$ polarization data in the analysis, neutrino lifetimes in the range $\tau_\nu = (2-16)\times 10^{9}\,\text{s} \, \left({m_\nu}/{ 0.05\,\text{eV} }\right)^3$ are preferred over neutrinos being purely stable with a $\sim$~1-2~$\sigma$  significance.

This paper is organized as follows. In Section~\ref{sec:NeutrinoDecay_model}, we consider a simple and generic model for invisible neutrino decays. In Section~\ref{sec:earlyUniverse}, we consider the production in the early Universe of beyond the Standard Model light neutrino decay products and set constraints on such production using BBN.  We also include a discussion of the applicability of the derived BBN constraints. In Section~\ref{sec:CMB}, we outline how we model the impact of neutrino decays upon cosmological perturbations and test the neutrino decay hypothesis with Planck 2018 data to set constraints on invisible neutrino decays. We summarize and discuss the main results of this work in Section~\ref{sec:conclusions}. Finally, in Section~\ref{sec:outlook}, we comment on how invisible neutrino decays are expected to be constrained in the future.

\vspace{0cm}
%%%%%%%%%%%%%%%%%%%%%%%%%%%%%%%%%%%%%%%%%%%%%%%%%%%%%%
\section{Invisible Neutrino Decays}\label{sec:NeutrinoDecay_model}
%%%%%%%%%%%%%%%%%%%%%%%%%%%%%%%%%%%%%%%%%%%%%%%%%%%%%%
Fast (i.e. $\tau_{\nu} \ll 10^{35}\, \text{yr}$) and invisible neutrino decays are a typical prediction of models in which \textit{global} lepton number is spontaneously broken so as to generate light Majorana neutrino masses. In such models, as a result, a massless Goldstone boson appears in the spectrum, the \textit{majoron}~\cite{Chikashige:1980ui,Gelmini:1980re,Schechter:1981cv}. 

Here, we shall consider the following effective interaction between neutrinos and a massless scalar $\phi$:
\begin{align}\label{eq:Lagrangian}
\mathcal{L}_{\rm int} = \lambda_{ij} \, i \, \phi \, \bar{\nu}_i \, \gamma_5 \, \nu_j + h.c. \, ,
\end{align}
where the $\nu_i$ correspond to the massive neutrino eigenstates, $i,\,j = 1,\,2,\,3$ and we shall assume neutrinos are Majorana particles\footnote{The applicability of the derived constraints will not significantly depend upon this assumption, as discussed in Section~\ref{sec:applicability}.}. $\lambda_{ij}$ are coupling constants, of which the off-diagonal elements with $i\neq j$ induce neutrino decay. 

Given the interactions above, the rate of neutrino decay $\nu_i \to \nu_j + \phi$ is:
\begin{align}\label{eq:rest_decay}
\scalebox{1.04}[1.1]{$\Gamma_{\nu_i \to \nu_j + \phi}  = \frac{\lambda_{ij}^2}{4\pi} \,\frac{(m_{\nu_i} - m_{\nu_j})^3(m_{\nu_i} + m_{\nu_j})}{m_{\nu_i}^3 } \simeq  \frac{\lambda_{ij}^2}{4\pi}m_{\nu_i}ƒ'' \,,$}
\end{align}
where in the last step we have assumed that $m_{\nu_i} \gg m_{\nu_j}$.

\vspace{-0.25cm}
%%%%%%%%%%%%%%%%%%%%%%%%%%%%%%%%%%%%%%%%%%%%%%%%%%%%%%
\section{Big Bang Nucleosynthesis Constraints}\label{sec:earlyUniverse}
%%%%%%%%%%%%%%%%%%%%%%%%%%%%%%%%%%%%%%%%%%%%%%%%%%%%%%

We place early Universe constraints on the invisible neutrino lifetime by exploiting the fact that the same interactions that allow for fast invisible neutrino decays also mean that processes of the type $\bar{\nu}\nu \to \phi \phi$ will be active in the early Universe (see Figure~\ref{fig:decay-annihilation} for an illustration of these processes). These processes can potentially lead to a thermal population of massless or very light $\phi$ species in the early Universe. This would thereby impact the primordial nuclei abundances and the number of effective neutrino species as inferred from CMB observations. 

In order to make a precise statement about the constraint on the coupling constant $\lambda_{ij}$, and therefore (via equation~\eqref{eq:rest_decay}) upon the lifetime of the neutrino, we need to calculate the abundance of massless $\phi$ particles in the early Universe. The presence of a thermal abundance of $\phi$ particles will only influence $N_{\rm eff}$ or the primordial element abundances if the $\phi$ population is generated prior to neutrino decoupling, at $T \gtrsim 2\,\text{MeV}$~\cite{Dolgov:2002wy}, corresponding to an era in the Universe in which neutrinos can be efficiently produced via $e^+e^- \to \bar{\nu} \nu$ annihilations. If there is a thermal population of $\phi$ particles prior to neutrino decoupling $\Delta N_\text{eff} \equiv N_{\rm eff} -N_{\rm eff}^{\rm SM} = 8/7 \times g_\phi/g_\gamma =  4/7 \simeq 0.57$, where $N_{\rm eff}^{\rm SM} = 3.045$~\cite{deSalas:2016ztq,Mangano:2005cc}. Note that $\Delta N_\text{eff} > 0.4$ is excluded at more than 95\% CL from current measurements of the primordial nuclei abundances~\cite{Pitrou:2018cgg}, see also~\cite{Cyburt:2015mya}.

\begin{figure}[t]
\centering
\includegraphics[width=0.48\textwidth]{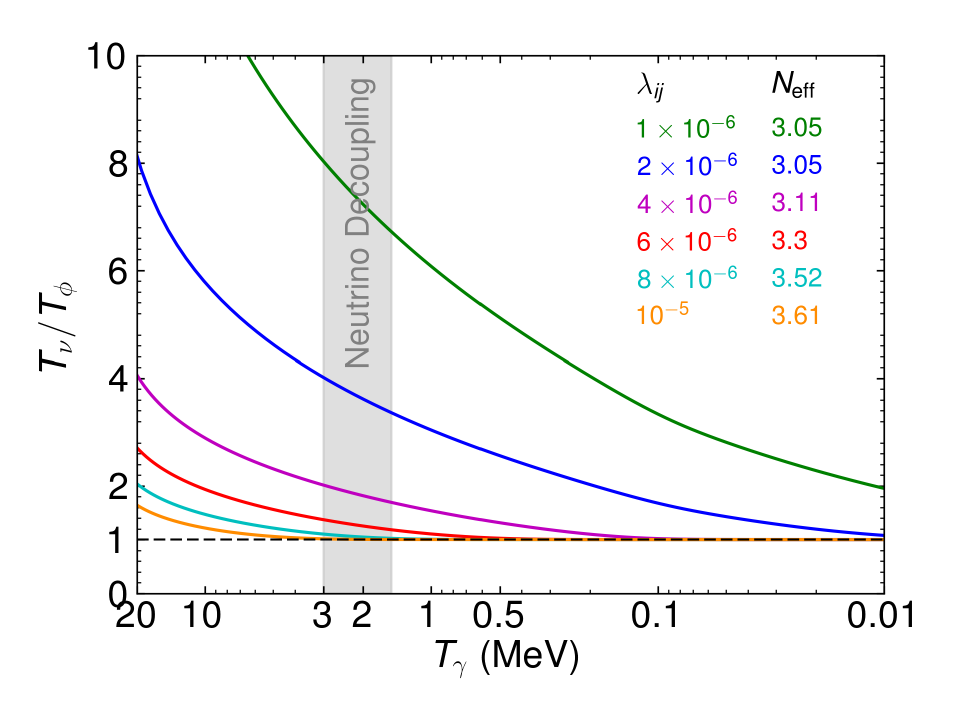}  
\vspace{-0.7cm}
\caption{Temperature evolution of $\phi$ as a result of $\bar{\nu}\nu \to \phi \phi $ interactions triggered by the neutrino decay interaction $\lambda_{ij}\, \phi\, \bar{\nu}_i\gamma_5 \nu_j$. The shaded region represents the epoch in the early Universe at which neutrinos decouple from the electromagnetic sector of the plasma. }\label{fig:T_evol}
\end{figure}

We will assume that all relevant species can be described by thermal distribution functions with negligible chemical potentials, and proceed as in~\cite{Escudero:2018mvt,Escudero:2019new}, to find the following temperature evolution equations:
\begin{subequations}\label{eq:T_evol}
\begin{align}
\frac{dT_\phi}{dt} &= - H \, T_\phi  + \frac{T_\phi}{4 \, \rho_\phi} \, \frac{\delta \rho_{\phi}}{\delta t}\, , \label{eq:T_evol_phi}\\
\frac{dT_\nu}{dt} &= - H \, T_\nu + \frac{T_\nu}{12 \, \rho_\nu} \left( \frac{\delta \rho_{\nu}^{\nu-e}}{\delta t} - \frac{\delta \rho_{\phi}}{\delta t} \right)  \, , \\
 \frac{dT_\gamma}{dt} &= - \frac{  4 H \rho_{\gamma} + 3 H \left( \rho_{e} + p_{e}\right) + 3 H \, T_\gamma \frac{dP_\text{int}}{dT_\gamma}+ \frac{\delta \rho_{\nu}^{\nu-e}}{\delta t} }{ \frac{\partial \rho_{\gamma}}{\partial T_\gamma} + \frac{\partial \rho_e}{\partial T_\gamma} +T_\gamma \frac{d^2 P_\text{int}}{dT_\gamma^2} } \,.
\end{align}
\end{subequations}
Where $\rho_i$, $p_i$ correspond to the energy density and pressure of a given species and their respective antiparticle. $H = \sqrt{8\pi \rho_{\rm total}/(3m_{\rm Pl}^2)}$ is the Hubble parameter with $m_{\rm Pl} = 1.22\times 10^{19}\,\text{GeV}$. $P_\text{int}$ and its derivatives take into account finite temperature corrections to the electromagnetic pressure and energy density, and $\delta \rho/\delta t$ are the energy density transfer rates, see~\cite{Escudero:2018mvt,Escudero:2019new} for details. The SM neutrino$\leftrightarrow$electron energy transfer rate, neglecting the electron mass, reads~\cite{Escudero:2019new}:
\begin{align}\label{eq:energyrates_nu_SM}
& \left. \frac{\delta \rho_{\nu}^{\nu-e}}{\delta t}  \right|_{\rm SM}^{\rm FD} = \frac{G_F^2}{\pi^5} \left( 3 - 4 s_W^2 + 24 s_W^4 \right) \times \\ \nonumber
& \left[ 32 \, f_a^{\rm FD} \,  \left( T_\gamma^9-T_{\nu}^9  \right) +  56 \,f_s^{\rm FD} \,   T_\gamma^4 \, T_{\nu}^4 \, \left( T_\gamma - T_{\nu}\right)\right] \,,
\end{align}
where $s_W^2 = 0.223$ is the $\sin^2$ of the Weinberg angle~\cite{pdg} and $G_F$ is Fermi's constant, $f_a^{\rm FD} = 0.884$ and $f_s^{\rm FD} = 0.829$, these two factors accounting for the Fermi-Dirac suppression of the rates.

\begin{figure}[t]
\centering
 \includegraphics[width=0.48\textwidth]{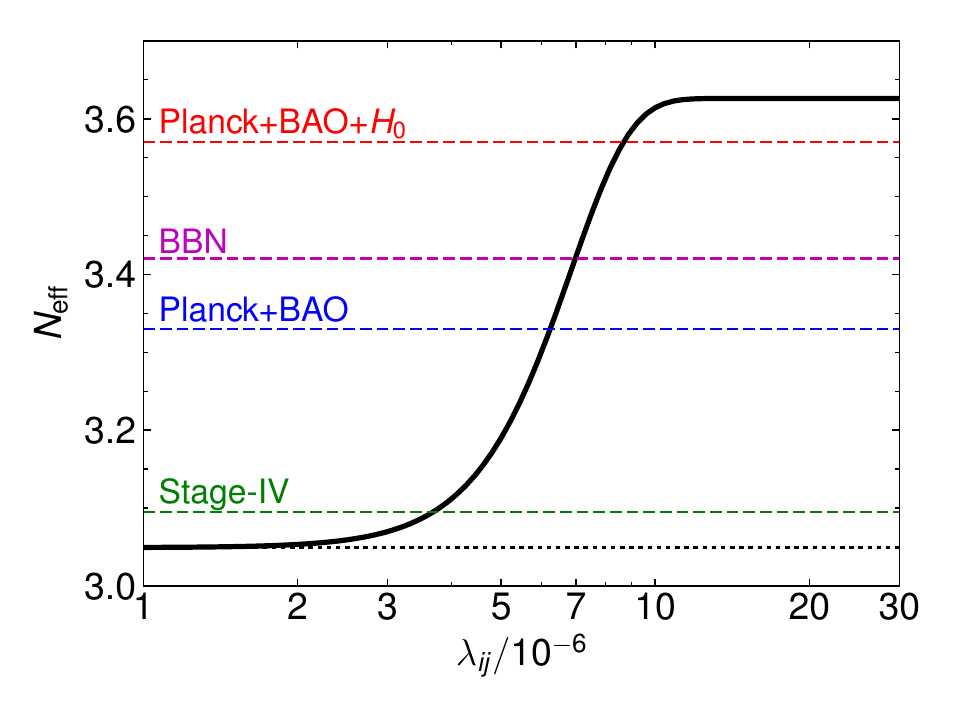}  
\vspace{-0.7cm}
\caption{$N_{\rm eff}$ as a function of the $\phi$-neutrino coupling constant, $\lambda_{ij}$ from equation~\eqref{eq:Lagrangian}. The horizontal dashed lines correspond to the $N_{\rm eff}$ + $2\sigma$ measurements from Planck~\cite{Aghanim:2018eyx} (within $\Lambda$CDM) and as inferred from the observed primordial nuclei abundances at the time of BBN~\cite{Pitrou:2018cgg}. We also show the expected sensitivity from Stage-IV CMB~\cite{Abazajian:2016yjj} experiments. }\label{fig:y_constraint}
\end{figure}

The neutrino-$\phi$ energy transfer rate takes into account the energy transfer resulting from the following processes $\bar{\nu} \nu \leftrightarrow \phi \phi$, $\nu \phi \leftrightarrow \nu \phi$, and $\nu_{i} \leftrightarrow \nu_j + \phi$. We have disregarded the scattering interactions since they are subdominant for massless species as compared to annihilations.  We also neglect the contribution from neutrino decays since the rate of neutrino decay $\Gamma \sim \lambda^2 m_\nu^2/T$ is not relevant for $T > 1\,\text{MeV}$ because it is tiny when compared to $2 \leftrightarrow 2$ processes, $\Gamma \sim \lambda^4 \, T$ because of neutrinos being highly boosted. Thus, the relevant energy transfer rate is given by annihilation processes, and reads (see Appendix~\ref{app:calculation} for the derivation):
\begin{align}\label{eq:energyrates_majoron}
\scalebox{1.07}[1.1]{$\frac{\delta \rho_\phi}{\delta t}  = \frac{ \sum_{i,j} \lambda_{ij}^4  }{ \pi ^5}\left[ T_\nu^5 \log \left(\frac{2 T_\nu}{m_\nu}\right) - T_\phi^5 \log \left(\frac{2 T_\phi}{m_\nu}\right) \right] \, .$}
\end{align}

We evolve the system of equations~\eqref{eq:T_evol} from $T_\gamma^0 = T_\nu^0 = 20\,\text{MeV}$ and $T_\phi^0 = T_\nu^0 \times 10^{-3} \times (\lambda_{ij}/10^{-6})$. $T_\phi^0 $ has been conservatively chosen to be at least a factor of 2 smaller than the one that can be obtained by integrating Eq~\eqref{eq:T_evol_phi} accounting only for $\bar{\nu}\nu \to \phi \phi$ from $T_\gamma = \infty$ until $T_\gamma^0 = 20\,\text{MeV}$. The temperature evolution for some values of the neutrino-$\phi$ coupling constant is displayed in Figure~\ref{fig:T_evol}. Notice that if $\lambda_{ij} > 8\times 10^{-6}$ a thermal population of $\phi$ particles will be produced at $T \gtrsim 3\,\text{MeV}$ which will yield $\Delta N_{\rm eff} \simeq 0.57$. Note also that for couplings $10^{-7} \lesssim \lambda_{ij} < 2\times 10^{-6}$, $\phi$ particles will thermalize with neutrinos but $N_{\rm eff}$ will be unaltered by entropy conservation. This can be appreciated from Figures~\ref{fig:T_evol} and~\ref{fig:y_constraint}.
 
In order to constrain the $\phi$-neutrino coupling, we shall use the latest constraints on $N_{\rm eff}$ as inferred from the measured primordial Helium and Deuterium abundances taken from the recent comprehensive analysis of Ref.~\cite{Pitrou:2018cgg} (see also~\cite{Cyburt:2015mya}). This analysis used $Y_\text{P} =0.2449\pm 0.0040 $~\cite{Aver:2015iza} and $\text{D}/\text{H} =(2.527\pm 0.030)\times 10^{-5}$~\cite{Cooke:2017cwo}. At 95.4\% CL, the $N_{\rm eff}$ constraint from BBN reads~\cite{Pitrou:2018cgg}:
\begin{align}\label{eq:Neff_Constraints}
N_{\rm eff} &= 2.88 \pm 0.54 \qquad 
\text{(BBN   at} \,\,\,\,\,95.4\%\,\text{CL})\,.
\end{align}

Note that within the neutrino decay scenario, $N_{\rm eff}$ is the same at the time of CMB formation and during BBN. This is because the $\phi$ population can only lead to a change in $N_{\rm eff}$ provided that it is generated before neutrino decoupling at $T > 2\,\text{MeV}$. Since the proton-to-neutron interactions freeze-out at $T_{p\to n}\sim 0.7\,\text{MeV}$, BBN occurs at $T_{\rm BBN} \sim 0.07\,\text{MeV} $~\cite{Sarkar:1995dd,Iocco:2008va,Pospelov:2010hj}, and recombination happens at $T_{\rm CMB} \sim 0.26 \,\text{eV}$, then this is clearly the case.
We show the resulting $N_{\rm eff}$ as a function of the value of the $\phi$-neutrino Yukawa coupling in Figure~\ref{fig:y_constraint}. The comparison between $N_{\rm eff}$ as a function of $\lambda_{ij}$ and that required for successful BBN results in the following constraint on $\lambda_{ij}$:
\begin{align}\label{eq:coup_constraints}
\lambda_{ij} &<  7.0\times 10^{-6}\, \qquad  \text{(BBN   at} \,\,\,\,\,95.4\%\,\text{CL})\,.
\end{align}

\begin{figure*}[t]
\centering
\begin{tabular}{cc}
\hspace{-0.4cm} \includegraphics[width=0.5\textwidth]{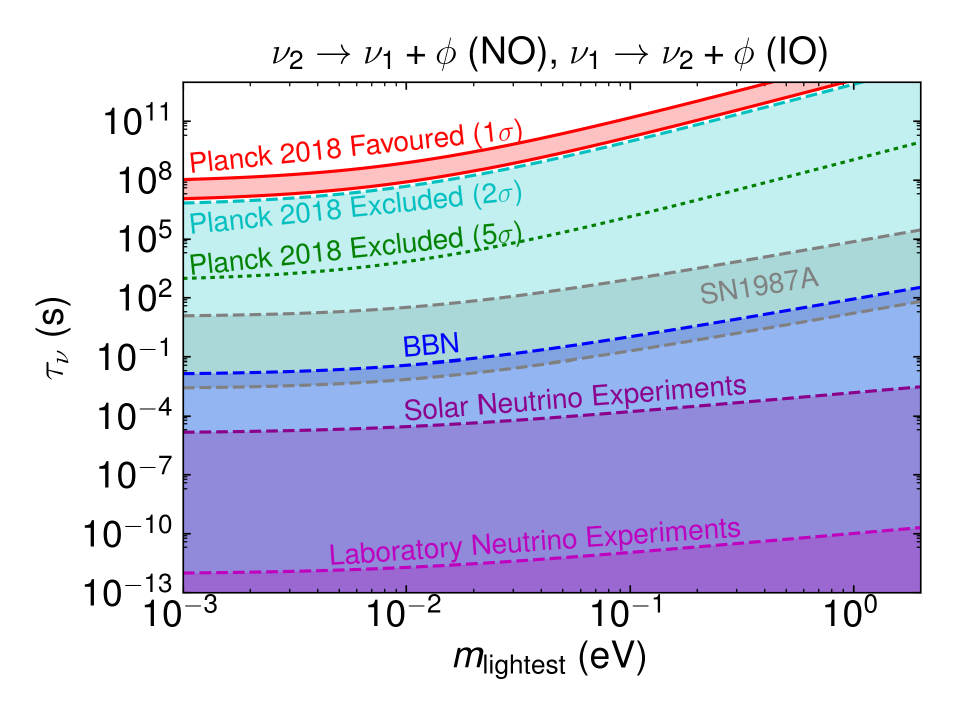}  & \hspace{-0.4cm} \includegraphics[width=0.5\textwidth]{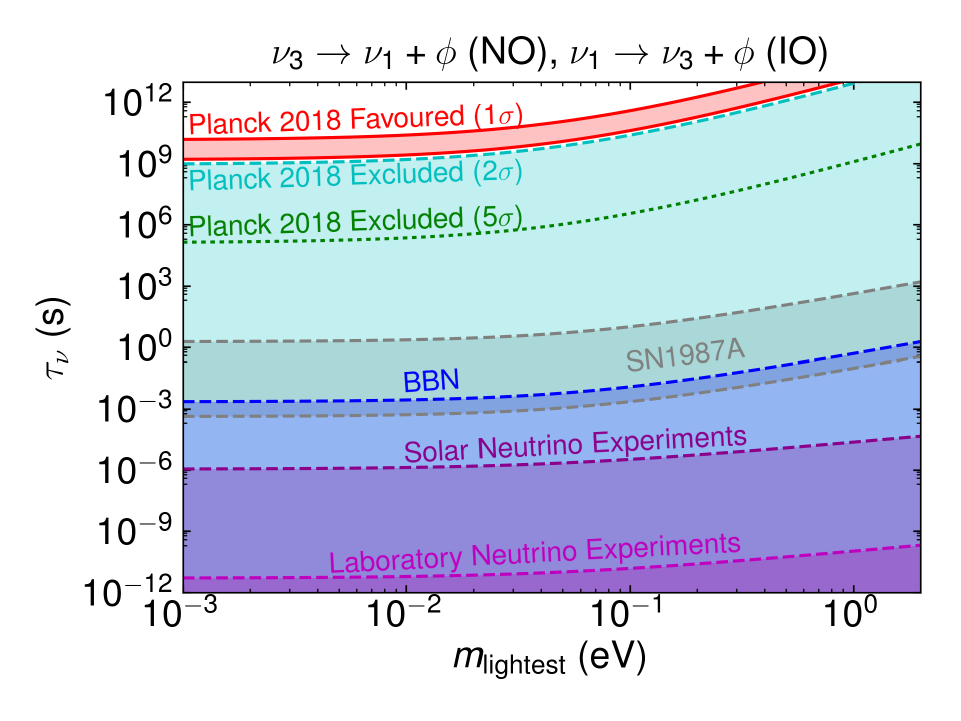} 
\end{tabular}
\vspace{-0.3cm}
\caption{Constraints on the lifetime of neutrino decay processes of the type $\nu_i \to \nu_j + \phi$, where $i,\,j$ label active neutrino mass eigenstates and $\phi$ is a massless sterile scalar. $m_{\rm lightest}$ corresponds to $m_{\nu_j}$. In the left panel the bounds are shown for $\nu_2 \to \nu_1 + \phi$ (NO) and $\nu_1 \to \nu_2 + \phi$ (IO) decay processes, while in the right panel we show the constraints for the $\nu_3 \to \nu_1 + \phi$ (NO) and $\nu_1 \to \nu_3 + \phi$ (IO) decay processes. The purple and magenta contours are ruled out by accelerator, long-baseline, and solar neutrino experiments~\cite{SNO_nudec,Berryman:2014qha,Funcke:2019grs,GonzalezGarcia:2008ru}. The grey area is excluded by SN1987A observations~\cite{Kachelriess:2000qc,Farzan:2002wx}. The blue contours correspond to the cosmological constraints obtained in this work by the requirement of successful BBN, see Section~\ref{sec:earlyUniverse}. The cyan contours correspond to the bounds obtained from the Planck 2018 CMB analysis, see Section~\ref{sec:CMB}. In addition, in red, we highlight the preferred region of parameter space by Planck 2018 CMB observations. }\label{fig:constraint_tau}
\end{figure*}

Finally, to translate the bound on the coupling into the neutrino decay lifetime, we need to specify the mass of one of the neutrinos in the decay process since only mass differences are known~\cite{Esteban:2018azc,deSalas:2017kay,Capozzi:2016rtj}. Therefore, our bound on $\tau_{\nu}$ depends upon the mass of one of the neutrinos in the process, and we choose this mass to be that of the final state neutrino $m_{\rm lightest}$. 

In Figure~\ref{fig:constraint_tau} we display the resulting constraint in the $\tau_{\nu}$-$m_{\rm lightest}$ plane for $\nu_2 \to \nu_1 + \phi$, $\nu_3 \to \nu_1 + \phi$ assuming NO and $\nu_1 \to \nu_3 + \phi$, $\nu_1 \to \nu_2 + \phi$ assuming IO, under the label BBN. We note that the constraints for $\nu_3 \to \nu_2 + \phi$ and $\nu_2 \to \nu_3 + \phi$ in the NO and IO respectively are of similar strength to those of $\nu_2 \to \nu_1 + \phi$ and $\nu_2 \to \nu_1 + \phi$. 
 
We therefore have shown that in order for a successful BBN, invisible neutrino decay modes of the type $\nu_i \to \nu_j + \phi$ (where $i,\,j$ represent massive neutrino states, and $\phi$ is a massless scalar) should have a lifetime
\begin{align}\label{eq:constraint_tau_BBN}
\tau_{\nu_i \to \nu_j + \phi} > 10^{-3}\,\text{s} \, \qquad  \text{(BBN   at} \,\,\,\,\,95.4\%\,\text{CL})\,.
\end{align}
This bound applies to any neutrino mass eigenstate (provided that the decay is kinematically accessible) and for both normal and inverted ordering. 

Supernova cooling can also be used to set constraints on the neutrino-$\phi$ coupling, and thereby on the neutrino decay lifetime. The agreement of SN1987A observations with supernova models excludes couplings in the range $3\times 10^{-7} \lesssim \lambda_{ij} \lesssim 2  \times 10^{-5} $ or $\lambda_{ij} \gtrsim 3\times 10^{-4}$~\cite{Kachelriess:2000qc,Farzan:2002wx}. This bound is shown in Figure~\ref{fig:constraint_tau} in grey. 

The bound of $\tau_{\nu_i \to \nu_j + \phi} > 10^{-3}\,\text{s}$ represents an improvement of 8 orders of magnitude as compared with constraints obtained from accelerator and long-baseline neutrino experiments~\cite{GonzalezGarcia:2008ru}. Separately, the bound of $\tau_{\nu_i \to \nu_j + \phi} > 10^{-3}\,\text{s}$ is still 2 orders of magnitude more stringent that those inferred from solar neutrino experiments~\cite{SNO_nudec,Berryman:2014qha,Funcke:2019grs}. However, in some regions of parameter space this BBN bound is less constraining than the bound that can be inferred from SN1987A observations~\cite{Kachelriess:2000qc,Farzan:2002wx}.

\newpage
%%%%%%%%%%%%%%%%%%%%%%%%%%%%%%%%%%%%%%%%%%%%%%%%%%%%%%
\subsection{Applicability of the BBN constraint}\label{sec:applicability}
%%%%%%%%%%%%%%%%%%%%%%%%%%%%%%%%%%%%%%%%%%%%%%%%%%%%%%
\vspace{-0.2cm}
%%%%%%%%%%%%%%%%%%%%%%%%%%%%%%%%%%%%%%%%%%%%%%%%%%%%%%
\subsubsection{Assumptions}\label{sec:assumptions_BBN}
%%%%%%%%%%%%%%%%%%%%%%%%%%%%%%%%%%%%%%%%%%%%%%%%%%%%%%
\vspace{-0.3cm}
Here we comment on how relaxing some of the assumptions that we made in order to obtain the constraint on the neutrino lifetime of $\tau_\nu > 10^{-3}\,\text{s} $~\eqref{eq:constraint_tau_BBN} from BBN could affect them, and we argue that they cannot be significantly altered.\vspace{-0.1cm}
\begin{enumerate}[leftmargin=0.5cm,itemsep=0pt]
\item {\bf Majorana-Dirac}: For a given neutrino decay rate, the annihilation cross section for Dirac neutrinos is $1/2$ that of Majorana neutrinos, since the neutrinos are not their antiparticles. Therefore, the constraint on $\lambda$ should be relaxed by a factor of $2^{1/4} \simeq 1.2$ in the Dirac case. And therefore, the constraint on the lifetime should naively be relaxed by a factor $\sqrt{2}$. However, if neutrinos are Dirac, the $\nu-\phi$ interaction will lead also to a thermal population of massless right handed neutrinos and $\Delta N_{\rm eff}$ will greatly exceed 0.57, which will result in an even tighter constraint. 
\item {\bf $\phi$ mass}: Regardless of what the mass of the $\phi$ scalar is, if the $\phi$ scalar is light enough to be in the neutrino decay final state, then its mass is negligible in the early Universe ($T \gtrsim 1\,\text{MeV}$) and therefore $m_\phi$ will not impact the annihilation rate. The mass may change the decay width at rest, however, the phase space suppression will be $\mathcal{O}(1)$ unless $m_\phi$ is very fine tuned $m_\phi \simeq m_{\nu_i}-m_{\nu_j}$. Hence, a non-negligible $m_\phi$ will not impact our conclusions. 
\end{enumerate}
\vspace{-0.69cm}
%%%%%%%%%%%%%%%%%%%%%%%%%%%%%%%%%%%%%%%%%%%%%%%%%%%%%%
\subsubsection{Other scenarios}\label{sec:scenarios_BBN}
%%%%%%%%%%%%%%%%%%%%%%%%%%%%%%%%%%%%%%%%%%%%%%%%%%%%%%
\vspace{-0.2cm}
Here we comment how the BBN constraint of $\tau_\nu > 10^{-3}\,\text{s} $~\eqref{eq:constraint_tau_BBN} applies to other particle physics scenarios in which the decay is not necessarily $\nu_i \to \nu_j +\phi$. \vspace{-0.1cm}
\begin{enumerate}[leftmargin=0.5cm,itemsep=0pt]
\item $\nu_i \to \nu_j + Z'$. Invisible neutrino decays also generically result from vector mediated neutrino self-interactions~\cite{He:1990pn,He:1991qd,Farzan:2015hkd,Babu:2017olk,Escudero:2019gzq}, provided that $m_{Z'} < |m_{\nu_3} -m_{\nu_1}| \simeq 0.05\,\text{eV}$. For such types of models, our bounds still apply since the presence of a thermal population of very light $Z'$s prior to neutrino decoupling would render $\Delta N_{\rm eff} = 1.71$, a value which is clearly excluded by CMB observations and successful BBN~\eqref{eq:Neff_Constraints}. In addition, as a result of processes of the type $\ell^+ \ell^- \to \gamma \, Z'$~\cite{Escudero:2019gzq}, coupling constants of $\mathcal{O}(10^{-8})$ would be ruled out for $m_{Z'} < |m_{\nu_3} -m_{\nu_1}|$ and hence $\tau_\nu/m_\nu \gtrsim \mathcal{O}(10^{4})\,\text{s}\,\text{eV}^{-1}$. 

\item $\nu_i \to \nu_4 + \phi$. If one of the light massive eigenstates decays into a scalar plus a fourth very light sterile neutrino (that has very small mixing with $\nu_{e,\, \mu,\, \tau}$), then $\tau_{\nu} >10^{-3}\,\text{s}$ since within this scenario $\Delta N_{\rm eff}$ could be as large as $\Delta N_{\rm eff} = 1.57$ at the time of BBN, which is again clearly excluded by current data~\eqref{eq:Neff_Constraints}.

\item $\nu_4 \to \nu_1 + \phi$. This scenario will be ruled out for $\tau_{\nu_4} < 10^{-3}\,\text{s}$ since the same interactions that trigger the 4th neutrino decay will render a thermal population of $\nu_4$ and $\phi$ particles, thereby rendering $\Delta N_{\rm eff} = 1.57$, which is incompatible with a successful BBN. Note that this bound will apply for $m_{\nu_4} \lesssim 1 \,\text{MeV}$.

\end{enumerate}

\begin{figure*}[t]
\centering
\begin{tabular}{cc}
\includegraphics[width=0.48\textwidth]{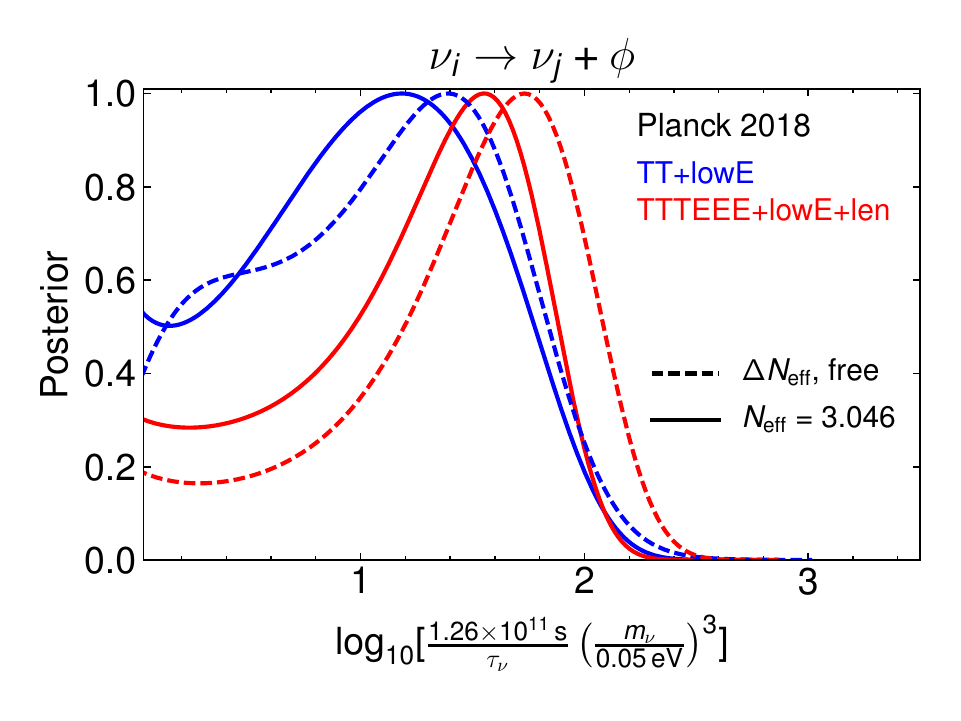}   & \includegraphics[width=0.48\textwidth]{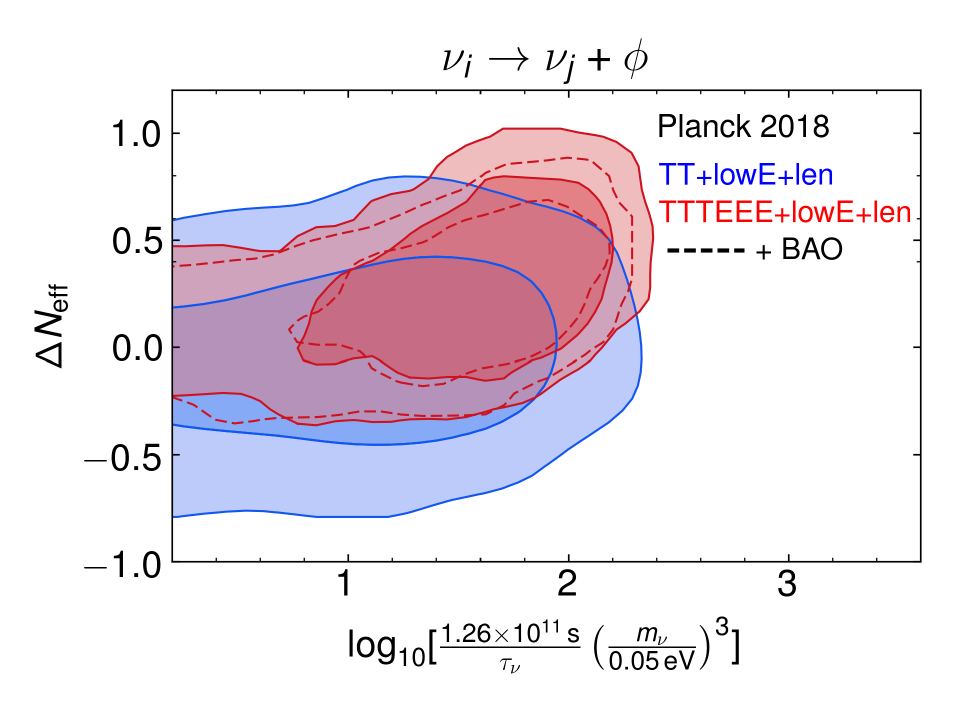}  \\ 
\includegraphics[width=0.48\textwidth]{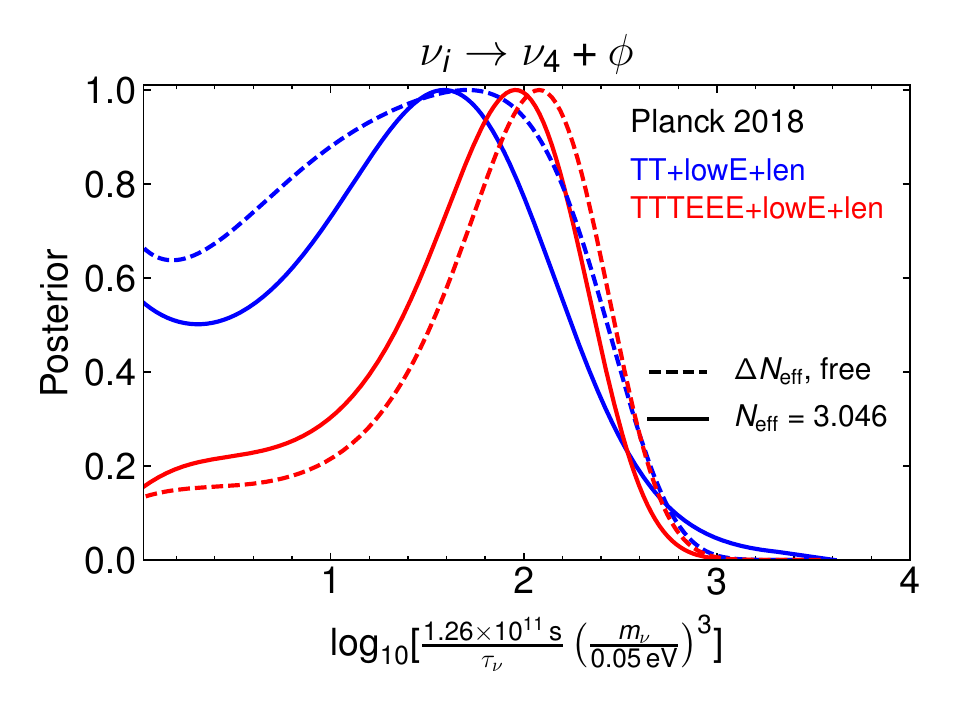}   & \includegraphics[width=0.48\textwidth]{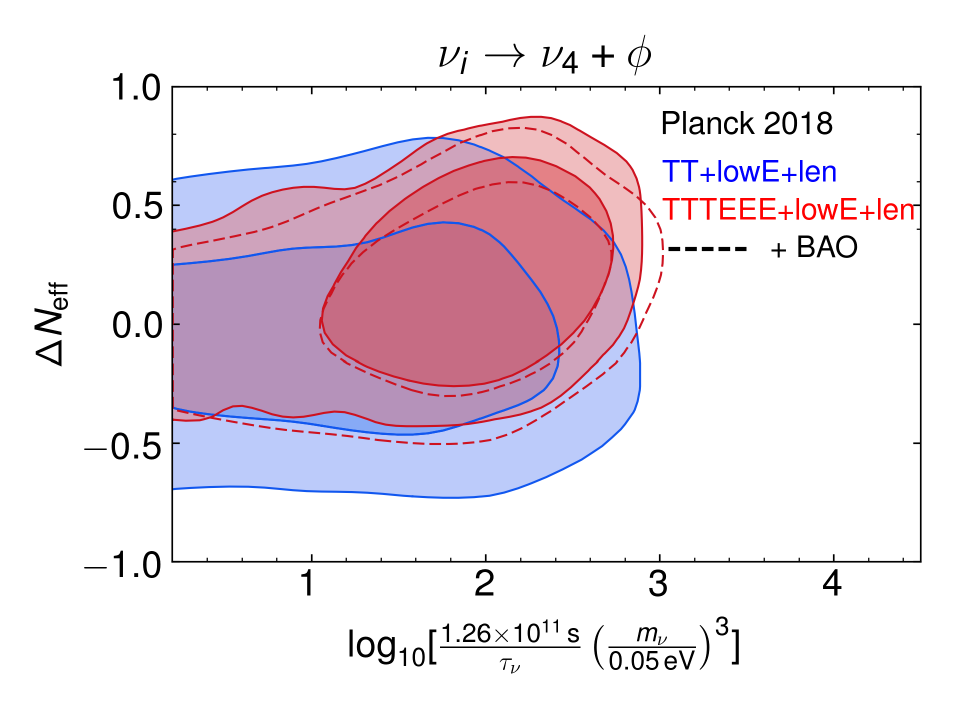} 
\end{tabular}
\vspace{-0.4cm}
\caption{Marginalized posterior distributions from the analysis to the \texttt{Planck 2018 TT+lowE+lensing} data (blue) and the \texttt{Planck 2018 TTTEEE+lowE+lensing} (red). The \textit{upper panels} correspond to the neutrino decay scenario $\nu_i \to \nu_j + \phi$, and the \textit{lower panels} correspond to the neutrino decay scenario $\nu_i \to \nu_4 +\phi$. The \textit{left panels} correspond to the one dimensional posterior of $\log_{10}\left[{1.26 \times 10^{11}\,\text{s}}/{\tau_\nu} \left({m_\nu}/{0.05\,\text{eV}}\right)^{3}\right]$. The \textit{right panels} correspond to the two dimensional posterior in the $\Delta N_{\rm eff}$-$\log_{10}\left[{1.26 \times 10^{11}\,\text{s}}/{\tau_\nu} \left({m_\nu}/{0.05\,\text{eV}}\right)^{3}\right]$ plane. The dashed lines in the right hand panels correspond to the contours including BAO data.  }\label{fig:Posterior}
\end{figure*}

%%%%%%%%%%%%%%%%%%%%%%%%%%%%%%%%%%%%%%%%%%%%%%%%%%%%%%
\section{CMB constraints}\label{sec:CMB}
%%%%%%%%%%%%%%%%%%%%%%%%%%%%%%%%%%%%%%%%%%%%%%%%%%%%%%
If neutrinos decay efficiently while still relativistic into other massless species, the decay process will effectively make the neutrino fluid no longer free-streaming~\cite{Hannestad:2004qu,Hannestad:2005ex}. In particular, neutrino decays will erase the neutrino anisotropic stress that otherwise arises in the course of expansion in a purely non-interacting massless fluid~\cite{Weinberg:2003ur,Ma:1995ey}. In this section, we describe how we implement the effect of neutrino decays upon the neutrino cosmological perturbations and use the latest public CMB measurements by the Planck satellite to set constraints on invisible neutrino decays.

%%%%%%%%%%%%%%%%%%%%%%%%%%%%%%%%%%%%%%%%%%%%%%%%%%%%%%
\subsection{Modeling neutrino decays}\label{sec:CMB_modelling}
%%%%%%%%%%%%%%%%%%%%%%%%%%%%%%%%%%%%%%%%%%%%%%%%%%%%%%
We follow Ref.~\cite{Hannestad:2005ex} in order to calculate the effective neutrino decay rate that erases the neutrino anisotropic stress, $\left< \Gamma \right>$. Ref.~\cite{Hannestad:2005ex} argues that $\left< \Gamma \right> \simeq \Gamma \, (m_\nu/E_\nu)^3$, which by thermally averaging $\left< m/E\right>$ and $\left< m^2/E^2\right>$ separately yields:
\begin{align}
\left< \Gamma \right> = \Gamma  \frac{1}{4}\left(\frac{m_\nu}{T_\nu}\right)^3\,.
\end{align}
Written as a function of the scale factor $a$, the neutrino lifetime and the neutrino mass, reads:
\begin{align}\label{eq:gamma_decay}
\scalebox{1.15}[1.15]{$\left< \Gamma \right> = \frac{4.4}{\text{Mpc}} \frac{1.26\times 10^{11}\,\text{s}}{\tau_\nu} \left(\frac{m_\nu}{ 0.05\,\text{eV} }\right)^3 \left(1073\, \frac{a}{a_0}\right)^3 \,,$}
\end{align}
where $a_0 = 1$. 

In order to account the effect of neutrino decays in the neutrino cosmological perturbations, we follow the relaxation time approximation for the neutrino collision term~\cite{Hannestad:2000gt}. This approximation amounts to modifying the massless neutrino Boltzmann hierarchy for the perturbed neutrino phase space in the following manner:
\begin{align}\label{eq:Perturbations}
\left. \frac{{F}_{\nu\,\ell}}{dt}\right|^{\nu-{\rm decay}} &= \left.\frac{{F}_{\nu\,\ell}}{dt}\right|^{\rm SM}  -   \left< \Gamma \right>  {F}_{\nu\,\ell}  \,, \,\, \ell \geq 2 \,,
\end{align}
where $t$ is proper time and ${F}_{\nu\,\ell}$ represents the contribution from the $\ell$th Legendre polynomial to the perturbed neutrino phase space distribution~\cite{Ma:1995ey}. The neutrino fluid is regarded as the neutrinos plus the massless species produced in the decay. We implement equation~\eqref{eq:Perturbations} in the cosmological Boltzmann code \texttt{CLASS}~\cite{Blas:2011rf,Lesgourgues:2011re}. For simplicity, we assume that neutrinos are massless since given Planck 2018 constraints~\cite{Aghanim:2018eyx} (see also~\cite{Vagnozzi:2019utt,Vagnozzi:2018jhn,Vagnozzi:2017ovm,Choudhury:2018byy}), $\sum m_\nu < 0.12$ at 95\% CL, and therefore neutrinos decay while relativistic for the relevant cosmological evolution considered in this study.  
 
 \begin{table*}
\begin{center}
{\def\arraystretch{1.2}
\begin{tabular}{l|l|ccc|ccc|c|c}
\hline\hline
& Parameter  &\multicolumn{3}{c|}{$\log_{10}\left(\Gamma_{\rm eff}\right)$}  & \multicolumn{3}{c|}{$\tau_\nu/(10^{9}\,s) \cdot  \left({m_\nu}/{ 0.05\,\text{eV} }\right)^{-3} $}  & $\Delta N_{\rm eff}$ & $\sqrt{\chi^2_{\Lambda {\rm CDM}} - \chi^2_{\nu}}$ \\ \hline
$\,$ Data &Scenario & $\,\,\,\,$ BF $\,\,\,$ & 68\% CL &  95\% CL   & $\,\,\,\,$ BF $\,\,\,$ & 68\% CL &  95\% CL  & 68\% CL & BF \\ \hline
\multirow{4}{*}{ \rotatebox[origin=c]{90}{ \texttt{Planck2018}}   $ \,$  \rotatebox[origin=c]{90}{ \texttt{TTTEEE+lowE}}}  $ \,$   %LCDM R 0.013 1384.54
&$\nu_{i} \to \nu_j + \phi$ 				& 1.41   & $1.2^{+0.7}_{-0.3}$   & $< 2.0$   & 4.9   & $7.7^{+8.5}_{-6.0} $ &  $>1.3$  &  - & 1.6  \\  % R 0.025 1383.21
&$\nu_{i} \to \nu_j + \phi$, +$\Delta N_{\rm eff}$ &  1.71  & $1.4^{+0.7}_{-0.3}$  & $<2.3$  &  2.4  & $4.6^{+4.0}_{-3.7} $ & $>0.6$   & $0.23\pm 0.29$  &1.3  \\  % R 0.035 1383.65
&$\nu_{i} \to \nu_4 + \phi$ 					& 2.01   & $1.6^{+0.8}_{-0.4}$  & $< 2.6$  & 1.2    & $3.0^{+3.8}_{-2.5}$   &  $> 0.3$   & - & 1.6  \\  % R 0.02 1383.24
&$\nu_{i} \to \nu_4 + \phi$, +$\Delta N_{\rm eff}$ &  1.68  & $1.3^{+0.8}_{-0.8}$   &  $< 2.5$ &  2.6  & $5.9^{+30}_{-4.9}$   & $> 0.4$   & $0.14\pm 0.28$ & 1.4  \\  % R 0.03 1383.62
  \hline 
   \multirow{4}{*}{ \rotatebox[origin=c]{90}{ \texttt{Planck2018}}   $ \,$  \rotatebox[origin=c]{90}{ \texttt{TT+lowE}}}  $ \,$    %LCDM R 0.03 594.83
&$\nu_{i} \to \nu_j + \phi$ 				  &  1.69   & $1.0^{+0.7}_{-0.5}$   & $< 1.8$   & 2.6   & $12^{+23}_{-9} $ &  $>1.8$  &  - &   1.1 \\  %R 0.05 594.19
&$\nu_{i} \to \nu_j + \phi$, +$\Delta N_{\rm eff}$ & 1.31  & $1.1^{+0.7}_{-0.5}$  & $<2.0$  &  6.1  & $11^{+25}_{-8} $ & $>1.4$   & $-0.05\pm 0.27$   & 1.5\\   % R 0.045 593.63
 &$\nu_{i} \to \nu_4 + \phi$ 				  & 1.51   & $1.4^{+0.9}_{-0.7}$  & $< 2.5$  & 3.9    & $5.4^{+20}_{-4.6}$   &  $> 0.4$   & - & 0.8   \\  %R 0.04 594.46
&$\nu_{i} \to \nu_4 + \phi$, +$\Delta N_{\rm eff}$   & 1.68  & $1.3^{+0.8}_{-0.8}$   &  $< 2.5$ &  2.6  & $5.9^{+30}_{-4.9}$   & $> 0.4$   & $-0.14\pm0.29$& 0.6   \\  %R 0.04 594.65
\hline
  \multirow{4}{*}{ \rotatebox[origin=c]{90}{ \texttt{Planck2018}}   $ \,$  \rotatebox[origin=c]{90}{ \texttt{+BAO}}}  $ \,$   %LCDM R 0.01 1387.19
&$\nu_{i} \to \nu_j + \phi$ 				& 1.60   & $1.2^{+0.7}_{-0.4}$   & $< 2.0$   & 3.1   & $8.1^{+11}_{-6.5} $ &  $>1.2$  &  - & 1.8  \\   % R 0.027 1385.59
&$\nu_{i} \to \nu_j + \phi$, +$\Delta N_{\rm eff}$ &  1.82  & $1.4^{+0.6}_{-0.3}$  & $<2.2$  &  1.9  & $5.0^{+4.0}_{-3.8} $ & $>0.8$   & $0.19\pm 0.24$  &2.0  \\   % R 0.01 1385.16
&$\nu_{i} \to \nu_4 + \phi$ 					& 1.90   & $1.6^{+0.8}_{-0.3}$  & $< 2.6$  & 1.6    & $2.9^{+3.1}_{-2.4}$   &  $> 0.3$   & - & 2.1  \\  % R 0.02 1384.89
&$\nu_{i} \to \nu_4 + \phi$, +$\Delta N_{\rm eff}$ &  1.60  & $1.8^{+0.7}_{-0.3}$   &  $< 2.7$ &  3.1  & $2.2^{+2.2}_{-1.8}$   & $> 0.2$   & $0.09\pm 0.25$ & 1.9  \\   % R 0.036 1385.37
  \hline 
  \hline \hline
\end{tabular}
}
\end{center}\vspace{-0.5cm}
\caption{Marginalized constraints from Planck 2018 CMB observations for $\log_{10}\left(\Gamma_{\rm eff}\right)$, $\Delta N_{\rm eff}$, and those inferred for the neutrino lifetime $\tau_\nu$ from equation~\eqref{eq:Gamma_eff}. Note that we also account for Planck 2018 lensing measurements. }\label{tab:summary_constraints_tau}
\end{table*}

%%%%%%%%%%%%%%%%%%%%%%%%%%%%%%%%%%%%%%%%%%%%%%%%%%%%%%
\subsection{CMB Analysis}\label{sec:CMB_modelling}
%%%%%%%%%%%%%%%%%%%%%%%%%%%%%%%%%%%%%%%%%%%%%%%%%%%%%%
In order to test the neutrino decay hypothesis with CMB observations we use the latest public CMB data from the Planck satellite~\cite{Aghanim:2018eyx,Aghanim:2019ame}\footnote{In Appendix~\ref{app:cosmo_full_2015} we consider the constraints that can be inferred from Planck 2015 data. We find very small differences between the results that can be inferred from Planck 2015 and 2018 data.}. In particular, we use both the high-$\ell$ Planck 2018 temperature and polarization spectra, the low-$\ell$ and low-E temperature and polarization spectra (we shall collectively call this data set combination lowE), and also the lensing measurements from the 2018 data release~\cite{Aghanim:2019ame}. We consider the following data set combinations \texttt{Planck 2018 TT+lowE+lensing} and \texttt{Planck 2018 TTTEEE+lowE+lensing}.

To perform the CMB analysis, since $\left< \Gamma \right> $ is the quantity that directly enters the Boltzmann hierarchy, we define
\begin{align}\label{eq:Gamma_eff}
\Gamma_{\rm eff} \equiv \frac{1.26\times 10^{11}\,\text{s}}{\tau_\nu} \left(\frac{m_\nu}{ 0.05\,\text{eV} }\right)^3 \,,
\end{align}
and we use a logarithmic prior on $\Gamma_{\rm eff}$ over the range $[10^0,10^8]$. Converting a constraint on $\Gamma_{\rm eff}$ into a constraint on the neutrino lifetime is trivial by using equation~\eqref{eq:Gamma_eff}. For the rest of the cosmological and nuisance parameters we use the same priors as the Planck collaboration in their 2018 base $\Lambda$CDM analysis~\cite{Aghanim:2018eyx,Aghanim:2019ame}.

In order to be maximally conservative, we consider several decay scenarios and also consider to which extent the presence of additional non-interacting massless species -- encoded in terms of $\Delta N_{\rm eff}$ -- can alter the invisible neutrino decay constraints. 

We consider the same decay scenario as in Section~\ref{sec:NeutrinoDecay_model} in which one active massive neutrino decays into another one by emitting a massless scalar particle $\phi$; namely, $\nu_i \to \nu_j + \phi$. Within this scenario, the number of interacting neutrino species is $N_{\rm int} = 2$, while the other neutrino simply free-streams. We consider another scenario in which an active neutrino decays into a sterile and very light neutrino $\nu_4$ by emitting a massless scalar field $\phi$; namely, $\nu_i \to \nu_4 +\phi$.  In this scenario the number of interacting neutrino species is $N_{\rm int} = 1$ while we consider the other two active neutrino species to be non-interacting and therefore purely free-streaming. We contrast both scenarios by varying $\Gamma_{\rm eff}$ and also $\Delta N_{\rm eff}$, for which we use a linear prior in the range $[-1,9]$\footnote{Note that an scenario with a negative $\Delta N_{\rm eff}$ will only correspond to a Universe with a very low reheating temperature, see e.g.~\cite{Kawasaki:2000en}, or with very light and electrophilic species in thermal equilibrium at the time of neutrino decoupling, see e.g.~\cite{Escudero:2018mvt}.}. We perform a Monte Carlo Markov Chain (MCMC) analysis using \texttt{MontePython-v3}~\cite{Brinckmann:2018cvx,Audren:2012wb} and we quote results of analyses in which the maximum Gelman-Rubin coefficient~\cite{Gelman:1992zz} for any parameter is $R-1 < 0.05$.

%%%%%%%%%%%%%%%%%%%%%%%%%%%%%%%%%%%%%%%%%%%%%%%%%%%%%%
\subsection{Planck 2018 Constraints}\label{sec:CMB_results}
%%%%%%%%%%%%%%%%%%%%%%%%%%%%%%%%%%%%%%%%%%%%%%%%%%%%%%
In the left panel of Figure~\ref{fig:Posterior}, we display the marginalized posterior distribution of the parameter $\Gamma_{\rm eff}$, which is directly related to the neutrino lifetime~\eqref{eq:Gamma_eff}. In the right panel of Figure~\ref{fig:Posterior}, we show the two-dimensional marginalized posterior between $\Gamma_{\rm eff}$ and $\Delta N_{\rm eff}$. It is obvious that the two parameters are not degenerate and from the left panel of Figure~\ref{fig:Posterior} we notice that the posterior distributions for both a varying $\Delta N_{\rm eff}$ and when it is fixed are fairly similar. 

In Table~\ref{tab:summary_constraints_tau} we quote the best fit, mean 68\% CL error bars and 95\% CL exclusions for the parameter $\Gamma_{\rm eff}$ and for the invisible neutrino decay lifetime. The reader is deferred to Table~\ref{tab:nudecay_constraints} in Appendix~\ref{app:cosmo_params} where we quote the mean and 68\% error bars for the standard cosmological parameters too. From Table~\ref{tab:summary_constraints_tau} we clearly appreciate that the derived limits from the \texttt{Planck 2018 TTTEEE+lowE+lensing} dataset are less stringent than those from the \texttt{Planck 2018 TT+lowE+lensing} dataset. This is essentially because when including high-$\ell$ polarization data there is a 1-2$\sigma$ preference for a non-infinite invisible neutrino decay lifetime. We therefore choose the \texttt{Planck 2018 TTTEEE+lowE+lensing} analysis to quote both 95\% CL upper and lower limits and $\pm$ 68 \% CL measurements. 

We show that Planck 2018 CMB observations bound the lifetime of neutrino decay processes like $\nu_i \to \nu_j + \phi$ to be \vspace{-0.3cm}
\begin{align}\label{eq:1to2_cons}
\tau_{\nu_i \to  \nu_j + \phi} >  1.3\times 10^{9}\,\text{s}  \, \left(\frac{m_{\nu_i}}{ 0.05\,\text{eV} }\right)^3 \qquad \text{Planck}\,,
\end{align}
at 95\% CL.
The lower bound on the neutrino lifetime of the decay mode of the type $\nu_i \to \nu_4 + \phi$, where $\nu_4$ is a very light and sterile neutrino, at 95\% CL reads:
\begin{align}\label{eq:1to4_cons}
\tau_{\nu_i \to \nu_4 + \phi} > 0.4 \times 10^{9}\,\text{s} \, \left(\frac{m_{\nu_i}}{ 0.05\,\text{eV} }\right)^3 \qquad \text{Planck}\,.
\end{align}
Furthermore, we also perform analyses allowing for an additional massless and non-interacting contribution to the energy density of the Universe, encoded in terms of $\Delta N_{\rm eff}$. We find that, when letting $\Delta N_{\rm eff} $ vary, the bounds are only slightly relaxed, and at 95\% CL read: 
\begin{align}
\tau_{\nu_i \to \nu_j + \phi} &> 0.6\times 10^{9}\,\text{s}  \, \left(\frac{m_{\nu_i}}{ 0.05\,\text{eV} }\right)^3\qquad \text{Planck}\,, \\
\tau_{\nu_i \to \nu_4+ \phi} &> 0.4\times 10^{9}\,\text{s} \, \left(\frac{m_{\nu_i}}{ 0.05\,\text{eV} }\right)^3\qquad \text{Planck}\,,
\end{align}
and hence are barely affected by an additional contribution to $N_{\rm eff}$ from massless non-interacting species.

The results from the \texttt{Planck 2018 TTTEEE+lowE+lensing} show a $\sim 1-2\,\sigma$ preference for invisibly decaying neutrinos. At 68\% CL the neutrino lifetimes are bounded to be
\begin{align}\label{eq:detection}
\tau_{\nu_i \to  \nu_j + \phi} &=7.7^{+8.5}_{-6.0}  \,\times 10^{9}\,\text{s}  \, \left(\frac{m_{\nu_i}}{ 0.05\,\text{eV} }\right)^3 \,,\\
\tau_{\nu_i \to \nu_4+\phi} &=3.0^{+3.8}_{-2.5}  \,\times 10^{9}\,\text{s}  \, \left(\frac{m_{\nu_i}}{ 0.05\,\text{eV} }\right)^3 \,.
\end{align}

Finally, in order to highlight the constraining power of Planck CMB observations on invisible neutrino decays, we study how much the fit to the Planck 2018 data is degraded in a scenario with a neutrino decay lifetime accessible to neutrino experiments via the observation of the next galactic supernova. 

Neutrino experiments should be sensitive to neutrino decays in a galactic supernova signal for 
\begin{align}
\tau_{\rm SN} \sim \frac{m_\nu}{E} \, D_{\rm SN} \sim 10^4\,\text{s} \, \frac{10\,\text{MeV}}{E}  \, \frac{m_\nu}{0.1\,\text{eV}} \, \frac{D_{\rm SN}}{10\,\text{kpc}}\,,
\end{align}
where $E \sim 10\,\text{MeV}$ is the typical mean energy of the neutrinos emitted and $D_{\rm SN}$ is the distance to the supernova.

We run a MCMC fixing $ \tau_\nu =  1.3\times 10^{5}\,\text{s} \, \left({m_\nu}/{ 0.05\,\text{eV} }\right)^3$ and allowing to vary the six standard cosmological parameters, the Planck nuisance parameters, and also $\Delta N_{\rm eff}$. For the \texttt{Planck 2018 TTTEEE+lowE+lensing} data set, we find that the best-fit points have a higher minimum $\chi^2$, as compared to $\Lambda$CDM, of:
\begin{align}% 1nudecay chi2/2 = 1388.66  2nudecay chi2/2 = 1405.12 
\sqrt{\chi^2_{\nu_i \to \nu_j +\phi}-\chi^2_{\Lambda {\rm CDM}}} = 6.4\, , \\
\sqrt{\chi^2_{\nu_i \to \nu_4+\phi}-\chi^2_{\Lambda {\rm CDM}}} = 2.9.
\end{align}
Within Gaussian statistics, these results demonstrate that fast neutrino decays at a rate of $\tau_\nu =  1.3\times 10^{5}\,\text{s} \, \left({m_\nu}/{ 0.05\,\text{eV} }\right)^3  $ are clearly disfavoured by Planck CMB observations with a $6.4\,\sigma$ and $2.9\,\sigma$ significance for $\nu_i \to \nu_j +\phi$ and $\nu_i \to \nu_4+\phi$ decays respectively.

%%%%%%%%%%%%%%%%%%%%%%%%%%%%%%%%%%%%%%%%%%%%%%%%%%%%%%
\subsection{Including BAO data}\label{sec:BAO_results}
%%%%%%%%%%%%%%%%%%%%%%%%%%%%%%%%%%%%%%%%%%%%%%%%%%%%%%
We have further considered the joint impact of Baryon Acoustic Oscillation (BAO) and Planck 2018 data on invisible neutrino decays. Although BAO data are not directly sensitive to the reduction of the neutrino anisotropic stress induced by neutrino decays, BAO data can be used to narrow down other cosmological parameters and to reduce degeneracies in the CMB fit. We consider the same BAO data as the Planck collaboration in their 2018 analysis~\cite{Aghanim:2018eyx}. Namely, we use: data from the 6dF Galaxy Survey~\cite{Beutler:2011hx}, the Main Galaxy Sample of SDSS~\cite{Ross:2014qpa} and DR12 of BOSS~\cite{Alam:2016hwk}. 

We combine BAO data with the full Planck data to form the \texttt{Planck 2018 TTTEEE+lowE+lensing+BAO} data set. As can be appreciated in Table~\ref{tab:summary_constraints_tau}, we find very similar constraints on $\tau_\nu$ as when only Planck data is considered. From the last column of Table~\ref{tab:summary_constraints_tau} we see that the significance of invisible neutrino decays is slightly enhanced as compared with Planck data alone. This results from the help of BAO data in breaking degeneracies in other cosmological parameters. 

Therefore, the inclusion of BAO data to Planck 2018 CMB observations does not significantly alter bounds we report on invisible neutrino decays and renders a slight enhancement of the significance of invisible neutrino decays. 

%%%%%%%%%%%%%%%%%%%%%%%%%%%%%%%%%%%%%%%%%%%%%%%%%%%%%%
\section{Summary and Discussion}\label{sec:conclusions}
%%%%%%%%%%%%%%%%%%%%%%%%%%%%%%%%%%%%%%%%%%%%%%%%%%%%%%

In this work we have revisited the cosmological constraints on invisible neutrino decay modes in light of the rather weak constraints from solar, atmospheric and long-baseline neutrino experiments. Collectively, we have exploited cosmological observations to place stringent constraints on invisible neutrino decays. See Ref.~\cite{Archidiacono:2013dua} for the previous CMB analysis. Figure~\ref{fig:constraint_tau} highlights the main constraints on invisible neutrino decays derived in this work. 

In summary, the main results obtained in this paper are:
\begin{enumerate}[leftmargin=0.77cm]
\item The invisible neutrino decay lifetime should be $\tau_{\nu} > 10^{-3} \, \text{s}$ at 95\% CL in order for the primordial elements to be synthesized successfully. In addition, we have discussed other neutrino decay scenarios for which it applies beyond $\nu_i \to \nu_j + \phi$ decays. 
\item Planck 2018 observations set stringent constraints on invisible neutrino decays. We have found that $\tau_{\nu_i \to \nu_j + \phi} >  1.3\times 10^{9}\,\text{s}  \, \left({m_{\nu_i}}/{ 0.05\,\text{eV} }\right)^3$ and that $\tau_{\nu_i \to \nu_4+ \phi} > 0.4\times 10^{9}\,\text{s} \, \left({m_{\nu_i}}/{ 0.05\,\text{eV} }\right)^3$, both at 95\% CL. The bound $\tau_{\nu_i \to \nu_j + \phi} >  1.3\times 10^{9}\,\text{s}  \, \left({m_{\nu_i}}/{ 0.05\,\text{eV} }\right)^3$ is only 10\% more stringent  than the one found in the analysis performed by Archidiacono and Hannestad~\cite{Archidiacono:2013dua}, that used Planck 2013 data. To derive this constraint, unlike in Ref.~\cite{Archidiacono:2013dua}, we did not assume that three neutrinos interact, but only the two that participate in the decay process $\nu_i \to \nu_j + \phi$. 
\item The CMB constraints on invisible neutrino decays are robust upon modifications of the cosmological model. In particular, we have shown that the bounds are barely affected by possible contributions to $N_{\rm eff}$ from non-interacting dark radiation. 
\item Invisible neutrino lifetimes {\small $\tau_\nu <   10^{5}\,\text{s} \, \left({m_\nu}/{ 0.05\,\text{eV} }\right)^3  $} that could be tested from the observations of the next galactic supernovae are highly disfavoured by Planck CMB observations. Neutrino decays $\nu_i \to \nu_j + \phi$ occurring at such a rate are excluded with a $6.4\,\sigma$ significance. 
\item The full Planck 2018 data set shows a mild preference for invisible neutrino decays. In particular, invisible neutrino lifetimes $\tau_{\nu_i \to  \nu_j + \phi} =7.7^{+8.5}_{-6.0}  \,\times 10^{9}\,\text{s}  \, \left({m_{\nu_i}}/{ 0.05\,\text{eV} }\right)^3$ and $\tau_{\nu_i \to \nu_4+\phi} =3.0^{+2.8}_{-2.5}  \,\times 10^{9}\,\text{s}  \, \left({m_{\nu_i}}/{ 0.05\,\text{eV} }\right)^3$ are preferred over $\Lambda$CDM with a significance of $\sim$ 1-2$\sigma$. 

\item We have shown that including current BAO data does not alter any of the conclusions we draw from Planck 2018 observations alone.
 
\end{enumerate}

%%%%%%%%%%%%%%%%%%%%%%%%%%%%%%%%%%%%%%%%%%%%%%%%%%%%%%
\section{Outlook}\label{sec:outlook}
%%%%%%%%%%%%%%%%%%%%%%%%%%%%%%%%%%%%%%%%%%%%%%%%%%%%%%

\begin{figure}[t]
\centering
\includegraphics[width=0.46\textwidth]{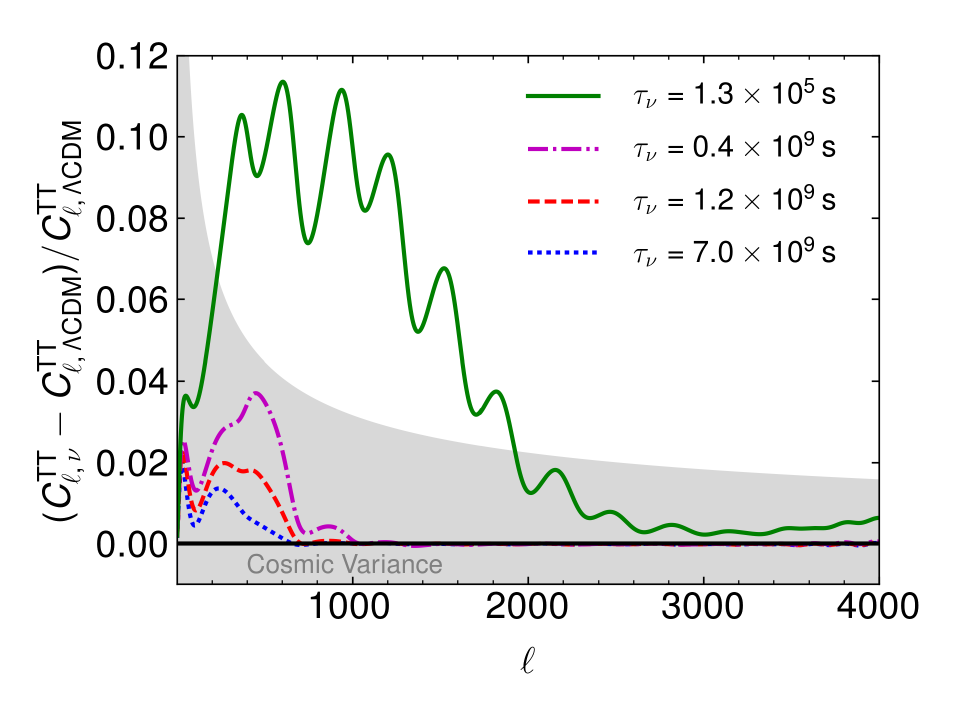}  
\vspace{-0.6cm}
\caption{TT power spectrum for various values of the neutrino decay lifetime as compared to $\Lambda$CDM, assuming the same cosmological parameters. We consider the neutrino decay process ${\nu_i \to  \nu_j + \phi} $ and we fix $m_{\nu_i} = 0.05\,\text{eV}$ for concreteness. The grey band indicates cosmic variance. }\label{fig:CL_TT}
\end{figure}

Cosmological constraints on invisible neutrino decays are typically orders of magnitude more stringent than those derived from laboratory and solar neutrino experiments. However, it must be noted, that in order to set cosmological constraints we have implicitly assumed that the neutrino interactions that trigger neutrino decays are time independent.  This is not the case in some models in which neutrinos do decay today, but would not have done so in the early Universe~\cite{Dvali:2016uhn}. Hence, all terrestrial, astrophysical and cosmological bounds are meaningful. 

Sensitivity to invisible neutrino decays is generically expected to improve in the future. Bounds from current and upcoming laboratory experiments, the next galactic supernovae and neutrino telescopes have been a subject of intense study, see e.g.~\cite{Beacom:2002vi,Ando:2004qe,Ando:2003ie,Abrahao:2015rba,Bustamante:2016ciw,Coloma:2017zpg,Choubey:2017dyu,Choubey:2017eyg,deSalas:2018kri,Ascencio-Sosa:2018lbk,Tang:2018rer,Denton:2018aml}. From the cosmological side, the positive detection of the neutrino energy density would represent a very strong constraint on the neutrino lifetime~\cite{Serpico:2007pt}. In addition, since baryon acoustic oscillations have now been shown~\cite{Baumann:2019tdh} to require the presence of free-streaming neutrino species, we think they could also be used to set constraints on the invisible neutrino lifetime, and could potentially reach $\tau_\nu \sim \mathcal{O}(10^{15})\,\text{s}$.

In this work we have focused on CMB constraints upon invisible neutrino decays. One may naively think that future CMB observations could help to tighten the constraints on invisible neutrino decays. However, we do not expect this to be the case. In Figure~\ref{fig:CL_TT}, we show the TT power spectrum for some neutrino decay scenarios as compared to $\Lambda$CDM. One can clearly appreciate that, for neutrino lifetimes that are not already excluded by Planck 2018 observations, the only modification to the power spectrum occurs for $\ell < 1000$, which corresponds to angular scales that have been measured already with cosmic variance error bars by the Planck satellite. This means that we expect constraints to improve only slightly, and particularly from future polarization measurements. 

To conclude, in this work we have shown that when the full Planck 2018 data is considered, neutrino decay lifetimes of $\tau_\nu = (2-16)\times 10^{9}\,\text{s} \, \left({m_\nu}/{ 0.05\,\text{eV} }\right)^3$ are preferred over neutrinos being stable with a 1-2 $\sigma$ significance. From the particle physics perspective, although beyond the scope of this work, it would be very interesting to work out a UV complete model that is capable of generating such neutrino lifetimes while being consistent with all other laboratory constraints, in particular those arising from the null searches of charged lepton-flavour violation processes.

\begin{acknowledgements}
We are supported by the European Research Council under the European Union's Horizon 2020 program (ERC Grant Agreement No 648680 DARKHORIZONS).  MF also receives funding from the STFC.
\end{acknowledgements}

\appendix   
%%%%%%%%%%%%%%%%%%%%%%%%%%%%%%%%%%%%%%%%%%%%%%%%%%%%%%
\section{Energy transfer rate for $1+ 2 \to 3 +4 $ annihilation processes}\label{app:calculation}
%%%%%%%%%%%%%%%%%%%%%%%%%%%%%%%%%%%%%%%%%%%%%%%%%%%%%%
Here we calculate the energy density transfer rate for a $1+ 2 \to 3 +4 $ annihilation process by closely following~\cite{Gondolo:1990dk,Edsjo:1997bg}. 
Neglecting statistical factors and assuming Maxwell-Boltzmann statistics for the distribution functions, the energy density transfer rate explicitly reads:
\begin{align}
\frac{\delta \rho}{\delta t} &= -  \sum_{\rm spins}\int d^3\tilde{p}_1 d^3\tilde{p}_2 d^3\tilde{p}_3 d^3\tilde{p}_4 \,(2\pi)^4 \times \\ \nonumber 
&\delta^4(p_1+p_2-p_3-p_4)\,|\mathcal{M}|^2 \, E_1 \,\left(f_1 f_2 - f_3 f_4\right) \, ,
\end{align}
where $d^3\tilde{p_i} =  d^3p_i/(2 E_i (2\pi)^3)$ and $\mathcal{M}$ is the amplitude for the $1+ 2 \to 3 +4 $ process,  where we assumed CP conservation. For our particular process of interest, $T_1 = T_2 = T $ and $T_3 = T_4 = T' $. And hence, within the MB approximation $f_1 f_2 = e^{-(E_1+E_2)/T}$ and $f_3 f_4 = e^{-(E_3+E_4)/T'}= e^{-(E_1+E_2)/T'}$ by detailed balance, which allows to reduce the phase space integrals from $12\to 1$ dimensions.

By following the integration procedure of Refs.~\cite{Gondolo:1990dk,Edsjo:1997bg}, and particularizing for $m_1 = m_2 = m_3 = m_4 =0$ we find
\begin{align}\label{eq:drho_dt_general}
\scalebox{1.06}[1.1]{$\frac{\delta \rho}{\delta t} =  \, \int_{s_{\rm min}}^{\infty} \frac{\sigma \, s^2\, ds }{64 \pi^4 }\,  \left[ T' \, K_2\left(\frac{\sqrt{s}}{T'}\right) -  T \, K_2\left(\frac{\sqrt{s}}{T}\right)  \right] ,$}
\end{align}
where $s_{\rm min} = \text{min}[(m_1+m_2)^2, \, (m_3+m_4)^2]$, and $\sigma$ is the usual cross section for the $1+2\to 3+4$ process.  
We are interested in applying this formula to the process $\phi \phi \to \bar{\nu}\nu$. At the energies of interest ($\sim \,\text{MeV}$) $m_1 = m_2 = m_3 = m_4 = 0$ is a very good approximation. The cross section for the $\phi \phi \to \bar{\nu}_i \nu_i$ process simply reads: 
\begin{align}
\sigma (s) \simeq  \frac{2 \sum_j \lambda_{ij}^4}{\pi s} \log\left(ƒ'' \frac{s}{m_\nu^2} \right)\,,
\end{align} 
where here we have for simplicity assumed that $m_{\nu_i} = m_{\nu_j}$\footnote{This hugely simplifies the cross section and we have explicitly checked that it is accurate at the 10\% level.}.
By using expression~\eqref{eq:drho_dt_general} we obtain the following analytical energy transfer rate:
\begin{align}
\frac{\delta \rho_\phi}{\delta t}  = \frac{ \sum_j \lambda_{ij}^4  }{ \pi ^5}\left[ T_\nu^5 \log \left(\frac{2 T_\nu}{m_\nu}\right) - T_\phi^5 \log \left(\frac{2 T_\phi}{m_\nu}\right) \right]ƒ'' \, .
\end{align} 
Note that this rate corresponds to $\phi \phi \leftrightarrow \bar{\nu} \nu$ for one single neutrino mass eigenstate in the final state.

%%%%%%%%%%%%%%%%%%%%%%%%%%%%%%%%%%%%%%%%%%%%%%%%%%%%%%
\section{Cosmological Parameters}\label{app:cosmo_params}
%%%%%%%%%%%%%%%%%%%%%%%%%%%%%%%%%%%%%%%%%%%%%%%%%%%%%%

In Table~\ref{tab:nudecay_constraints} we quote the mean $\pm$ 68\% CL intervals for all the relevant cosmological parameters as derived in each analysis performed in this study.

\begin{table*}
{\def\arraystretch{1.3}
\begin{tabular*}{0.973\textwidth}{{c|c|c|cc|c|cc}} 
\toprule 
&                     & \multicolumn{5}{c}{Scenario} \\ \hline
Data & $\qquad $  Parameter   $ \qquad $                  & $\qquad \nu_i \to \nu_4 + \phi \qquad $       & $ \,\,\, \nu_i \to \nu_4 + \phi$, + $\Delta N_{\rm eff}  \,\,\,$  && $\qquad \nu_i \to \nu_j + \phi \qquad$       & $\,\,\, \nu_i \to \nu_j + \phi$, + $\Delta N_{\rm eff} \,\,\,$ \\ \hline \hline
\multirow{9}{*}{ \rotatebox[origin=c]{90}{ \texttt{Planck2018 TTTEEE+lowE+len}}    }  $ \,$  &$\log_{10}(\Gamma_{\rm eff})$ & $<2.6$    &  $<2.5$                     && $<2.0$   & $ <2.3$   \\ 
&$\Delta N_{\rm eff}$          & -                   &  $0.14\pm 0.28$                       && -                   &  $0.23\pm0.29$  \\ \cline{2-7} 
&$\tau_\nu \cdot  \left({m_\nu}/{ 0.05\,\text{eV} }\right)^{-3} $ & $>0.3 \times 10^{9}\,\text{s}$    &  $>0.4 \times 10^{9}\,\text{s}$                      && $>1.3 \times 10^{9}\,\text{s}$   & $ >0.6 \times 10^{9}\,\text{s}$   \\ \cline{2-7} 
&$\Omega_{\rm b} h^2$          & $0.02240\pm 0.00017$  &  $0.0225\pm 0.0003$                   && $0.02230\pm 0.00018$  &  $0.0226\pm 0.0003$\\ 
&$\Omega_{\rm cdm} h^2$        & $0.1187 \pm 0.0014$   & $0.121 \pm 0.004$                     && $0.1186 \pm 0.0014$   & $0.123 \pm 0.005$ \\ 
&$\log({10^{10}A_{\rm s}})$    & $3.08 \pm 0.03 $    & $3.09 \pm 0.03 $                      &&  $3.08 \pm 0.03 $   & $3.09 \pm 0.03 $\\ 
&$n_{\rm s}$                   & $0.970 \pm 0.005$   &  $0.976 \pm 0.010$                    &&  $0.972 \pm 0.005$  &  $0.981 \pm 0.015$\\ 
&$\tau_{\rm reio }$            & $0.050\pm 0.008 $   &  $0.051\pm 0.008 $                    && $0.050\pm 0.007 $   &  $0.050\pm 0.008 $ \\ 
&$H_0 \,[{\rm km/s/Mpc}]$          & $67.9 \pm 0.6 $     & $69.2 \pm 2.0 $                       && $68.0 \pm 0.06 $     & $69.7 \pm 2.0 $\\ 
\hline \hline 
\multirow{9}{*}{ \rotatebox[origin=c]{90}{ \texttt{Planck2018 TT+lowE+len}}    }  $ \,$  &$\log_{10}(\Gamma_{\rm eff})$ & $<2.5$    &  $<2.5$                     && $<2.0$   & $ <1.8$   \\ 
&$\Delta N_{\rm eff}$          & -                   &  $-0.14\pm 0.29$                       && -                   &  $-0.05\pm0.27$  \\ \cline{2-7} 
&$\tau_\nu \cdot  \left({m_\nu}/{ 0.05\,\text{eV} }\right)^{-3} $ & $>0.4 \times 10^{9}\,\text{s}$    &  $>0.4 \times 10^{9}\,\text{s}$                      && $>1.8 \times 10^{9}\,\text{s}$   & $ >1.4 \times 10^{9}\,\text{s}$   \\ \cline{2-7} 
&$\Omega_{\rm b} h^2$          & $0.02211\pm 0.00023$  &  $0.02212\pm 0.0003$                   && $0.0220\pm 0.0002$  &  $0.0220\pm 0.0003$\\ 
&$\Omega_{\rm cdm} h^2$        & $0.1197 \pm 0.0017$   & $0.120 \pm 0.004$                       && $0.1203 \pm 0.0017$      & $0.120 \pm 0.005$ \\ 
&$\log({10^{10}A_{\rm s}})$      & $3.037 \pm 0.014 $    & $3.03 \pm 0.02 $                                 &&  $3.035\pm 0.015 $           & $3.03 \pm 0.02 $\\ 
&$n_{\rm s}$                 	     & $0.964 \pm 0.005$   &  $0.963 \pm 0.013$                            &&  $0.966 \pm 0.005$           &  $0.965 \pm 0.012$\\ 
&$\tau_{\rm reio }$                  & $0.052\pm 0.008 $   &  $0.051\pm 0.008 $                          && $0.052\pm 0.008 $            &  $0.051\pm 0.008 $ \\ 
&$H_0 \,[{\rm km/s/Mpc}]$          & $67.0 \pm 0.7 $     & $66.9 \pm 2.5 $                                    && $67.0 \pm 0.8 $               & $66.8 \pm 2.2 $\\ 
\hline \hline 
\multirow{9}{*}{ \rotatebox[origin=c]{90}{ \texttt{Planck2018+BAO}}    }  $ \,$  &$\log_{10}(\Gamma_{\rm eff})$ & $<2.6$    &  $<2.7$                     && $<2.0$   & $ <2.0$   \\ 
&$\Delta N_{\rm eff}$          & -                   &  $0.09\pm 0.25$                       && -                   &  $0.19\pm0.24$  \\ \cline{2-7} 
&$\tau_\nu \cdot  \left({m_\nu}/{ 0.05\,\text{eV} }\right)^{-3} $ & $>0.3 \times 10^{9}\,\text{s}$    &  $>0.2 \times 10^{9}\,\text{s}$                      && $>1.2 \times 10^{9}\,\text{s}$   & $ >0.8 \times 10^{9}\,\text{s}$   \\ \cline{2-7} 
&$\Omega_{\rm b} h^2$          & $0.02237\pm 0.00016$  &  $0.0225\pm 0.0002$                   && $0.02234\pm 0.00017$  &  $0.0224\pm 0.0003$\\ 
&$\Omega_{\rm cdm} h^2$        & $0.1192 \pm 0.0012$   & $0.121 \pm 0.004$                     && $0.1190 \pm 0.0013$   & $0.123 \pm 0.003$ \\ 
&$\log({10^{10}A_{\rm s}})$    & $3.07 \pm 0.03 $    & $3.08 \pm 0.03 $                      &&  $3.08 \pm 0.03 $   & $3.09 \pm 0.03 $\\ 
&$n_{\rm s}$                   & $0.969 \pm 0.005$   &  $0.973 \pm 0.010$                    &&  $0.971 \pm 0.005$  &  $0.979 \pm 0.010$\\ 
&$\tau_{\rm reio }$            & $0.048\pm 0.008 $   &  $0.050\pm 0.008 $                    && $0.050\pm 0.008 $   &  $0.050\pm 0.008 $ \\ 
&$H_0 \,[{\rm km/s/Mpc}]$          & $67.7 \pm 0.6 $     & $68.5 \pm 1.7 $                       && $67.8 \pm 0.6 $     & $69.1 \pm 1.7 $\\ 
\hline \hline 
\end{tabular*}
}
\vspace{-0.1cm}
\caption{Marginalized posteriors for the standard cosmological parameters plus the neutrino decay lifetime and $\Delta N_{\rm eff}$ from the the analysis to Planck 2018 data. $i,\,j$ label active neutrino mass eigenstates. The rows correspond to the mean and $\pm$ $1\sigma$ errors but for the case of the neutrino decay parameters in which we quote a 95\% CL bound. }
\label{tab:nudecay_constraints}
\end{table*} 

%%%%%%%%%%%%%%%%%%%%%%%%%%%%%%%%%%%%%%%%%%%%%%%%%%%%%%
\section{Planck 2015 Constraints}\label{app:cosmo_full_2015}
\begin{figure*}[t]
\centering
\begin{tabular}{cc}
\includegraphics[width=0.48\textwidth]{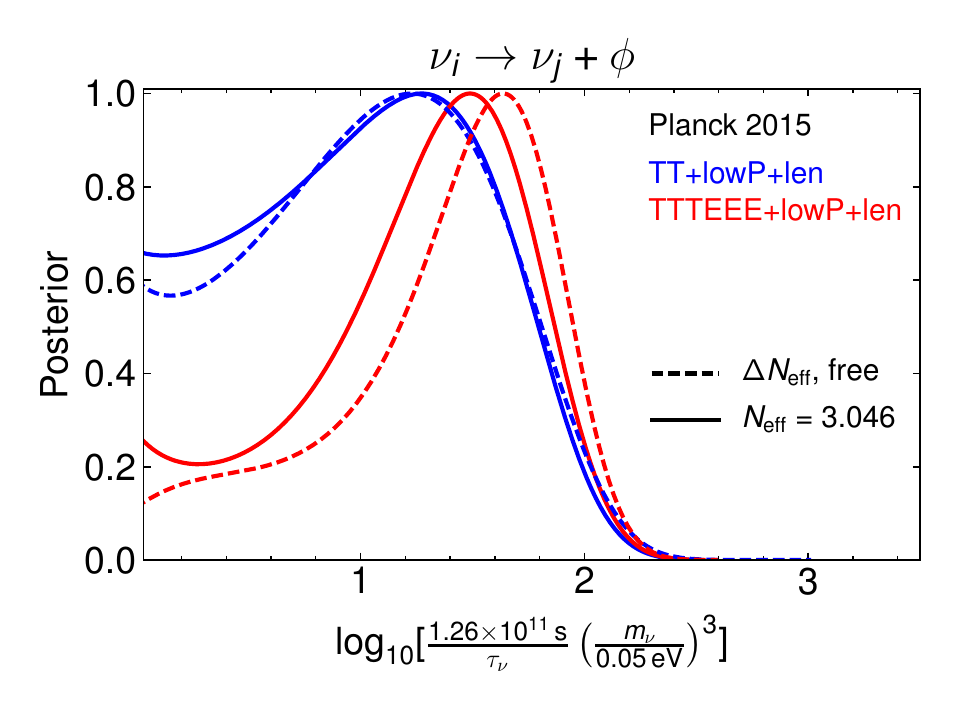}   & \includegraphics[width=0.48\textwidth]{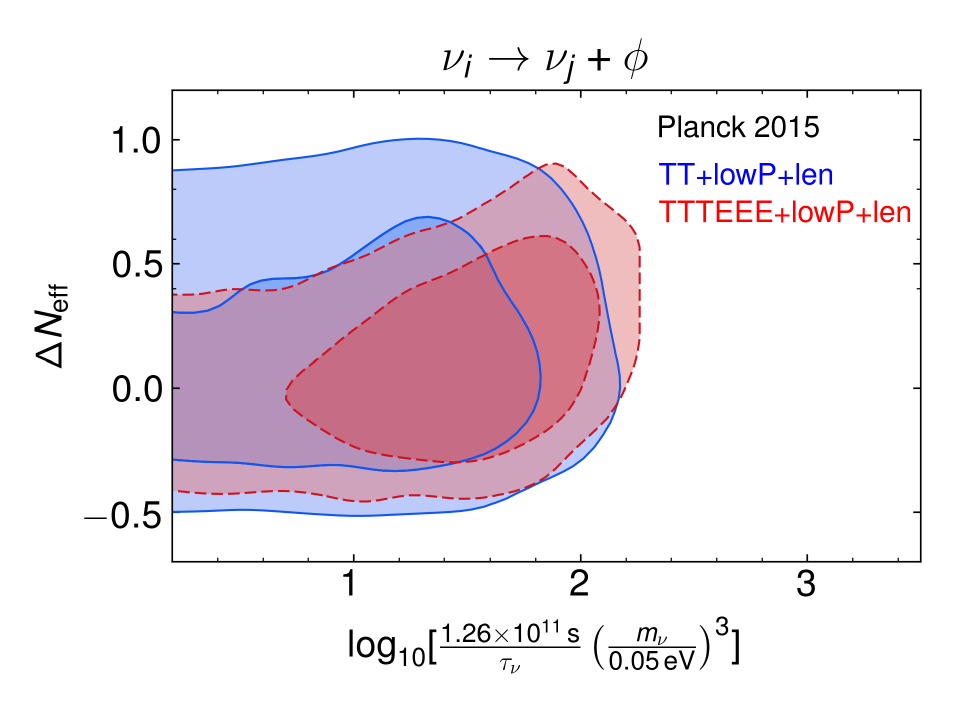} \\
\includegraphics[width=0.48\textwidth]{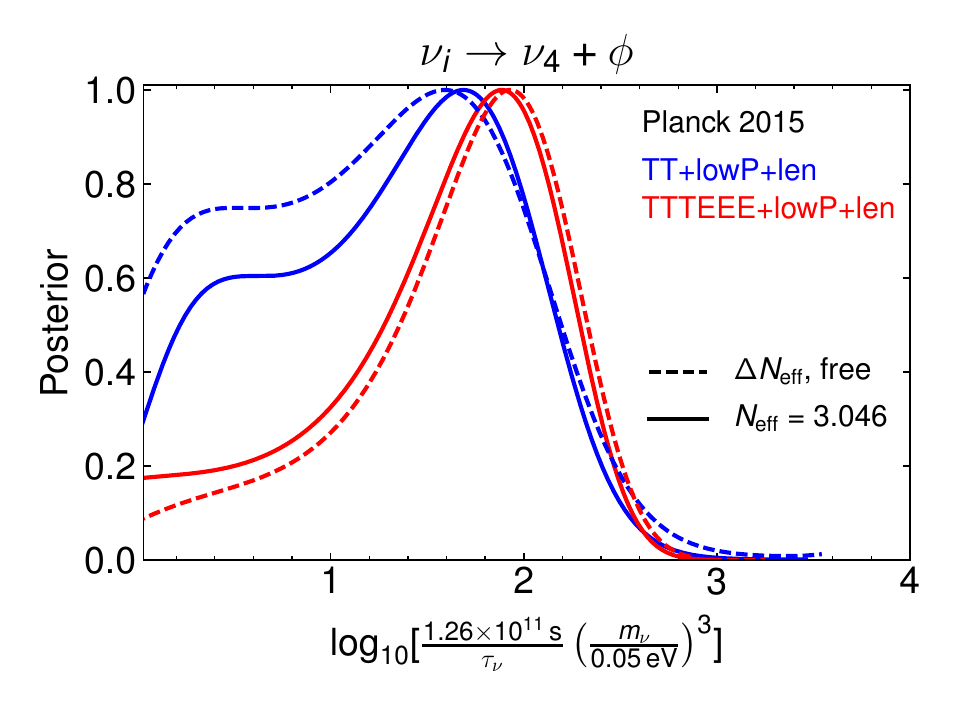}   & \includegraphics[width=0.48\textwidth]{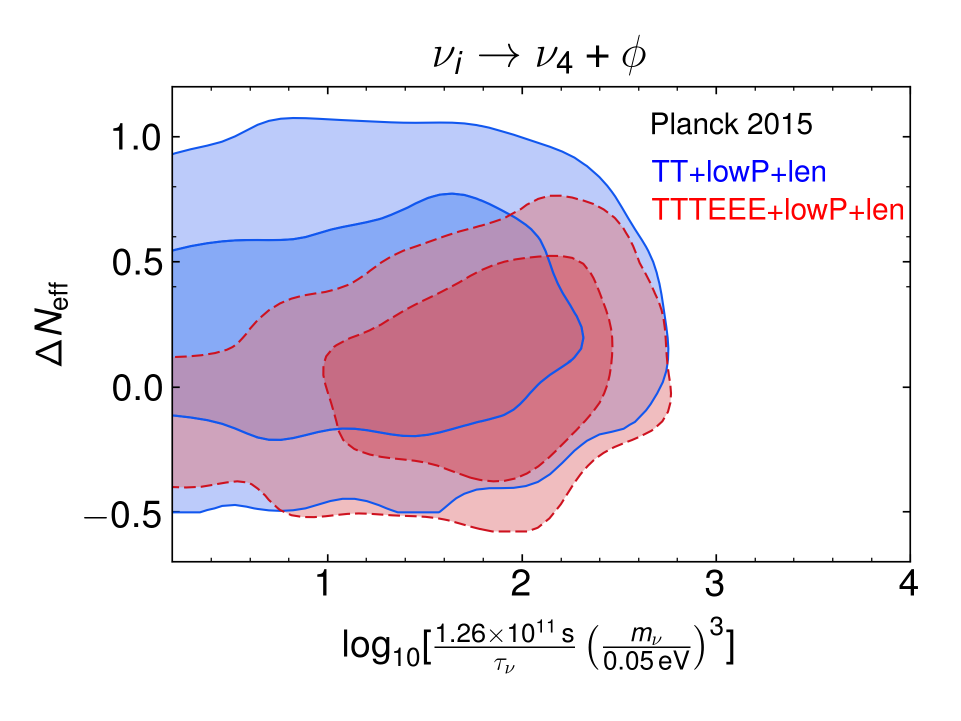} 
\end{tabular}
\vspace{-0.4cm}
\caption{Marginalized posterior distributions from the analysis to the \texttt{Planck 2015 TT+lowP+lensing} data (blue) and the \texttt{Planck 2015 TTTEEE+lowP+lensing} (red). The \textit{upper panels} correspond to the neutrino decay scenario $\nu_i \to \nu_j' + \phi$, and the \textit{lower panels} correspond to the neutrino decay scenario $\nu_i \to \nu_4 +\phi$. The \textit{left panels} correspond to the one dimensional posterior of $\log_{10}\left[{1.26 \times 10^{11}\,\text{s}}/{\tau_\nu} \left({m_\nu}/{0.05\,\text{eV}}\right)^{3}\right]$. The \textit{right panels} correspond to the two dimensional posterior in the $\Delta N_{\rm eff}$-$\log_{10}\left[{1.26 \times 10^{11}\,\text{s}}/{\tau_\nu} \left({m_\nu}/{0.05\,\text{eV}}\right)^{3}\right]$ plane.   }\label{fig:Posterior_2015}
\end{figure*}

%%%%%%%%%%%%%%%%%%%%%%%%%%%%%%%%%%%%%%%%%%%%%%%%%%%%%%
Here we outline the constraints on invisible neutrino decays that can be derived from Planck 2015 CMB observations~\cite{Adam:2015rua,Ade:2015xua}. We consider the following data set combinations \texttt{Planck 2015 TT+lowP+lensing} and \texttt{Planck 2015 TTTEEE+lowP+lensing}. In Table~\ref{tab:summary_constraints_tau_2015} we quote the best fit, mean 68\% CL error bars and 95\% CL exclusions for the parameter $\Gamma_{\rm eff}$ and for the invisible neutrino decay lifetime. In Figure~\ref{fig:Posterior_2015} we display the marginalized 1D and 2D posterior distribution of the parameter $\Gamma_{\rm eff}$ and $\Delta N_{\rm eff} - \Gamma_{\rm eff}$ respectively. 

Finally, when considering a lifetime of $\tau_\nu =  1.3\times 10^{5}\,\text{s} \, \left({m_\nu}/{ 0.05\,\text{eV} }\right)^3  $, we find that the fit to Planck 2015 data is degraded with respect to $\Lambda$CDM at the level of 
\begin{align}
\scalebox{1.0}[0.95]{$\sqrt{\chi^2_{\nu_i \to \nu_j +\phi}-\chi^2_{\Lambda {\rm CDM}}} = 5.4\,$} , \\
\scalebox{1.0}[0.95]{$\sqrt{\chi^2_{\nu_i \to \nu_4+\phi}-\chi^2_{\Lambda {\rm CDM}}} = 3.1$}\,.
\end{align}
when considering the \texttt{Planck 2015 TT+lowP+lensing} data set. Similarly, for the \texttt{Planck 2015 TTTEEE+lowP+lensing} data set we find:
\begin{align}
\scalebox{1.0}[0.95]{$\sqrt{\chi^2_{\nu_i \to \nu_j +\phi}-\chi^2_{\Lambda {\rm CDM}}} = 7.7\, $}, \\
\scalebox{1.0}[0.95]{$\sqrt{\chi^2_{\nu_i \to \nu_4+\phi}-\chi^2_{\Lambda {\rm CDM}}} = 3.9$}\,.
\end{align}

\begin{table*}
\begin{center}
{\def\arraystretch{1.15}
\begin{tabular}{l|l|ccc|ccc|c|c}
\hline\hline
& Parameter  &\multicolumn{3}{c|}{$\log_{10}\left(\Gamma_{\rm eff}\right)$}  & \multicolumn{3}{c|}{$\tau_\nu/(10^{9}\,s) \cdot  \left({m_\nu}/{ 0.05\,\text{eV} }\right)^{-3} $}  & $\Delta N_{\rm eff}$ & $\sqrt{\chi^2_{\Lambda {\rm CDM}} - \chi^2_{\nu}}$ \\ \hline
$\,$ Data &Scenario & $\,\,\,\,$ BF $\,\,\,$ & 68\% CL &  95\% CL   & $\,\,\,\,$ BF $\,\,\,$ & 68\% CL &  95\% CL  & 68\% CL & BF \\ \hline
\multirow{4}{*}{ \rotatebox[origin=c]{90}{ \texttt{Planck2015}}   $ \,$  \rotatebox[origin=c]{90}{ \texttt{TTTEEE+lowP}}}  $ \,$   
&$\nu_{i} \to \nu_j + \phi$ 					& 1.61   & $1.3^{+0.7}_{-0.3}$  & $< 2.0$  & 3.0    & $7.0^{+7.2}_{-5.5}$   &  $> 1.2$   & - & 1.8  \\  
&$\nu_{i} \to \nu_j + \phi$, +$\Delta N_{\rm eff}$ &  1.58  & $1.4^{+0.6}_{-0.3}$   &  $< 2.2$ &  3.3  & $5.5^{+5.0}_{-4.2}$   & $> 0.9$   & $0.02\pm 0.27$ & 2.2  \\  
&$\nu_{i} \to \nu_4 + \phi$ 				& 1.18   & $1.6^{+0.8}_{-0.4}$   & $< 2.5$   & 8.1   & $3.4^{+3.7}_{-2.8} $ &  $>0.4$  &  - & 1.8  \\  
&$\nu_{i} \to \nu_4 + \phi$, +$\Delta N_{\rm eff}$ &  1.98  & $1.7^{+0.7}_{-0.3}$  & $<2.6$  &  1.3  & $2.8^{+3.3}_{-2.2} $ & $>0.3$   & $-0.04\pm 0.26$  &1.7  \\  
  \hline 
    \multirow{4}{*}{ \rotatebox[origin=c]{90}{ \texttt{Planck2015}}   $ \,$  \rotatebox[origin=c]{90}{ \texttt{TT+lowP}}}  $ \,$   
 &$\nu_{i} \to \nu_j + \phi$ 				  & 1.00   & $1.0^{+0.7}_{-0.5}$  & $< 1.9$  & 13    & $13^{+31}_{-10}$   &  $> 1.7$   & - & 1.2   \\  
&$\nu_{i} \to \nu_j + \phi$, +$\Delta N_{\rm eff}$   & 1.35  & $1.0^{+0.7}_{-0.5}$   &  $< 1.9$ &  5.6  & $12^{+28}_{-9}$   & $> 1.7$   & $0.23\pm0.38$& 1.2   \\  
&$\nu_{i} \to \nu_4 + \phi$ 				  &  2.15   & $1.3^{+0.8}_{-0.7}$   & $< 2.3$   & 77   & $7^{+23}_{-5} $ &  $>0.6$  &  - &   1.0 \\  
&$\nu_{i} \to \nu_4 + \phi$, +$\Delta N_{\rm eff}$ & 1.38  & $1.2^{+0.7}_{-0.8}$  & $<2.4$  &  5.3  & $7^{+35}_{-6} $ & $>0.5$   & $0.26\pm 0.35$   & 0.1\\  
  \hline \hline
\end{tabular}
}
\end{center}\vspace{-0.5cm}
\caption{Marginalized constraints from Planck 2015 CMB observations for $\log_{10}\left(\Gamma_{\rm eff}\right)$, $\Delta N_{\rm eff}$, and those inferred for the neutrino lifetime $\tau_\nu$ from equation~\eqref{eq:Gamma_eff}. Note that we also account for Planck 2015 lensing measurements. }\label{tab:summary_constraints_tau_2015}
\end{table*}

%%%%%%%%%%%%%%%%%%%%%%%%%%%%%%%%%%%%%%%%%%%%%%%%%%%%%%
%\bibliographystyle{JHEP}
\bibliography{biblio}

\begin{thebibliography}{109}%
\makeatletter
\providecommand \@ifxundefined [1]{%
 \@ifx{#1\undefined}
}%
\providecommand \@ifnum [1]{%
 \ifnum #1\expandafter \@firstoftwo
 \else \expandafter \@secondoftwo
 \fi
}%
\providecommand \@ifx [1]{%
 \ifx #1\expandafter \@firstoftwo
 \else \expandafter \@secondoftwo
 \fi
}%
\providecommand \natexlab [1]{#1}%
\providecommand \enquote  [1]{``#1''}%
\providecommand \bibnamefont  [1]{#1}%
\providecommand \bibfnamefont [1]{#1}%
\providecommand \citenamefont [1]{#1}%
\providecommand \href@noop [0]{\@secondoftwo}%
\providecommand \href [0]{\begingroup \@sanitize@url \@href}%
\providecommand \@href[1]{\@@startlink{#1}\@@href}%
\providecommand \@@href[1]{\endgroup#1\@@endlink}%
\providecommand \@sanitize@url [0]{\catcode `\\12\catcode `\$12\catcode
  `\&12\catcode `\#12\catcode `\^12\catcode `\_12\catcode `\%12\relax}%
\providecommand \@@startlink[1]{}%
\providecommand \@@endlink[0]{}%
\providecommand \url  [0]{\begingroup\@sanitize@url \@url }%
\providecommand \@url [1]{\endgroup\@href {#1}{\urlprefix }}%
\providecommand \urlprefix  [0]{URL }%
\providecommand \Eprint [0]{\href }%
\providecommand \doibase [0]{http://dx.doi.org/}%
\providecommand \selectlanguage [0]{\@gobble}%
\providecommand \bibinfo  [0]{\@secondoftwo}%
\providecommand \bibfield  [0]{\@secondoftwo}%
\providecommand \translation [1]{[#1]}%
\providecommand \BibitemOpen [0]{}%
\providecommand \bibitemStop [0]{}%
\providecommand \bibitemNoStop [0]{.\EOS\space}%
\providecommand \EOS [0]{\spacefactor3000\relax}%
\providecommand \BibitemShut  [1]{\csname bibitem#1\endcsname}%
\let\auto@bib@innerbib\@empty
%</preamble>
\bibitem [{\citenamefont {Esteban}\ \emph {et~al.}(2019)\citenamefont
  {Esteban}, \citenamefont {Gonzalez-Garcia}, \citenamefont
  {Hernandez-Cabezudo}, \citenamefont {Maltoni},\ and\ \citenamefont
  {Schwetz}}]{Esteban:2018azc}%
  \BibitemOpen
  \bibfield  {author} {\bibinfo {author} {\bibfnamefont {I.}~\bibnamefont
  {Esteban}}, \bibinfo {author} {\bibfnamefont {M.~C.}\ \bibnamefont
  {Gonzalez-Garcia}}, \bibinfo {author} {\bibfnamefont {A.}~\bibnamefont
  {Hernandez-Cabezudo}}, \bibinfo {author} {\bibfnamefont {M.}~\bibnamefont
  {Maltoni}}, \ and\ \bibinfo {author} {\bibfnamefont {T.}~\bibnamefont
  {Schwetz}},\ }\href {\doibase 10.1007/JHEP01(2019)106} {\bibfield  {journal}
  {\bibinfo  {journal} {JHEP}\ }\textbf {\bibinfo {volume} {01}},\ \bibinfo
  {pages} {106} (\bibinfo {year} {2019})},\ \Eprint
  {http://arxiv.org/abs/1811.05487} {arXiv:1811.05487 [hep-ph]} \BibitemShut
  {NoStop}%
%%CITATION = ARXIV:1811.05487;%%
\bibitem [{\citenamefont {de~Salas}\ \emph {et~al.}(2018)\citenamefont
  {de~Salas}, \citenamefont {Forero}, \citenamefont {Ternes}, \citenamefont
  {Tortola},\ and\ \citenamefont {Valle}}]{deSalas:2017kay}%
  \BibitemOpen
  \bibfield  {author} {\bibinfo {author} {\bibfnamefont {P.~F.}\ \bibnamefont
  {de~Salas}}, \bibinfo {author} {\bibfnamefont {D.~V.}\ \bibnamefont
  {Forero}}, \bibinfo {author} {\bibfnamefont {C.~A.}\ \bibnamefont {Ternes}},
  \bibinfo {author} {\bibfnamefont {M.}~\bibnamefont {Tortola}}, \ and\
  \bibinfo {author} {\bibfnamefont {J.~W.~F.}\ \bibnamefont {Valle}},\ }\href
  {\doibase 10.1016/j.physletb.2018.06.019} {\bibfield  {journal} {\bibinfo
  {journal} {Phys. Lett.}\ }\textbf {\bibinfo {volume} {B782}},\ \bibinfo
  {pages} {633} (\bibinfo {year} {2018})},\ \Eprint
  {http://arxiv.org/abs/1708.01186} {arXiv:1708.01186 [hep-ph]} \BibitemShut
  {NoStop}%
%%CITATION = ARXIV:1708.01186;%%
\bibitem [{\citenamefont {Capozzi}\ \emph {et~al.}(2016)\citenamefont
  {Capozzi}, \citenamefont {Lisi}, \citenamefont {Marrone}, \citenamefont
  {Montanino},\ and\ \citenamefont {Palazzo}}]{Capozzi:2016rtj}%
  \BibitemOpen
  \bibfield  {author} {\bibinfo {author} {\bibfnamefont {F.}~\bibnamefont
  {Capozzi}}, \bibinfo {author} {\bibfnamefont {E.}~\bibnamefont {Lisi}},
  \bibinfo {author} {\bibfnamefont {A.}~\bibnamefont {Marrone}}, \bibinfo
  {author} {\bibfnamefont {D.}~\bibnamefont {Montanino}}, \ and\ \bibinfo
  {author} {\bibfnamefont {A.}~\bibnamefont {Palazzo}},\ }\href {\doibase
  10.1016/j.nuclphysb.2016.02.016} {\bibfield  {journal} {\bibinfo  {journal}
  {Nucl. Phys.}\ }\textbf {\bibinfo {volume} {B908}},\ \bibinfo {pages} {218}
  (\bibinfo {year} {2016})},\ \Eprint {http://arxiv.org/abs/1601.07777}
  {arXiv:1601.07777 [hep-ph]} \BibitemShut {NoStop}%
%%CITATION = ARXIV:1601.07777;%%
\bibitem [{\citenamefont {Petcov}(1977)}]{Petcov:1976ff}%
  \BibitemOpen
  \bibfield  {author} {\bibinfo {author} {\bibfnamefont {S.~T.}\ \bibnamefont
  {Petcov}},\ }\href@noop {} {\bibfield  {journal} {\bibinfo  {journal} {Sov.
  J. Nucl. Phys.}\ }\textbf {\bibinfo {volume} {25}},\ \bibinfo {pages} {340}
  (\bibinfo {year} {1977})},\ \bibinfo {note} {[Erratum: Yad.
  Fiz.25,1336(1977)]}\BibitemShut {NoStop}%
%%CITATION = SJNCA,25,340;%%
\bibitem [{\citenamefont {Hosotani}(1981)}]{HOSOTANI1981411}%
  \BibitemOpen
  \bibfield  {author} {\bibinfo {author} {\bibfnamefont {Y.}~\bibnamefont
  {Hosotani}},\ }\href {\doibase https://doi.org/10.1016/0550-3213(81)90306-0}
  {\bibfield  {journal} {\bibinfo  {journal} {Nuclear Physics B}\ }\textbf
  {\bibinfo {volume} {191}},\ \bibinfo {pages} {411 } (\bibinfo {year}
  {1981})}\BibitemShut {NoStop}%
\bibitem [{\citenamefont {Pal}\ and\ \citenamefont
  {Wolfenstein}(1982)}]{PhysRevD.25.766}%
  \BibitemOpen
  \bibfield  {author} {\bibinfo {author} {\bibfnamefont {P.~B.}\ \bibnamefont
  {Pal}}\ and\ \bibinfo {author} {\bibfnamefont {L.}~\bibnamefont
  {Wolfenstein}},\ }\href {\doibase 10.1103/PhysRevD.25.766} {\bibfield
  {journal} {\bibinfo  {journal} {Phys. Rev. D}\ }\textbf {\bibinfo {volume}
  {25}},\ \bibinfo {pages} {766} (\bibinfo {year} {1982})}\BibitemShut
  {NoStop}%
\bibitem [{\citenamefont {Chikashige}\ \emph {et~al.}(1981)\citenamefont
  {Chikashige}, \citenamefont {Mohapatra},\ and\ \citenamefont
  {Peccei}}]{Chikashige:1980ui}%
  \BibitemOpen
  \bibfield  {author} {\bibinfo {author} {\bibfnamefont {Y.}~\bibnamefont
  {Chikashige}}, \bibinfo {author} {\bibfnamefont {R.~N.}\ \bibnamefont
  {Mohapatra}}, \ and\ \bibinfo {author} {\bibfnamefont {R.~D.}\ \bibnamefont
  {Peccei}},\ }\href {\doibase 10.1016/0370-2693(81)90011-3} {\bibfield
  {journal} {\bibinfo  {journal} {Phys. Lett.}\ }\textbf {\bibinfo {volume}
  {B98}},\ \bibinfo {pages} {265} (\bibinfo {year} {1981})}\BibitemShut
  {NoStop}%
%%CITATION = PHLTA,B98,265;%%
\bibitem [{\citenamefont {Gelmini}\ and\ \citenamefont
  {Roncadelli}(1981)}]{Gelmini:1980re}%
  \BibitemOpen
  \bibfield  {author} {\bibinfo {author} {\bibfnamefont {G.~B.}\ \bibnamefont
  {Gelmini}}\ and\ \bibinfo {author} {\bibfnamefont {M.}~\bibnamefont
  {Roncadelli}},\ }\href {\doibase 10.1016/0370-2693(81)90559-1} {\bibfield
  {journal} {\bibinfo  {journal} {Phys. Lett.}\ }\textbf {\bibinfo {volume}
  {99B}},\ \bibinfo {pages} {411} (\bibinfo {year} {1981})}\BibitemShut
  {NoStop}%
%%CITATION = PHLTA,99B,411;%%
\bibitem [{\citenamefont {Schechter}\ and\ \citenamefont
  {Valle}(1982)}]{Schechter:1981cv}%
  \BibitemOpen
  \bibfield  {author} {\bibinfo {author} {\bibfnamefont {J.}~\bibnamefont
  {Schechter}}\ and\ \bibinfo {author} {\bibfnamefont {J.~W.~F.}\ \bibnamefont
  {Valle}},\ }\href {\doibase 10.1103/PhysRevD.25.774} {\bibfield  {journal}
  {\bibinfo  {journal} {Phys. Rev.}\ }\textbf {\bibinfo {volume} {D25}},\
  \bibinfo {pages} {774} (\bibinfo {year} {1982})}\BibitemShut {NoStop}%
%%CITATION = PHRVA,D25,774;%%
\bibitem [{\citenamefont {Lee}\ and\ \citenamefont
  {Shrock}(1977)}]{Lee:1977tib}%
  \BibitemOpen
  \bibfield  {author} {\bibinfo {author} {\bibfnamefont {B.~W.}\ \bibnamefont
  {Lee}}\ and\ \bibinfo {author} {\bibfnamefont {R.~E.}\ \bibnamefont
  {Shrock}},\ }\href {\doibase 10.1103/PhysRevD.16.1444} {\bibfield  {journal}
  {\bibinfo  {journal} {Phys. Rev.}\ }\textbf {\bibinfo {volume} {D16}},\
  \bibinfo {pages} {1444} (\bibinfo {year} {1977})}\BibitemShut {NoStop}%
%%CITATION = PHRVA,D16,1444;%%
\bibitem [{\citenamefont {Berezhiani}\ and\ \citenamefont
  {Khlopov}(1990{\natexlab{a}})}]{Berezhiani:1990wn}%
  \BibitemOpen
  \bibfield  {author} {\bibinfo {author} {\bibfnamefont {Z.~G.}\ \bibnamefont
  {Berezhiani}}\ and\ \bibinfo {author} {\bibfnamefont {M.~{\relax Yu}.}\
  \bibnamefont {Khlopov}},\ }\href@noop {} {\bibfield  {journal} {\bibinfo
  {journal} {Sov. J. Nucl. Phys.}\ }\textbf {\bibinfo {volume} {51}},\ \bibinfo
  {pages} {739} (\bibinfo {year} {1990}{\natexlab{a}})},\ \bibinfo {note}
  {[Yad. Fiz.51,1157(1990)]}\BibitemShut {NoStop}%
%%CITATION = SJNCA,51,739;%%
\bibitem [{\citenamefont {Berezhiani}\ and\ \citenamefont
  {Khlopov}(1990{\natexlab{b}})}]{Berezhiani:1990jj}%
  \BibitemOpen
  \bibfield  {author} {\bibinfo {author} {\bibfnamefont {Z.~G.}\ \bibnamefont
  {Berezhiani}}\ and\ \bibinfo {author} {\bibfnamefont {M.~{\relax Yu}.}\
  \bibnamefont {Khlopov}},\ }\href@noop {} {\bibfield  {journal} {\bibinfo
  {journal} {Sov. J. Nucl. Phys.}\ }\textbf {\bibinfo {volume} {51}},\ \bibinfo
  {pages} {935} (\bibinfo {year} {1990}{\natexlab{b}})},\ \bibinfo {note}
  {[Yad. Fiz.51,1479(1990)]}\BibitemShut {NoStop}%
%%CITATION = SJNCA,51,935;%%
\bibitem [{\citenamefont {Joshipura}\ and\ \citenamefont
  {Rindani}(1992)}]{Joshipura:1992vn}%
  \BibitemOpen
  \bibfield  {author} {\bibinfo {author} {\bibfnamefont {A.~S.}\ \bibnamefont
  {Joshipura}}\ and\ \bibinfo {author} {\bibfnamefont {S.~D.}\ \bibnamefont
  {Rindani}},\ }\href {\doibase 10.1103/PhysRevD.46.3000} {\bibfield  {journal}
  {\bibinfo  {journal} {Phys. Rev.}\ }\textbf {\bibinfo {volume} {D46}},\
  \bibinfo {pages} {3000} (\bibinfo {year} {1992})},\ \Eprint
  {http://arxiv.org/abs/hep-ph/9205220} {arXiv:hep-ph/9205220 [hep-ph]}
  \BibitemShut {NoStop}%
%%CITATION = HEP-PH/9205220;%%
\bibitem [{\citenamefont {Burgess}\ and\ \citenamefont
  {Cline}(1993)}]{Burgess:1992dt}%
  \BibitemOpen
  \bibfield  {author} {\bibinfo {author} {\bibfnamefont {C.~P.}\ \bibnamefont
  {Burgess}}\ and\ \bibinfo {author} {\bibfnamefont {J.~M.}\ \bibnamefont
  {Cline}},\ }\href {\doibase 10.1016/0370-2693(93)91720-8} {\bibfield
  {journal} {\bibinfo  {journal} {Phys. Lett.}\ }\textbf {\bibinfo {volume}
  {B298}},\ \bibinfo {pages} {141} (\bibinfo {year} {1993})},\ \Eprint
  {http://arxiv.org/abs/hep-ph/9209299} {arXiv:hep-ph/9209299 [hep-ph]}
  \BibitemShut {NoStop}%
%%CITATION = HEP-PH/9209299;%%
\bibitem [{\citenamefont {Berezhiani}\ \emph {et~al.}(1992)\citenamefont
  {Berezhiani}, \citenamefont {Smirnov},\ and\ \citenamefont
  {Valle}}]{Berezhiani:1992cd}%
  \BibitemOpen
  \bibfield  {author} {\bibinfo {author} {\bibfnamefont {Z.~G.}\ \bibnamefont
  {Berezhiani}}, \bibinfo {author} {\bibfnamefont {A.~{\relax Yu}.}\
  \bibnamefont {Smirnov}}, \ and\ \bibinfo {author} {\bibfnamefont {J.~W.~F.}\
  \bibnamefont {Valle}},\ }\href {\doibase 10.1016/0370-2693(92)90126-O}
  {\bibfield  {journal} {\bibinfo  {journal} {Phys. Lett.}\ }\textbf {\bibinfo
  {volume} {B291}},\ \bibinfo {pages} {99} (\bibinfo {year} {1992})},\ \Eprint
  {http://arxiv.org/abs/hep-ph/9207209} {arXiv:hep-ph/9207209 [hep-ph]}
  \BibitemShut {NoStop}%
%%CITATION = HEP-PH/9207209;%%
\bibitem [{\citenamefont {Shrock}(1974)}]{PhysRevD.9.743}%
  \BibitemOpen
  \bibfield  {author} {\bibinfo {author} {\bibfnamefont {R.}~\bibnamefont
  {Shrock}},\ }\href {\doibase 10.1103/PhysRevD.9.743} {\bibfield  {journal}
  {\bibinfo  {journal} {Phys. Rev. D}\ }\textbf {\bibinfo {volume} {9}},\
  \bibinfo {pages} {743} (\bibinfo {year} {1974})}\BibitemShut {NoStop}%
\bibitem [{\citenamefont {Georgi}\ and\ \citenamefont
  {Randall}(1990)}]{GEORGI1990196}%
  \BibitemOpen
  \bibfield  {author} {\bibinfo {author} {\bibfnamefont {H.}~\bibnamefont
  {Georgi}}\ and\ \bibinfo {author} {\bibfnamefont {L.}~\bibnamefont
  {Randall}},\ }\href {\doibase https://doi.org/10.1016/0370-2693(90)90055-B}
  {\bibfield  {journal} {\bibinfo  {journal} {Physics Letters B}\ }\textbf
  {\bibinfo {volume} {244}},\ \bibinfo {pages} {196 } (\bibinfo {year}
  {1990})}\BibitemShut {NoStop}%
\bibitem [{\citenamefont {Davidson}\ \emph {et~al.}(2005)\citenamefont
  {Davidson}, \citenamefont {Gorbahn},\ and\ \citenamefont
  {Santamaria}}]{Davidson:2005cs}%
  \BibitemOpen
  \bibfield  {author} {\bibinfo {author} {\bibfnamefont {S.}~\bibnamefont
  {Davidson}}, \bibinfo {author} {\bibfnamefont {M.}~\bibnamefont {Gorbahn}}, \
  and\ \bibinfo {author} {\bibfnamefont {A.}~\bibnamefont {Santamaria}},\
  }\href {\doibase 10.1016/j.physletb.2005.08.086} {\bibfield  {journal}
  {\bibinfo  {journal} {Phys. Lett.}\ }\textbf {\bibinfo {volume} {B626}},\
  \bibinfo {pages} {151} (\bibinfo {year} {2005})},\ \Eprint
  {http://arxiv.org/abs/hep-ph/0506085} {arXiv:hep-ph/0506085 [hep-ph]}
  \BibitemShut {NoStop}%
%%CITATION = HEP-PH/0506085;%%
\bibitem [{\citenamefont {Bell}\ \emph {et~al.}(2005)\citenamefont {Bell},
  \citenamefont {Cirigliano}, \citenamefont {Ramsey-Musolf}, \citenamefont
  {Vogel},\ and\ \citenamefont {Wise}}]{Bell:2005kz}%
  \BibitemOpen
  \bibfield  {author} {\bibinfo {author} {\bibfnamefont {N.~F.}\ \bibnamefont
  {Bell}}, \bibinfo {author} {\bibfnamefont {V.}~\bibnamefont {Cirigliano}},
  \bibinfo {author} {\bibfnamefont {M.~J.}\ \bibnamefont {Ramsey-Musolf}},
  \bibinfo {author} {\bibfnamefont {P.}~\bibnamefont {Vogel}}, \ and\ \bibinfo
  {author} {\bibfnamefont {M.~B.}\ \bibnamefont {Wise}},\ }\href {\doibase
  10.1103/PhysRevLett.95.151802} {\bibfield  {journal} {\bibinfo  {journal}
  {Phys. Rev. Lett.}\ }\textbf {\bibinfo {volume} {95}},\ \bibinfo {pages}
  {151802} (\bibinfo {year} {2005})},\ \Eprint
  {http://arxiv.org/abs/hep-ph/0504134} {arXiv:hep-ph/0504134 [hep-ph]}
  \BibitemShut {NoStop}%
%%CITATION = HEP-PH/0504134;%%
\bibitem [{\citenamefont {Lindner}\ \emph {et~al.}(2017)\citenamefont
  {Lindner}, \citenamefont {Radov?i?},\ and\ \citenamefont
  {Welter}}]{Lindner:2017uvt}%
  \BibitemOpen
  \bibfield  {author} {\bibinfo {author} {\bibfnamefont {M.}~\bibnamefont
  {Lindner}}, \bibinfo {author} {\bibfnamefont {B.}~\bibnamefont {Radov?i?}}, \
  and\ \bibinfo {author} {\bibfnamefont {J.}~\bibnamefont {Welter}},\ }\href
  {\doibase 10.1007/JHEP07(2017)139} {\bibfield  {journal} {\bibinfo  {journal}
  {JHEP}\ }\textbf {\bibinfo {volume} {07}},\ \bibinfo {pages} {139} (\bibinfo
  {year} {2017})},\ \Eprint {http://arxiv.org/abs/1706.02555} {arXiv:1706.02555
  [hep-ph]} \BibitemShut {NoStop}%
%%CITATION = ARXIV:1706.02555;%%
\bibitem [{\citenamefont {Dvali}\ and\ \citenamefont
  {Funcke}(2016)}]{Dvali:2016uhn}%
  \BibitemOpen
  \bibfield  {author} {\bibinfo {author} {\bibfnamefont {G.}~\bibnamefont
  {Dvali}}\ and\ \bibinfo {author} {\bibfnamefont {L.}~\bibnamefont {Funcke}},\
  }\href {\doibase 10.1103/PhysRevD.93.113002} {\bibfield  {journal} {\bibinfo
  {journal} {Phys. Rev.}\ }\textbf {\bibinfo {volume} {D93}},\ \bibinfo {pages}
  {113002} (\bibinfo {year} {2016})},\ \Eprint
  {http://arxiv.org/abs/1602.03191} {arXiv:1602.03191 [hep-ph]} \BibitemShut
  {NoStop}%
%%CITATION = ARXIV:1602.03191;%%
\bibitem [{\citenamefont {Fujikawa}\ and\ \citenamefont
  {Shrock}(1980)}]{Fujikawa:1980yx}%
  \BibitemOpen
  \bibfield  {author} {\bibinfo {author} {\bibfnamefont {K.}~\bibnamefont
  {Fujikawa}}\ and\ \bibinfo {author} {\bibfnamefont {R.}~\bibnamefont
  {Shrock}},\ }\href {\doibase 10.1103/PhysRevLett.45.963} {\bibfield
  {journal} {\bibinfo  {journal} {Phys. Rev. Lett.}\ }\textbf {\bibinfo
  {volume} {45}},\ \bibinfo {pages} {963} (\bibinfo {year} {1980})}\BibitemShut
  {NoStop}%
%%CITATION = PRLTA,45,963;%%
\bibitem [{\citenamefont {Beda}\ \emph {et~al.}(2013)\citenamefont {Beda},
  \citenamefont {Brudanin}, \citenamefont {Egorov}, \citenamefont {Medvedev},
  \citenamefont {Pogosov}, \citenamefont {Shevchik}, \citenamefont
  {Shirchenko}, \citenamefont {Starostin},\ and\ \citenamefont
  {Zhitnikov}}]{Beda:2013mta}%
  \BibitemOpen
  \bibfield  {author} {\bibinfo {author} {\bibfnamefont {A.~G.}\ \bibnamefont
  {Beda}}, \bibinfo {author} {\bibfnamefont {V.~B.}\ \bibnamefont {Brudanin}},
  \bibinfo {author} {\bibfnamefont {V.~G.}\ \bibnamefont {Egorov}}, \bibinfo
  {author} {\bibfnamefont {D.~V.}\ \bibnamefont {Medvedev}}, \bibinfo {author}
  {\bibfnamefont {V.~S.}\ \bibnamefont {Pogosov}}, \bibinfo {author}
  {\bibfnamefont {E.~A.}\ \bibnamefont {Shevchik}}, \bibinfo {author}
  {\bibfnamefont {M.~V.}\ \bibnamefont {Shirchenko}}, \bibinfo {author}
  {\bibfnamefont {A.~S.}\ \bibnamefont {Starostin}}, \ and\ \bibinfo {author}
  {\bibfnamefont {I.~V.}\ \bibnamefont {Zhitnikov}},\ }\href {\doibase
  10.1134/S1547477113020027} {\bibfield  {journal} {\bibinfo  {journal} {Phys.
  Part. Nucl. Lett.}\ }\textbf {\bibinfo {volume} {10}},\ \bibinfo {pages}
  {139} (\bibinfo {year} {2013})}\BibitemShut {NoStop}%
%%CITATION = 00438,10,139;%%
\bibitem [{\citenamefont {Agostini}\ \emph {et~al.}(2017)\citenamefont
  {Agostini} \emph {et~al.}}]{Borexino:2017fbd}%
  \BibitemOpen
  \bibfield  {author} {\bibinfo {author} {\bibfnamefont {M.}~\bibnamefont
  {Agostini}} \emph {et~al.} (\bibinfo {collaboration} {Borexino}),\ }\href
  {\doibase 10.1103/PhysRevD.96.091103} {\bibfield  {journal} {\bibinfo
  {journal} {Phys. Rev.}\ }\textbf {\bibinfo {volume} {D96}},\ \bibinfo {pages}
  {091103} (\bibinfo {year} {2017})},\ \Eprint
  {http://arxiv.org/abs/1707.09355} {arXiv:1707.09355 [hep-ex]} \BibitemShut
  {NoStop}%
%%CITATION = ARXIV:1707.09355;%%
\bibitem [{\citenamefont {Mirizzi}\ \emph {et~al.}(2007)\citenamefont
  {Mirizzi}, \citenamefont {Montanino},\ and\ \citenamefont
  {Serpico}}]{Mirizzi:2007jd}%
  \BibitemOpen
  \bibfield  {author} {\bibinfo {author} {\bibfnamefont {A.}~\bibnamefont
  {Mirizzi}}, \bibinfo {author} {\bibfnamefont {D.}~\bibnamefont {Montanino}},
  \ and\ \bibinfo {author} {\bibfnamefont {P.~D.}\ \bibnamefont {Serpico}},\
  }\href {\doibase 10.1103/PhysRevD.76.053007} {\bibfield  {journal} {\bibinfo
  {journal} {Phys. Rev.}\ }\textbf {\bibinfo {volume} {D76}},\ \bibinfo {pages}
  {053007} (\bibinfo {year} {2007})},\ \Eprint {http://arxiv.org/abs/0705.4667}
  {arXiv:0705.4667 [hep-ph]} \BibitemShut {NoStop}%
%%CITATION = ARXIV:0705.4667;%%
\bibitem [{\citenamefont {Aalberts}\ \emph {et~al.}(2018)\citenamefont
  {Aalberts} \emph {et~al.}}]{Aalberts:2018obr}%
  \BibitemOpen
  \bibfield  {author} {\bibinfo {author} {\bibfnamefont {J.~L.}\ \bibnamefont
  {Aalberts}} \emph {et~al.},\ }\href {\doibase 10.1103/PhysRevD.98.023001}
  {\bibfield  {journal} {\bibinfo  {journal} {Phys. Rev.}\ }\textbf {\bibinfo
  {volume} {D98}},\ \bibinfo {pages} {023001} (\bibinfo {year} {2018})},\
  \Eprint {http://arxiv.org/abs/1803.00588} {arXiv:1803.00588 [astro-ph.CO]}
  \BibitemShut {NoStop}%
%%CITATION = ARXIV:1803.00588;%%
\bibitem [{\citenamefont {Chianese}\ \emph {et~al.}(2019)\citenamefont
  {Chianese}, \citenamefont {Di~Bari}, \citenamefont {Farrag},\ and\
  \citenamefont {Samanta}}]{Chianese:2018luo}%
  \BibitemOpen
  \bibfield  {author} {\bibinfo {author} {\bibfnamefont {M.}~\bibnamefont
  {Chianese}}, \bibinfo {author} {\bibfnamefont {P.}~\bibnamefont {Di~Bari}},
  \bibinfo {author} {\bibfnamefont {K.}~\bibnamefont {Farrag}}, \ and\ \bibinfo
  {author} {\bibfnamefont {R.}~\bibnamefont {Samanta}},\ }\href {\doibase
  10.1016/j.physletb.2018.09.040} {\bibfield  {journal} {\bibinfo  {journal}
  {Phys. Lett.}\ }\textbf {\bibinfo {volume} {B790}},\ \bibinfo {pages} {64}
  (\bibinfo {year} {2019})},\ \Eprint {http://arxiv.org/abs/1805.11717}
  {arXiv:1805.11717 [hep-ph]} \BibitemShut {NoStop}%
%%CITATION = ARXIV:1805.11717;%%
\bibitem [{\citenamefont {Raffelt}(1990)}]{PhysRevLett.64.2856}%
  \BibitemOpen
  \bibfield  {author} {\bibinfo {author} {\bibfnamefont {G.~G.}\ \bibnamefont
  {Raffelt}},\ }\href {\doibase 10.1103/PhysRevLett.64.2856} {\bibfield
  {journal} {\bibinfo  {journal} {Phys. Rev. Lett.}\ }\textbf {\bibinfo
  {volume} {64}},\ \bibinfo {pages} {2856} (\bibinfo {year}
  {1990})}\BibitemShut {NoStop}%
\bibitem [{\citenamefont {Arceo-Díaz}\ \emph {et~al.}(2015)\citenamefont
  {Arceo-Díaz}, \citenamefont {Schröder}, \citenamefont {Zuber},\ and\
  \citenamefont {Jack}}]{ARCEODIAZ20151}%
  \BibitemOpen
  \bibfield  {author} {\bibinfo {author} {\bibfnamefont {S.}~\bibnamefont
  {Arceo-Díaz}}, \bibinfo {author} {\bibfnamefont {K.-P.}\ \bibnamefont
  {Schröder}}, \bibinfo {author} {\bibfnamefont {K.}~\bibnamefont {Zuber}},
  \ and\ \bibinfo {author} {\bibfnamefont {D.}~\bibnamefont {Jack}},\ }\href
  {\doibase https://doi.org/10.1016/j.astropartphys.2015.03.006} {\bibfield
  {journal} {\bibinfo  {journal} {Astroparticle Physics}\ }\textbf {\bibinfo
  {volume} {70}},\ \bibinfo {pages} {1 } (\bibinfo {year} {2015})}\BibitemShut
  {NoStop}%
\bibitem [{\citenamefont {Raffelt}(1999)}]{RAFFELT1999319}%
  \BibitemOpen
  \bibfield  {author} {\bibinfo {author} {\bibfnamefont {G.~G.}\ \bibnamefont
  {Raffelt}},\ }\href {\doibase https://doi.org/10.1016/S0370-1573(99)00074-5}
  {\bibfield  {journal} {\bibinfo  {journal} {Physics Reports}\ }\textbf
  {\bibinfo {volume} {320}},\ \bibinfo {pages} {319 } (\bibinfo {year}
  {1999})}\BibitemShut {NoStop}%
\bibitem [{\citenamefont {Beacom}\ and\ \citenamefont
  {Bell}(2002)}]{Beacom:2002cb}%
  \BibitemOpen
  \bibfield  {author} {\bibinfo {author} {\bibfnamefont {J.~F.}\ \bibnamefont
  {Beacom}}\ and\ \bibinfo {author} {\bibfnamefont {N.~F.}\ \bibnamefont
  {Bell}},\ }\href {\doibase 10.1103/PhysRevD.65.113009} {\bibfield  {journal}
  {\bibinfo  {journal} {Phys. Rev.}\ }\textbf {\bibinfo {volume} {D65}},\
  \bibinfo {pages} {113009} (\bibinfo {year} {2002})},\ \Eprint
  {http://arxiv.org/abs/hep-ph/0204111} {arXiv:hep-ph/0204111 [hep-ph]}
  \BibitemShut {NoStop}%
%%CITATION = HEP-PH/0204111;%%
\bibitem [{\citenamefont {Picoreti}\ \emph {et~al.}(2016)\citenamefont
  {Picoreti}, \citenamefont {Guzzo}, \citenamefont {de~Holanda},\ and\
  \citenamefont {Peres}}]{Picoreti:2015ika}%
  \BibitemOpen
  \bibfield  {author} {\bibinfo {author} {\bibfnamefont {R.}~\bibnamefont
  {Picoreti}}, \bibinfo {author} {\bibfnamefont {M.~M.}\ \bibnamefont {Guzzo}},
  \bibinfo {author} {\bibfnamefont {P.~C.}\ \bibnamefont {de~Holanda}}, \ and\
  \bibinfo {author} {\bibfnamefont {O.~L.~G.}\ \bibnamefont {Peres}},\ }\href
  {\doibase 10.1016/j.physletb.2016.08.007} {\bibfield  {journal} {\bibinfo
  {journal} {Phys. Lett.}\ }\textbf {\bibinfo {volume} {B761}},\ \bibinfo
  {pages} {70} (\bibinfo {year} {2016})},\ \Eprint
  {http://arxiv.org/abs/1506.08158} {arXiv:1506.08158 [hep-ph]} \BibitemShut
  {NoStop}%
%%CITATION = ARXIV:1506.08158;%%
\bibitem [{\citenamefont {Aharmim}\ \emph {et~al.}(2018)\citenamefont {Aharmim}
  \emph {et~al.}}]{SNO_nudec}%
  \BibitemOpen
  \bibfield  {author} {\bibinfo {author} {\bibfnamefont {B.}~\bibnamefont
  {Aharmim}} \emph {et~al.} (\bibinfo {collaboration} {SNO}),\ }\href@noop {}
  {\  (\bibinfo {year} {2018})},\ \Eprint {http://arxiv.org/abs/1812.01088}
  {arXiv:1812.01088 [hep-ex]} \BibitemShut {NoStop}%
%%CITATION = ARXIV:1812.01088;%%
\bibitem [{\citenamefont {Berryman}\ \emph {et~al.}(2015)\citenamefont
  {Berryman}, \citenamefont {de~Gouvea},\ and\ \citenamefont
  {Hernandez}}]{Berryman:2014qha}%
  \BibitemOpen
  \bibfield  {author} {\bibinfo {author} {\bibfnamefont {J.~M.}\ \bibnamefont
  {Berryman}}, \bibinfo {author} {\bibfnamefont {A.}~\bibnamefont {de~Gouvea}},
  \ and\ \bibinfo {author} {\bibfnamefont {D.}~\bibnamefont {Hernandez}},\
  }\href {\doibase 10.1103/PhysRevD.92.073003} {\bibfield  {journal} {\bibinfo
  {journal} {Phys. Rev.}\ }\textbf {\bibinfo {volume} {D92}},\ \bibinfo {pages}
  {073003} (\bibinfo {year} {2015})},\ \Eprint {http://arxiv.org/abs/1411.0308}
  {arXiv:1411.0308 [hep-ph]} \BibitemShut {NoStop}%
%%CITATION = ARXIV:1411.0308;%%
\bibitem [{\citenamefont {Funcke}\ \emph {et~al.}(2019)\citenamefont {Funcke},
  \citenamefont {Raffelt},\ and\ \citenamefont {Vitagliano}}]{Funcke:2019grs}%
  \BibitemOpen
  \bibfield  {author} {\bibinfo {author} {\bibfnamefont {L.}~\bibnamefont
  {Funcke}}, \bibinfo {author} {\bibfnamefont {G.}~\bibnamefont {Raffelt}}, \
  and\ \bibinfo {author} {\bibfnamefont {E.}~\bibnamefont {Vitagliano}},\
  }\href@noop {} {\  (\bibinfo {year} {2019})},\ \Eprint
  {http://arxiv.org/abs/1905.01264} {arXiv:1905.01264 [hep-ph]} \BibitemShut
  {NoStop}%
%%CITATION = ARXIV:1905.01264;%%
\bibitem [{\citenamefont {Gonzalez-Garcia}\ and\ \citenamefont
  {Maltoni}(2008)}]{GonzalezGarcia:2008ru}%
  \BibitemOpen
  \bibfield  {author} {\bibinfo {author} {\bibfnamefont {M.~C.}\ \bibnamefont
  {Gonzalez-Garcia}}\ and\ \bibinfo {author} {\bibfnamefont {M.}~\bibnamefont
  {Maltoni}},\ }\href {\doibase 10.1016/j.physletb.2008.04.041} {\bibfield
  {journal} {\bibinfo  {journal} {Phys. Lett.}\ }\textbf {\bibinfo {volume}
  {B663}},\ \bibinfo {pages} {405} (\bibinfo {year} {2008})},\ \Eprint
  {http://arxiv.org/abs/0802.3699} {arXiv:0802.3699 [hep-ph]} \BibitemShut
  {NoStop}%
%%CITATION = ARXIV:0802.3699;%%
\bibitem [{\citenamefont {Gomes}\ \emph {et~al.}(2015)\citenamefont {Gomes},
  \citenamefont {Gomes},\ and\ \citenamefont {Peres}}]{Gomes:2014yua}%
  \BibitemOpen
  \bibfield  {author} {\bibinfo {author} {\bibfnamefont {R.~A.}\ \bibnamefont
  {Gomes}}, \bibinfo {author} {\bibfnamefont {A.~L.~G.}\ \bibnamefont {Gomes}},
  \ and\ \bibinfo {author} {\bibfnamefont {O.~L.~G.}\ \bibnamefont {Peres}},\
  }\href {\doibase 10.1016/j.physletb.2014.12.014} {\bibfield  {journal}
  {\bibinfo  {journal} {Phys. Lett.}\ }\textbf {\bibinfo {volume} {B740}},\
  \bibinfo {pages} {345} (\bibinfo {year} {2015})},\ \Eprint
  {http://arxiv.org/abs/1407.5640} {arXiv:1407.5640 [hep-ph]} \BibitemShut
  {NoStop}%
%%CITATION = ARXIV:1407.5640;%%
\bibitem [{\citenamefont {Gago}\ \emph {et~al.}(2017)\citenamefont {Gago},
  \citenamefont {Gomes}, \citenamefont {Gomes}, \citenamefont {Jones-Perez},\
  and\ \citenamefont {Peres}}]{Gago:2017zzy}%
  \BibitemOpen
  \bibfield  {author} {\bibinfo {author} {\bibfnamefont {A.~M.}\ \bibnamefont
  {Gago}}, \bibinfo {author} {\bibfnamefont {R.~A.}\ \bibnamefont {Gomes}},
  \bibinfo {author} {\bibfnamefont {A.~L.~G.}\ \bibnamefont {Gomes}}, \bibinfo
  {author} {\bibfnamefont {J.}~\bibnamefont {Jones-Perez}}, \ and\ \bibinfo
  {author} {\bibfnamefont {O.~L.~G.}\ \bibnamefont {Peres}},\ }\href {\doibase
  10.1007/JHEP11(2017)022} {\bibfield  {journal} {\bibinfo  {journal} {JHEP}\
  }\textbf {\bibinfo {volume} {11}},\ \bibinfo {pages} {022} (\bibinfo {year}
  {2017})},\ \Eprint {http://arxiv.org/abs/1705.03074} {arXiv:1705.03074
  [hep-ph]} \BibitemShut {NoStop}%
%%CITATION = ARXIV:1705.03074;%%
\bibitem [{\citenamefont {Choubey}\ \emph
  {et~al.}(2018{\natexlab{a}})\citenamefont {Choubey}, \citenamefont {Dutta},\
  and\ \citenamefont {Pramanik}}]{Choubey:2018cfz}%
  \BibitemOpen
  \bibfield  {author} {\bibinfo {author} {\bibfnamefont {S.}~\bibnamefont
  {Choubey}}, \bibinfo {author} {\bibfnamefont {D.}~\bibnamefont {Dutta}}, \
  and\ \bibinfo {author} {\bibfnamefont {D.}~\bibnamefont {Pramanik}},\ }\href
  {\doibase 10.1007/JHEP08(2018)141} {\bibfield  {journal} {\bibinfo  {journal}
  {JHEP}\ }\textbf {\bibinfo {volume} {08}},\ \bibinfo {pages} {141} (\bibinfo
  {year} {2018}{\natexlab{a}})},\ \Eprint {http://arxiv.org/abs/1805.01848}
  {arXiv:1805.01848 [hep-ph]} \BibitemShut {NoStop}%
%%CITATION = ARXIV:1805.01848;%%
\bibitem [{\citenamefont {Hannestad}(2005)}]{Hannestad:2004qu}%
  \BibitemOpen
  \bibfield  {author} {\bibinfo {author} {\bibfnamefont {S.}~\bibnamefont
  {Hannestad}},\ }\href {\doibase 10.1088/1475-7516/2005/02/011} {\bibfield
  {journal} {\bibinfo  {journal} {JCAP}\ }\textbf {\bibinfo {volume} {0502}},\
  \bibinfo {pages} {011} (\bibinfo {year} {2005})},\ \Eprint
  {http://arxiv.org/abs/astro-ph/0411475} {arXiv:astro-ph/0411475 [astro-ph]}
  \BibitemShut {NoStop}%
%%CITATION = ASTRO-PH/0411475;%%
\bibitem [{\citenamefont {Hannestad}\ and\ \citenamefont
  {Raffelt}(2005)}]{Hannestad:2005ex}%
  \BibitemOpen
  \bibfield  {author} {\bibinfo {author} {\bibfnamefont {S.}~\bibnamefont
  {Hannestad}}\ and\ \bibinfo {author} {\bibfnamefont {G.}~\bibnamefont
  {Raffelt}},\ }\href {\doibase 10.1103/PhysRevD.72.103514} {\bibfield
  {journal} {\bibinfo  {journal} {Phys. Rev.}\ }\textbf {\bibinfo {volume}
  {D72}},\ \bibinfo {pages} {103514} (\bibinfo {year} {2005})},\ \Eprint
  {http://arxiv.org/abs/hep-ph/0509278} {arXiv:hep-ph/0509278 [hep-ph]}
  \BibitemShut {NoStop}%
%%CITATION = HEP-PH/0509278;%%
\bibitem [{\citenamefont {Basboll}\ \emph {et~al.}(2009)\citenamefont
  {Basboll}, \citenamefont {Bjaelde}, \citenamefont {Hannestad},\ and\
  \citenamefont {Raffelt}}]{Basboll:2008fx}%
  \BibitemOpen
  \bibfield  {author} {\bibinfo {author} {\bibfnamefont {A.}~\bibnamefont
  {Basboll}}, \bibinfo {author} {\bibfnamefont {O.~E.}\ \bibnamefont
  {Bjaelde}}, \bibinfo {author} {\bibfnamefont {S.}~\bibnamefont {Hannestad}},
  \ and\ \bibinfo {author} {\bibfnamefont {G.~G.}\ \bibnamefont {Raffelt}},\
  }\href {\doibase 10.1103/PhysRevD.79.043512} {\bibfield  {journal} {\bibinfo
  {journal} {Phys. Rev.}\ }\textbf {\bibinfo {volume} {D79}},\ \bibinfo {pages}
  {043512} (\bibinfo {year} {2009})},\ \Eprint {http://arxiv.org/abs/0806.1735}
  {arXiv:0806.1735 [astro-ph]} \BibitemShut {NoStop}%
%%CITATION = ARXIV:0806.1735;%%
\bibitem [{\citenamefont {Archidiacono}\ and\ \citenamefont
  {Hannestad}(2014)}]{Archidiacono:2013dua}%
  \BibitemOpen
  \bibfield  {author} {\bibinfo {author} {\bibfnamefont {M.}~\bibnamefont
  {Archidiacono}}\ and\ \bibinfo {author} {\bibfnamefont {S.}~\bibnamefont
  {Hannestad}},\ }\href {\doibase 10.1088/1475-7516/2014/07/046} {\bibfield
  {journal} {\bibinfo  {journal} {JCAP}\ }\textbf {\bibinfo {volume} {1407}},\
  \bibinfo {pages} {046} (\bibinfo {year} {2014})},\ \Eprint
  {http://arxiv.org/abs/1311.3873} {arXiv:1311.3873 [astro-ph.CO]} \BibitemShut
  {NoStop}%
%%CITATION = ARXIV:1311.3873;%%
\bibitem [{\citenamefont {Bell}\ \emph {et~al.}(2006)\citenamefont {Bell},
  \citenamefont {Pierpaoli},\ and\ \citenamefont {Sigurdson}}]{Bell:2005dr}%
  \BibitemOpen
  \bibfield  {author} {\bibinfo {author} {\bibfnamefont {N.~F.}\ \bibnamefont
  {Bell}}, \bibinfo {author} {\bibfnamefont {E.}~\bibnamefont {Pierpaoli}}, \
  and\ \bibinfo {author} {\bibfnamefont {K.}~\bibnamefont {Sigurdson}},\ }\href
  {\doibase 10.1103/PhysRevD.73.063523} {\bibfield  {journal} {\bibinfo
  {journal} {Phys. Rev.}\ }\textbf {\bibinfo {volume} {D73}},\ \bibinfo {pages}
  {063523} (\bibinfo {year} {2006})},\ \Eprint
  {http://arxiv.org/abs/astro-ph/0511410} {arXiv:astro-ph/0511410 [astro-ph]}
  \BibitemShut {NoStop}%
%%CITATION = ASTRO-PH/0511410;%%
\bibitem [{\citenamefont {Kolb}\ \emph {et~al.}(1991)\citenamefont {Kolb},
  \citenamefont {Turner}, \citenamefont {Chakravorty},\ and\ \citenamefont
  {Schramm}}]{Kolb:1991sn}%
  \BibitemOpen
  \bibfield  {author} {\bibinfo {author} {\bibfnamefont {E.~W.}\ \bibnamefont
  {Kolb}}, \bibinfo {author} {\bibfnamefont {M.~S.}\ \bibnamefont {Turner}},
  \bibinfo {author} {\bibfnamefont {A.}~\bibnamefont {Chakravorty}}, \ and\
  \bibinfo {author} {\bibfnamefont {D.~N.}\ \bibnamefont {Schramm}},\ }\href
  {\doibase 10.1103/PhysRevLett.67.533} {\bibfield  {journal} {\bibinfo
  {journal} {Phys. Rev. Lett.}\ }\textbf {\bibinfo {volume} {67}},\ \bibinfo
  {pages} {533} (\bibinfo {year} {1991})},\ \bibinfo {note}
  {[,272(1991)]}\BibitemShut {NoStop}%
%%CITATION = PRLTA,67,533;%%
\bibitem [{\citenamefont {Dolgov}\ and\ \citenamefont
  {Rothstein}(1993)}]{Dolgov:1993uz}%
  \BibitemOpen
  \bibfield  {author} {\bibinfo {author} {\bibfnamefont {A.~D.}\ \bibnamefont
  {Dolgov}}\ and\ \bibinfo {author} {\bibfnamefont {I.~Z.}\ \bibnamefont
  {Rothstein}},\ }\href {\doibase 10.1103/PhysRevLett.71.476} {\bibfield
  {journal} {\bibinfo  {journal} {Phys. Rev. Lett.}\ }\textbf {\bibinfo
  {volume} {71}},\ \bibinfo {pages} {476} (\bibinfo {year} {1993})},\ \Eprint
  {http://arxiv.org/abs/hep-ph/9302292} {arXiv:hep-ph/9302292 [hep-ph]}
  \BibitemShut {NoStop}%
%%CITATION = HEP-PH/9302292;%%
\bibitem [{\citenamefont {Kawasaki}\ \emph {et~al.}(1994)\citenamefont
  {Kawasaki}, \citenamefont {Kernan}, \citenamefont {Kang}, \citenamefont
  {Scherrer}, \citenamefont {Steigman},\ and\ \citenamefont
  {Walker}}]{Kawasaki:1993gz}%
  \BibitemOpen
  \bibfield  {author} {\bibinfo {author} {\bibfnamefont {M.}~\bibnamefont
  {Kawasaki}}, \bibinfo {author} {\bibfnamefont {P.}~\bibnamefont {Kernan}},
  \bibinfo {author} {\bibfnamefont {H.-S.}\ \bibnamefont {Kang}}, \bibinfo
  {author} {\bibfnamefont {R.~J.}\ \bibnamefont {Scherrer}}, \bibinfo {author}
  {\bibfnamefont {G.}~\bibnamefont {Steigman}}, \ and\ \bibinfo {author}
  {\bibfnamefont {T.~P.}\ \bibnamefont {Walker}},\ }\href {\doibase
  10.1016/0550-3213(94)90359-X} {\bibfield  {journal} {\bibinfo  {journal}
  {Nucl. Phys.}\ }\textbf {\bibinfo {volume} {B419}},\ \bibinfo {pages} {105}
  (\bibinfo {year} {1994})}\BibitemShut {NoStop}%
%%CITATION = NUPHA,B419,105;%%
\bibitem [{\citenamefont {Dodelson}\ \emph {et~al.}(1994)\citenamefont
  {Dodelson}, \citenamefont {Gyuk},\ and\ \citenamefont
  {Turner}}]{Dodelson:1994it}%
  \BibitemOpen
  \bibfield  {author} {\bibinfo {author} {\bibfnamefont {S.}~\bibnamefont
  {Dodelson}}, \bibinfo {author} {\bibfnamefont {G.}~\bibnamefont {Gyuk}}, \
  and\ \bibinfo {author} {\bibfnamefont {M.~S.}\ \bibnamefont {Turner}},\
  }\href {\doibase 10.1103/PhysRevLett.72.3754} {\bibfield  {journal} {\bibinfo
   {journal} {Phys. Rev. Lett.}\ }\textbf {\bibinfo {volume} {72}},\ \bibinfo
  {pages} {3754} (\bibinfo {year} {1994})},\ \Eprint
  {http://arxiv.org/abs/astro-ph/9402028} {arXiv:astro-ph/9402028 [astro-ph]}
  \BibitemShut {NoStop}%
%%CITATION = ASTRO-PH/9402028;%%
\bibitem [{\citenamefont {Dolgov}\ \emph {et~al.}(1997)\citenamefont {Dolgov},
  \citenamefont {Pastor}, \citenamefont {Romao},\ and\ \citenamefont
  {Valle}}]{Dolgov:1996fp}%
  \BibitemOpen
  \bibfield  {author} {\bibinfo {author} {\bibfnamefont {A.~D.}\ \bibnamefont
  {Dolgov}}, \bibinfo {author} {\bibfnamefont {S.}~\bibnamefont {Pastor}},
  \bibinfo {author} {\bibfnamefont {J.~C.}\ \bibnamefont {Romao}}, \ and\
  \bibinfo {author} {\bibfnamefont {J.~W.~F.}\ \bibnamefont {Valle}},\ }\href
  {\doibase 10.1016/S0550-3213(97)00213-7} {\bibfield  {journal} {\bibinfo
  {journal} {Nucl. Phys.}\ }\textbf {\bibinfo {volume} {B496}},\ \bibinfo
  {pages} {24} (\bibinfo {year} {1997})},\ \Eprint
  {http://arxiv.org/abs/hep-ph/9610507} {arXiv:hep-ph/9610507 [hep-ph]}
  \BibitemShut {NoStop}%
%%CITATION = HEP-PH/9610507;%%
\bibitem [{\citenamefont {Hannestad}(1998)}]{Hannestad:1997ar}%
  \BibitemOpen
  \bibfield  {author} {\bibinfo {author} {\bibfnamefont {S.}~\bibnamefont
  {Hannestad}},\ }\href {\doibase 10.1103/PhysRevD.57.2213} {\bibfield
  {journal} {\bibinfo  {journal} {Phys. Rev.}\ }\textbf {\bibinfo {volume}
  {D57}},\ \bibinfo {pages} {2213} (\bibinfo {year} {1998})},\ \Eprint
  {http://arxiv.org/abs/hep-ph/9711249} {arXiv:hep-ph/9711249 [hep-ph]}
  \BibitemShut {NoStop}%
%%CITATION = HEP-PH/9711249;%%
\bibitem [{\citenamefont {Adam}\ \emph {et~al.}(2016)\citenamefont {Adam} \emph
  {et~al.}}]{Adam:2015rua}%
  \BibitemOpen
  \bibfield  {author} {\bibinfo {author} {\bibfnamefont {R.}~\bibnamefont
  {Adam}} \emph {et~al.} (\bibinfo {collaboration} {Planck}),\ }\href {\doibase
  10.1051/0004-6361/201527101} {\bibfield  {journal} {\bibinfo  {journal}
  {Astron. Astrophys.}\ }\textbf {\bibinfo {volume} {594}},\ \bibinfo {pages}
  {A1} (\bibinfo {year} {2016})},\ \Eprint {http://arxiv.org/abs/1502.01582}
  {arXiv:1502.01582 [astro-ph.CO]} \BibitemShut {NoStop}%
%%CITATION = ARXIV:1502.01582;%%
\bibitem [{\citenamefont {Ade}\ \emph {et~al.}(2016)\citenamefont {Ade} \emph
  {et~al.}}]{Ade:2015xua}%
  \BibitemOpen
  \bibfield  {author} {\bibinfo {author} {\bibfnamefont {P.~A.~R.}\
  \bibnamefont {Ade}} \emph {et~al.} (\bibinfo {collaboration} {Planck}),\
  }\href {\doibase 10.1051/0004-6361/201525830} {\bibfield  {journal} {\bibinfo
   {journal} {Astron. Astrophys.}\ }\textbf {\bibinfo {volume} {594}},\
  \bibinfo {pages} {A13} (\bibinfo {year} {2016})},\ \Eprint
  {http://arxiv.org/abs/1502.01589} {arXiv:1502.01589 [astro-ph.CO]}
  \BibitemShut {NoStop}%
%%CITATION = ARXIV:1502.01589;%%
\bibitem [{\citenamefont {Bashinsky}\ and\ \citenamefont
  {Seljak}(2004)}]{Bashinsky:2003tk}%
  \BibitemOpen
  \bibfield  {author} {\bibinfo {author} {\bibfnamefont {S.}~\bibnamefont
  {Bashinsky}}\ and\ \bibinfo {author} {\bibfnamefont {U.}~\bibnamefont
  {Seljak}},\ }\href {\doibase 10.1103/PhysRevD.69.083002} {\bibfield
  {journal} {\bibinfo  {journal} {Phys. Rev.}\ }\textbf {\bibinfo {volume}
  {D69}},\ \bibinfo {pages} {083002} (\bibinfo {year} {2004})},\ \Eprint
  {http://arxiv.org/abs/astro-ph/0310198} {arXiv:astro-ph/0310198 [astro-ph]}
  \BibitemShut {NoStop}%
%%CITATION = ASTRO-PH/0310198;%%
\bibitem [{\citenamefont {Trotta}\ and\ \citenamefont
  {Melchiorri}(2005)}]{Trotta:2004ty}%
  \BibitemOpen
  \bibfield  {author} {\bibinfo {author} {\bibfnamefont {R.}~\bibnamefont
  {Trotta}}\ and\ \bibinfo {author} {\bibfnamefont {A.}~\bibnamefont
  {Melchiorri}},\ }\href {\doibase 10.1103/PhysRevLett.95.011305} {\bibfield
  {journal} {\bibinfo  {journal} {Phys. Rev. Lett.}\ }\textbf {\bibinfo
  {volume} {95}},\ \bibinfo {pages} {011305} (\bibinfo {year} {2005})},\
  \Eprint {http://arxiv.org/abs/astro-ph/0412066} {arXiv:astro-ph/0412066
  [astro-ph]} \BibitemShut {NoStop}%
%%CITATION = ASTRO-PH/0412066;%%
\bibitem [{\citenamefont {de~Salas}\ \emph {et~al.}(2019)\citenamefont
  {de~Salas}, \citenamefont {Pastor}, \citenamefont {Ternes}, \citenamefont
  {Thakore},\ and\ \citenamefont {Tórtola}}]{deSalas:2018kri}%
  \BibitemOpen
  \bibfield  {author} {\bibinfo {author} {\bibfnamefont {P.~F.}\ \bibnamefont
  {de~Salas}}, \bibinfo {author} {\bibfnamefont {S.}~\bibnamefont {Pastor}},
  \bibinfo {author} {\bibfnamefont {C.~A.}\ \bibnamefont {Ternes}}, \bibinfo
  {author} {\bibfnamefont {T.}~\bibnamefont {Thakore}}, \ and\ \bibinfo
  {author} {\bibfnamefont {M.}~\bibnamefont {Tórtola}},\ }\href {\doibase
  10.1016/j.physletb.2018.12.066} {\bibfield  {journal} {\bibinfo  {journal}
  {Phys. Lett.}\ }\textbf {\bibinfo {volume} {B789}},\ \bibinfo {pages} {472}
  (\bibinfo {year} {2019})},\ \Eprint {http://arxiv.org/abs/1810.10916}
  {arXiv:1810.10916 [hep-ph]} \BibitemShut {NoStop}%
%%CITATION = ARXIV:1810.10916;%%
\bibitem [{\citenamefont {Denton}\ and\ \citenamefont
  {Tamborra}(2018)}]{Denton:2018aml}%
  \BibitemOpen
  \bibfield  {author} {\bibinfo {author} {\bibfnamefont {P.~B.}\ \bibnamefont
  {Denton}}\ and\ \bibinfo {author} {\bibfnamefont {I.}~\bibnamefont
  {Tamborra}},\ }\href {\doibase 10.1103/PhysRevLett.121.121802} {\bibfield
  {journal} {\bibinfo  {journal} {Phys. Rev. Lett.}\ }\textbf {\bibinfo
  {volume} {121}},\ \bibinfo {pages} {121802} (\bibinfo {year} {2018})},\
  \Eprint {http://arxiv.org/abs/1805.05950} {arXiv:1805.05950 [hep-ph]}
  \BibitemShut {NoStop}%
%%CITATION = ARXIV:1805.05950;%%
\bibitem [{\citenamefont {Dolgov}(2002)}]{Dolgov:2002wy}%
  \BibitemOpen
  \bibfield  {author} {\bibinfo {author} {\bibfnamefont {A.~D.}\ \bibnamefont
  {Dolgov}},\ }\href {\doibase 10.1016/S0370-1573(02)00139-4} {\bibfield
  {journal} {\bibinfo  {journal} {Phys. Rept.}\ }\textbf {\bibinfo {volume}
  {370}},\ \bibinfo {pages} {333} (\bibinfo {year} {2002})},\ \Eprint
  {http://arxiv.org/abs/hep-ph/0202122} {arXiv:hep-ph/0202122 [hep-ph]}
  \BibitemShut {NoStop}%
%%CITATION = HEP-PH/0202122;%%
\bibitem [{\citenamefont {de~Salas}\ and\ \citenamefont
  {Pastor}(2016)}]{deSalas:2016ztq}%
  \BibitemOpen
  \bibfield  {author} {\bibinfo {author} {\bibfnamefont {P.~F.}\ \bibnamefont
  {de~Salas}}\ and\ \bibinfo {author} {\bibfnamefont {S.}~\bibnamefont
  {Pastor}},\ }\href {\doibase 10.1088/1475-7516/2016/07/051} {\bibfield
  {journal} {\bibinfo  {journal} {JCAP}\ }\textbf {\bibinfo {volume} {1607}},\
  \bibinfo {pages} {051} (\bibinfo {year} {2016})},\ \Eprint
  {http://arxiv.org/abs/1606.06986} {arXiv:1606.06986 [hep-ph]} \BibitemShut
  {NoStop}%
%%CITATION = ARXIV:1606.06986;%%
\bibitem [{\citenamefont {Mangano}\ \emph {et~al.}(2005)\citenamefont
  {Mangano}, \citenamefont {Miele}, \citenamefont {Pastor}, \citenamefont
  {Pinto}, \citenamefont {Pisanti},\ and\ \citenamefont
  {Serpico}}]{Mangano:2005cc}%
  \BibitemOpen
  \bibfield  {author} {\bibinfo {author} {\bibfnamefont {G.}~\bibnamefont
  {Mangano}}, \bibinfo {author} {\bibfnamefont {G.}~\bibnamefont {Miele}},
  \bibinfo {author} {\bibfnamefont {S.}~\bibnamefont {Pastor}}, \bibinfo
  {author} {\bibfnamefont {T.}~\bibnamefont {Pinto}}, \bibinfo {author}
  {\bibfnamefont {O.}~\bibnamefont {Pisanti}}, \ and\ \bibinfo {author}
  {\bibfnamefont {P.~D.}\ \bibnamefont {Serpico}},\ }\href {\doibase
  10.1016/j.nuclphysb.2005.09.041} {\bibfield  {journal} {\bibinfo  {journal}
  {Nucl. Phys.}\ }\textbf {\bibinfo {volume} {B729}},\ \bibinfo {pages} {221}
  (\bibinfo {year} {2005})},\ \Eprint {http://arxiv.org/abs/hep-ph/0506164}
  {arXiv:hep-ph/0506164 [hep-ph]} \BibitemShut {NoStop}%
%%CITATION = HEP-PH/0506164;%%
\bibitem [{\citenamefont {Pitrou}\ \emph {et~al.}(2018)\citenamefont {Pitrou},
  \citenamefont {Coc}, \citenamefont {Uzan},\ and\ \citenamefont
  {Vangioni}}]{Pitrou:2018cgg}%
  \BibitemOpen
  \bibfield  {author} {\bibinfo {author} {\bibfnamefont {C.}~\bibnamefont
  {Pitrou}}, \bibinfo {author} {\bibfnamefont {A.}~\bibnamefont {Coc}},
  \bibinfo {author} {\bibfnamefont {J.-P.}\ \bibnamefont {Uzan}}, \ and\
  \bibinfo {author} {\bibfnamefont {E.}~\bibnamefont {Vangioni}},\ }\href
  {\doibase 10.1016/j.physrep.2018.04.005} {\bibfield  {journal} {\bibinfo
  {journal} {Phys. Rept.}\ }\textbf {\bibinfo {volume} {754}},\ \bibinfo
  {pages} {1} (\bibinfo {year} {2018})},\ \Eprint
  {http://arxiv.org/abs/1801.08023} {arXiv:1801.08023 [astro-ph.CO]}
  \BibitemShut {NoStop}%
%%CITATION = ARXIV:1801.08023;%%
\bibitem [{\citenamefont {Cyburt}\ \emph {et~al.}(2016)\citenamefont {Cyburt},
  \citenamefont {Fields}, \citenamefont {Olive},\ and\ \citenamefont
  {Yeh}}]{Cyburt:2015mya}%
  \BibitemOpen
  \bibfield  {author} {\bibinfo {author} {\bibfnamefont {R.~H.}\ \bibnamefont
  {Cyburt}}, \bibinfo {author} {\bibfnamefont {B.~D.}\ \bibnamefont {Fields}},
  \bibinfo {author} {\bibfnamefont {K.~A.}\ \bibnamefont {Olive}}, \ and\
  \bibinfo {author} {\bibfnamefont {T.-H.}\ \bibnamefont {Yeh}},\ }\href
  {\doibase 10.1103/RevModPhys.88.015004} {\bibfield  {journal} {\bibinfo
  {journal} {Rev. Mod. Phys.}\ }\textbf {\bibinfo {volume} {88}},\ \bibinfo
  {pages} {015004} (\bibinfo {year} {2016})},\ \Eprint
  {http://arxiv.org/abs/1505.01076} {arXiv:1505.01076 [astro-ph.CO]}
  \BibitemShut {NoStop}%
%%CITATION = ARXIV:1505.01076;%%
\bibitem [{\citenamefont {Escudero}(2019)}]{Escudero:2018mvt}%
  \BibitemOpen
  \bibfield  {author} {\bibinfo {author} {\bibfnamefont {M.}~\bibnamefont
  {Escudero}},\ }\href {\doibase 10.1088/1475-7516/2019/02/007} {\bibfield
  {journal} {\bibinfo  {journal} {JCAP}\ }\textbf {\bibinfo {volume} {1902}},\
  \bibinfo {pages} {007} (\bibinfo {year} {2019})},\ \Eprint
  {http://arxiv.org/abs/1812.05605} {arXiv:1812.05605 [hep-ph]} \BibitemShut
  {NoStop}%
%%CITATION = ARXIV:1812.05605;%%
\bibitem [{\citenamefont {Escudero}(2020)}]{Escudero:2019new}%
  \BibitemOpen
  \bibfield  {author} {\bibinfo {author} {\bibfnamefont {M.}~\bibnamefont
  {Escudero}},\ }\href@noop {} {\  (\bibinfo {year} {2020})},\ \Eprint
  {http://arxiv.org/abs/2001.04466} {arXiv:2001.04466 [hep-ph]} \BibitemShut
  {NoStop}%
%%CITATION = ARXIV:2001.04466;%%
\bibitem [{\citenamefont {Tanabashi}\ \emph {et~al.}(2018)\citenamefont
  {Tanabashi} \emph {et~al.}}]{pdg}%
  \BibitemOpen
  \bibfield  {author} {\bibinfo {author} {\bibfnamefont {M.}~\bibnamefont
  {Tanabashi}} \emph {et~al.} (\bibinfo {collaboration} {ParticleDataGroup}),\
  }\href {\doibase 10.1103/PhysRevD.98.030001} {\bibfield  {journal} {\bibinfo
  {journal} {Phys. Rev.}\ }\textbf {\bibinfo {volume} {D98}},\ \bibinfo {pages}
  {030001} (\bibinfo {year} {2018})}\BibitemShut {NoStop}%
%%CITATION = PHRVA,D98,030001;%%
\bibitem [{\citenamefont {Aghanim}\ \emph {et~al.}(2018)\citenamefont {Aghanim}
  \emph {et~al.}}]{Aghanim:2018eyx}%
  \BibitemOpen
  \bibfield  {author} {\bibinfo {author} {\bibfnamefont {N.}~\bibnamefont
  {Aghanim}} \emph {et~al.} (\bibinfo {collaboration} {Planck}),\ }\href@noop
  {} {\  (\bibinfo {year} {2018})},\ \Eprint {http://arxiv.org/abs/1807.06209}
  {arXiv:1807.06209 [astro-ph.CO]} \BibitemShut {NoStop}%
%%CITATION = ARXIV:1807.06209;%%
\bibitem [{\citenamefont {Abazajian}\ \emph {et~al.}(2016)\citenamefont
  {Abazajian} \emph {et~al.}}]{Abazajian:2016yjj}%
  \BibitemOpen
  \bibfield  {author} {\bibinfo {author} {\bibfnamefont {K.~N.}\ \bibnamefont
  {Abazajian}} \emph {et~al.} (\bibinfo {collaboration} {CMB-S4}),\ }\href@noop
  {} {\  (\bibinfo {year} {2016})},\ \Eprint {http://arxiv.org/abs/1610.02743}
  {arXiv:1610.02743 [astro-ph.CO]} \BibitemShut {NoStop}%
%%CITATION = ARXIV:1610.02743;%%
\bibitem [{\citenamefont {Aver}\ \emph {et~al.}(2015)\citenamefont {Aver},
  \citenamefont {Olive},\ and\ \citenamefont {Skillman}}]{Aver:2015iza}%
  \BibitemOpen
  \bibfield  {author} {\bibinfo {author} {\bibfnamefont {E.}~\bibnamefont
  {Aver}}, \bibinfo {author} {\bibfnamefont {K.~A.}\ \bibnamefont {Olive}}, \
  and\ \bibinfo {author} {\bibfnamefont {E.~D.}\ \bibnamefont {Skillman}},\
  }\href {\doibase 10.1088/1475-7516/2015/07/011} {\bibfield  {journal}
  {\bibinfo  {journal} {JCAP}\ }\textbf {\bibinfo {volume} {1507}},\ \bibinfo
  {pages} {011} (\bibinfo {year} {2015})},\ \Eprint
  {http://arxiv.org/abs/1503.08146} {arXiv:1503.08146 [astro-ph.CO]}
  \BibitemShut {NoStop}%
%%CITATION = ARXIV:1503.08146;%%
\bibitem [{\citenamefont {Cooke}\ \emph {et~al.}(2018)\citenamefont {Cooke},
  \citenamefont {Pettini},\ and\ \citenamefont {Steidel}}]{Cooke:2017cwo}%
  \BibitemOpen
  \bibfield  {author} {\bibinfo {author} {\bibfnamefont {R.~J.}\ \bibnamefont
  {Cooke}}, \bibinfo {author} {\bibfnamefont {M.}~\bibnamefont {Pettini}}, \
  and\ \bibinfo {author} {\bibfnamefont {C.~C.}\ \bibnamefont {Steidel}},\
  }\href {\doibase 10.3847/1538-4357/aaab53} {\bibfield  {journal} {\bibinfo
  {journal} {Astrophys. J.}\ }\textbf {\bibinfo {volume} {855}},\ \bibinfo
  {pages} {102} (\bibinfo {year} {2018})},\ \Eprint
  {http://arxiv.org/abs/1710.11129} {arXiv:1710.11129 [astro-ph.CO]}
  \BibitemShut {NoStop}%
%%CITATION = ARXIV:1710.11129;%%
\bibitem [{\citenamefont {Sarkar}(1996)}]{Sarkar:1995dd}%
  \BibitemOpen
  \bibfield  {author} {\bibinfo {author} {\bibfnamefont {S.}~\bibnamefont
  {Sarkar}},\ }\href {\doibase 10.1088/0034-4885/59/12/001} {\bibfield
  {journal} {\bibinfo  {journal} {Rept. Prog. Phys.}\ }\textbf {\bibinfo
  {volume} {59}},\ \bibinfo {pages} {1493} (\bibinfo {year} {1996})},\ \Eprint
  {http://arxiv.org/abs/hep-ph/9602260} {arXiv:hep-ph/9602260 [hep-ph]}
  \BibitemShut {NoStop}%
%%CITATION = HEP-PH/9602260;%%
\bibitem [{\citenamefont {Iocco}\ \emph {et~al.}(2009)\citenamefont {Iocco},
  \citenamefont {Mangano}, \citenamefont {Miele}, \citenamefont {Pisanti},\
  and\ \citenamefont {Serpico}}]{Iocco:2008va}%
  \BibitemOpen
  \bibfield  {author} {\bibinfo {author} {\bibfnamefont {F.}~\bibnamefont
  {Iocco}}, \bibinfo {author} {\bibfnamefont {G.}~\bibnamefont {Mangano}},
  \bibinfo {author} {\bibfnamefont {G.}~\bibnamefont {Miele}}, \bibinfo
  {author} {\bibfnamefont {O.}~\bibnamefont {Pisanti}}, \ and\ \bibinfo
  {author} {\bibfnamefont {P.~D.}\ \bibnamefont {Serpico}},\ }\href {\doibase
  10.1016/j.physrep.2009.02.002} {\bibfield  {journal} {\bibinfo  {journal}
  {Phys. Rept.}\ }\textbf {\bibinfo {volume} {472}},\ \bibinfo {pages} {1}
  (\bibinfo {year} {2009})},\ \Eprint {http://arxiv.org/abs/0809.0631}
  {arXiv:0809.0631 [astro-ph]} \BibitemShut {NoStop}%
%%CITATION = ARXIV:0809.0631;%%
\bibitem [{\citenamefont {Pospelov}\ and\ \citenamefont
  {Pradler}(2010)}]{Pospelov:2010hj}%
  \BibitemOpen
  \bibfield  {author} {\bibinfo {author} {\bibfnamefont {M.}~\bibnamefont
  {Pospelov}}\ and\ \bibinfo {author} {\bibfnamefont {J.}~\bibnamefont
  {Pradler}},\ }\href {\doibase 10.1146/annurev.nucl.012809.104521} {\bibfield
  {journal} {\bibinfo  {journal} {Ann. Rev. Nucl. Part. Sci.}\ }\textbf
  {\bibinfo {volume} {60}},\ \bibinfo {pages} {539} (\bibinfo {year} {2010})},\
  \Eprint {http://arxiv.org/abs/1011.1054} {arXiv:1011.1054 [hep-ph]}
  \BibitemShut {NoStop}%
%%CITATION = ARXIV:1011.1054;%%
\bibitem [{\citenamefont {Kachelriess}\ \emph {et~al.}(2000)\citenamefont
  {Kachelriess}, \citenamefont {Tomas},\ and\ \citenamefont
  {Valle}}]{Kachelriess:2000qc}%
  \BibitemOpen
  \bibfield  {author} {\bibinfo {author} {\bibfnamefont {M.}~\bibnamefont
  {Kachelriess}}, \bibinfo {author} {\bibfnamefont {R.}~\bibnamefont {Tomas}},
  \ and\ \bibinfo {author} {\bibfnamefont {J.~W.~F.}\ \bibnamefont {Valle}},\
  }\href {\doibase 10.1103/PhysRevD.62.023004} {\bibfield  {journal} {\bibinfo
  {journal} {Phys. Rev.}\ }\textbf {\bibinfo {volume} {D62}},\ \bibinfo {pages}
  {023004} (\bibinfo {year} {2000})},\ \Eprint
  {http://arxiv.org/abs/hep-ph/0001039} {arXiv:hep-ph/0001039 [hep-ph]}
  \BibitemShut {NoStop}%
%%CITATION = HEP-PH/0001039;%%
\bibitem [{\citenamefont {Farzan}(2003)}]{Farzan:2002wx}%
  \BibitemOpen
  \bibfield  {author} {\bibinfo {author} {\bibfnamefont {Y.}~\bibnamefont
  {Farzan}},\ }\href {\doibase 10.1103/PhysRevD.67.073015} {\bibfield
  {journal} {\bibinfo  {journal} {Phys. Rev.}\ }\textbf {\bibinfo {volume}
  {D67}},\ \bibinfo {pages} {073015} (\bibinfo {year} {2003})},\ \Eprint
  {http://arxiv.org/abs/hep-ph/0211375} {arXiv:hep-ph/0211375 [hep-ph]}
  \BibitemShut {NoStop}%
%%CITATION = HEP-PH/0211375;%%
\bibitem [{\citenamefont {He}\ \emph {et~al.}(1991{\natexlab{a}})\citenamefont
  {He}, \citenamefont {Joshi}, \citenamefont {Lew},\ and\ \citenamefont
  {Volkas}}]{He:1990pn}%
  \BibitemOpen
  \bibfield  {author} {\bibinfo {author} {\bibfnamefont {X.~G.}\ \bibnamefont
  {He}}, \bibinfo {author} {\bibfnamefont {G.~C.}\ \bibnamefont {Joshi}},
  \bibinfo {author} {\bibfnamefont {H.}~\bibnamefont {Lew}}, \ and\ \bibinfo
  {author} {\bibfnamefont {R.~R.}\ \bibnamefont {Volkas}},\ }\href {\doibase
  10.1103/PhysRevD.43.R22} {\bibfield  {journal} {\bibinfo  {journal} {Phys.
  Rev.}\ }\textbf {\bibinfo {volume} {D43}},\ \bibinfo {pages} {22} (\bibinfo
  {year} {1991}{\natexlab{a}})}\BibitemShut {NoStop}%
%%CITATION = PHRVA,D43,22;%%
\bibitem [{\citenamefont {He}\ \emph {et~al.}(1991{\natexlab{b}})\citenamefont
  {He}, \citenamefont {Joshi}, \citenamefont {Lew},\ and\ \citenamefont
  {Volkas}}]{He:1991qd}%
  \BibitemOpen
  \bibfield  {author} {\bibinfo {author} {\bibfnamefont {X.-G.}\ \bibnamefont
  {He}}, \bibinfo {author} {\bibfnamefont {G.~C.}\ \bibnamefont {Joshi}},
  \bibinfo {author} {\bibfnamefont {H.}~\bibnamefont {Lew}}, \ and\ \bibinfo
  {author} {\bibfnamefont {R.~R.}\ \bibnamefont {Volkas}},\ }\href {\doibase
  10.1103/PhysRevD.44.2118} {\bibfield  {journal} {\bibinfo  {journal} {Phys.
  Rev.}\ }\textbf {\bibinfo {volume} {D44}},\ \bibinfo {pages} {2118} (\bibinfo
  {year} {1991}{\natexlab{b}})}\BibitemShut {NoStop}%
%%CITATION = PHRVA,D44,2118;%%
\bibitem [{\citenamefont {Farzan}\ and\ \citenamefont
  {Shoemaker}(2016)}]{Farzan:2015hkd}%
  \BibitemOpen
  \bibfield  {author} {\bibinfo {author} {\bibfnamefont {Y.}~\bibnamefont
  {Farzan}}\ and\ \bibinfo {author} {\bibfnamefont {I.~M.}\ \bibnamefont
  {Shoemaker}},\ }\href {\doibase 10.1007/JHEP07(2016)033} {\bibfield
  {journal} {\bibinfo  {journal} {JHEP}\ }\textbf {\bibinfo {volume} {07}},\
  \bibinfo {pages} {033} (\bibinfo {year} {2016})},\ \Eprint
  {http://arxiv.org/abs/1512.09147} {arXiv:1512.09147 [hep-ph]} \BibitemShut
  {NoStop}%
%%CITATION = ARXIV:1512.09147;%%
\bibitem [{\citenamefont {Babu}\ \emph {et~al.}(2017)\citenamefont {Babu},
  \citenamefont {Friedland}, \citenamefont {Machado},\ and\ \citenamefont
  {Mocioiu}}]{Babu:2017olk}%
  \BibitemOpen
  \bibfield  {author} {\bibinfo {author} {\bibfnamefont {K.~S.}\ \bibnamefont
  {Babu}}, \bibinfo {author} {\bibfnamefont {A.}~\bibnamefont {Friedland}},
  \bibinfo {author} {\bibfnamefont {P.~A.~N.}\ \bibnamefont {Machado}}, \ and\
  \bibinfo {author} {\bibfnamefont {I.}~\bibnamefont {Mocioiu}},\ }\href
  {\doibase 10.1007/JHEP12(2017)096} {\bibfield  {journal} {\bibinfo  {journal}
  {JHEP}\ }\textbf {\bibinfo {volume} {12}},\ \bibinfo {pages} {096} (\bibinfo
  {year} {2017})},\ \Eprint {http://arxiv.org/abs/1705.01822} {arXiv:1705.01822
  [hep-ph]} \BibitemShut {NoStop}%
%%CITATION = ARXIV:1705.01822;%%
\bibitem [{\citenamefont {Escudero}\ \emph {et~al.}(2019)\citenamefont
  {Escudero}, \citenamefont {Hooper}, \citenamefont {Krnjaic},\ and\
  \citenamefont {Pierre}}]{Escudero:2019gzq}%
  \BibitemOpen
  \bibfield  {author} {\bibinfo {author} {\bibfnamefont {M.}~\bibnamefont
  {Escudero}}, \bibinfo {author} {\bibfnamefont {D.}~\bibnamefont {Hooper}},
  \bibinfo {author} {\bibfnamefont {G.}~\bibnamefont {Krnjaic}}, \ and\
  \bibinfo {author} {\bibfnamefont {M.}~\bibnamefont {Pierre}},\ }\href
  {\doibase 10.1007/JHEP03(2019)071} {\bibfield  {journal} {\bibinfo  {journal}
  {JHEP}\ }\textbf {\bibinfo {volume} {03}},\ \bibinfo {pages} {071} (\bibinfo
  {year} {2019})},\ \Eprint {http://arxiv.org/abs/1901.02010} {arXiv:1901.02010
  [hep-ph]} \BibitemShut {NoStop}%
%%CITATION = ARXIV:1901.02010;%%
\bibitem [{\citenamefont {Weinberg}(2004)}]{Weinberg:2003ur}%
  \BibitemOpen
  \bibfield  {author} {\bibinfo {author} {\bibfnamefont {S.}~\bibnamefont
  {Weinberg}},\ }\href {\doibase 10.1103/PhysRevD.69.023503} {\bibfield
  {journal} {\bibinfo  {journal} {Phys. Rev.}\ }\textbf {\bibinfo {volume}
  {D69}},\ \bibinfo {pages} {023503} (\bibinfo {year} {2004})},\ \Eprint
  {http://arxiv.org/abs/astro-ph/0306304} {arXiv:astro-ph/0306304 [astro-ph]}
  \BibitemShut {NoStop}%
%%CITATION = ASTRO-PH/0306304;%%
\bibitem [{\citenamefont {Ma}\ and\ \citenamefont
  {Bertschinger}(1995)}]{Ma:1995ey}%
  \BibitemOpen
  \bibfield  {author} {\bibinfo {author} {\bibfnamefont {C.-P.}\ \bibnamefont
  {Ma}}\ and\ \bibinfo {author} {\bibfnamefont {E.}~\bibnamefont
  {Bertschinger}},\ }\href {\doibase 10.1086/176550} {\bibfield  {journal}
  {\bibinfo  {journal} {Astrophys. J.}\ }\textbf {\bibinfo {volume} {455}},\
  \bibinfo {pages} {7} (\bibinfo {year} {1995})},\ \Eprint
  {http://arxiv.org/abs/astro-ph/9506072} {arXiv:astro-ph/9506072 [astro-ph]}
  \BibitemShut {NoStop}%
%%CITATION = ASTRO-PH/9506072;%%
\bibitem [{\citenamefont {Hannestad}\ and\ \citenamefont
  {Scherrer}(2000)}]{Hannestad:2000gt}%
  \BibitemOpen
  \bibfield  {author} {\bibinfo {author} {\bibfnamefont {S.}~\bibnamefont
  {Hannestad}}\ and\ \bibinfo {author} {\bibfnamefont {R.~J.}\ \bibnamefont
  {Scherrer}},\ }\href {\doibase 10.1103/PhysRevD.62.043522} {\bibfield
  {journal} {\bibinfo  {journal} {Phys. Rev.}\ }\textbf {\bibinfo {volume}
  {D62}},\ \bibinfo {pages} {043522} (\bibinfo {year} {2000})},\ \Eprint
  {http://arxiv.org/abs/astro-ph/0003046} {arXiv:astro-ph/0003046 [astro-ph]}
  \BibitemShut {NoStop}%
%%CITATION = ASTRO-PH/0003046;%%
\bibitem [{\citenamefont {Blas}\ \emph {et~al.}(2011)\citenamefont {Blas},
  \citenamefont {Lesgourgues},\ and\ \citenamefont {Tram}}]{Blas:2011rf}%
  \BibitemOpen
  \bibfield  {author} {\bibinfo {author} {\bibfnamefont {D.}~\bibnamefont
  {Blas}}, \bibinfo {author} {\bibfnamefont {J.}~\bibnamefont {Lesgourgues}}, \
  and\ \bibinfo {author} {\bibfnamefont {T.}~\bibnamefont {Tram}},\ }\href
  {\doibase 10.1088/1475-7516/2011/07/034} {\bibfield  {journal} {\bibinfo
  {journal} {JCAP}\ }\textbf {\bibinfo {volume} {1107}},\ \bibinfo {pages}
  {034} (\bibinfo {year} {2011})},\ \Eprint {http://arxiv.org/abs/1104.2933}
  {arXiv:1104.2933 [astro-ph.CO]} \BibitemShut {NoStop}%
%%CITATION = ARXIV:1104.2933;%%
\bibitem [{\citenamefont {Lesgourgues}(2011)}]{Lesgourgues:2011re}%
  \BibitemOpen
  \bibfield  {author} {\bibinfo {author} {\bibfnamefont {J.}~\bibnamefont
  {Lesgourgues}},\ }\href@noop {} {\  (\bibinfo {year} {2011})},\ \Eprint
  {http://arxiv.org/abs/1104.2932} {arXiv:1104.2932 [astro-ph.IM]} \BibitemShut
  {NoStop}%
%%CITATION = ARXIV:1104.2932;%%
\bibitem [{\citenamefont {Vagnozzi}(2019)}]{Vagnozzi:2019utt}%
  \BibitemOpen
  \bibfield  {author} {\bibinfo {author} {\bibfnamefont {S.}~\bibnamefont
  {Vagnozzi}},\ }\href@noop {} {\  (\bibinfo {year} {2019})},\ \Eprint
  {http://arxiv.org/abs/1907.08010} {arXiv:1907.08010 [astro-ph.CO]}
  \BibitemShut {NoStop}%
%%CITATION = ARXIV:1907.08010;%%
\bibitem [{\citenamefont {Vagnozzi}\ \emph {et~al.}(2018)\citenamefont
  {Vagnozzi}, \citenamefont {Dhawan}, \citenamefont {Gerbino}, \citenamefont
  {Freese}, \citenamefont {Goobar},\ and\ \citenamefont
  {Mena}}]{Vagnozzi:2018jhn}%
  \BibitemOpen
  \bibfield  {author} {\bibinfo {author} {\bibfnamefont {S.}~\bibnamefont
  {Vagnozzi}}, \bibinfo {author} {\bibfnamefont {S.}~\bibnamefont {Dhawan}},
  \bibinfo {author} {\bibfnamefont {M.}~\bibnamefont {Gerbino}}, \bibinfo
  {author} {\bibfnamefont {K.}~\bibnamefont {Freese}}, \bibinfo {author}
  {\bibfnamefont {A.}~\bibnamefont {Goobar}}, \ and\ \bibinfo {author}
  {\bibfnamefont {O.}~\bibnamefont {Mena}},\ }\href {\doibase
  10.1103/PhysRevD.98.083501} {\bibfield  {journal} {\bibinfo  {journal} {Phys.
  Rev.}\ }\textbf {\bibinfo {volume} {D98}},\ \bibinfo {pages} {083501}
  (\bibinfo {year} {2018})},\ \Eprint {http://arxiv.org/abs/1801.08553}
  {arXiv:1801.08553 [astro-ph.CO]} \BibitemShut {NoStop}%
%%CITATION = ARXIV:1801.08553;%%
\bibitem [{\citenamefont {Vagnozzi}\ \emph {et~al.}(2017)\citenamefont
  {Vagnozzi}, \citenamefont {Giusarma}, \citenamefont {Mena}, \citenamefont
  {Freese}, \citenamefont {Gerbino}, \citenamefont {Ho},\ and\ \citenamefont
  {Lattanzi}}]{Vagnozzi:2017ovm}%
  \BibitemOpen
  \bibfield  {author} {\bibinfo {author} {\bibfnamefont {S.}~\bibnamefont
  {Vagnozzi}}, \bibinfo {author} {\bibfnamefont {E.}~\bibnamefont {Giusarma}},
  \bibinfo {author} {\bibfnamefont {O.}~\bibnamefont {Mena}}, \bibinfo {author}
  {\bibfnamefont {K.}~\bibnamefont {Freese}}, \bibinfo {author} {\bibfnamefont
  {M.}~\bibnamefont {Gerbino}}, \bibinfo {author} {\bibfnamefont
  {S.}~\bibnamefont {Ho}}, \ and\ \bibinfo {author} {\bibfnamefont
  {M.}~\bibnamefont {Lattanzi}},\ }\href {\doibase 10.1103/PhysRevD.96.123503}
  {\bibfield  {journal} {\bibinfo  {journal} {Phys. Rev.}\ }\textbf {\bibinfo
  {volume} {D96}},\ \bibinfo {pages} {123503} (\bibinfo {year} {2017})},\
  \Eprint {http://arxiv.org/abs/1701.08172} {arXiv:1701.08172 [astro-ph.CO]}
  \BibitemShut {NoStop}%
%%CITATION = ARXIV:1701.08172;%%
\bibitem [{\citenamefont {Roy~Choudhury}\ and\ \citenamefont
  {Choubey}(2018)}]{Choudhury:2018byy}%
  \BibitemOpen
  \bibfield  {author} {\bibinfo {author} {\bibfnamefont {S.}~\bibnamefont
  {Roy~Choudhury}}\ and\ \bibinfo {author} {\bibfnamefont {S.}~\bibnamefont
  {Choubey}},\ }\href {\doibase 10.1088/1475-7516/2018/09/017} {\bibfield
  {journal} {\bibinfo  {journal} {JCAP}\ }\textbf {\bibinfo {volume} {1809}},\
  \bibinfo {pages} {017} (\bibinfo {year} {2018})},\ \Eprint
  {http://arxiv.org/abs/1806.10832} {arXiv:1806.10832 [astro-ph.CO]}
  \BibitemShut {NoStop}%
%%CITATION = ARXIV:1806.10832;%%
\bibitem [{\citenamefont {Aghanim}\ \emph {et~al.}(2019)\citenamefont {Aghanim}
  \emph {et~al.}}]{Aghanim:2019ame}%
  \BibitemOpen
  \bibfield  {author} {\bibinfo {author} {\bibfnamefont {N.}~\bibnamefont
  {Aghanim}} \emph {et~al.} (\bibinfo {collaboration} {Planck}),\ }\href@noop
  {} {\  (\bibinfo {year} {2019})},\ \Eprint {http://arxiv.org/abs/1907.12875}
  {arXiv:1907.12875 [astro-ph.CO]} \BibitemShut {NoStop}%
%%CITATION = ARXIV:1907.12875;%%
\bibitem [{\citenamefont {Kawasaki}\ \emph {et~al.}(2000)\citenamefont
  {Kawasaki}, \citenamefont {Kohri},\ and\ \citenamefont
  {Sugiyama}}]{Kawasaki:2000en}%
  \BibitemOpen
  \bibfield  {author} {\bibinfo {author} {\bibfnamefont {M.}~\bibnamefont
  {Kawasaki}}, \bibinfo {author} {\bibfnamefont {K.}~\bibnamefont {Kohri}}, \
  and\ \bibinfo {author} {\bibfnamefont {N.}~\bibnamefont {Sugiyama}},\ }\href
  {\doibase 10.1103/PhysRevD.62.023506} {\bibfield  {journal} {\bibinfo
  {journal} {Phys. Rev.}\ }\textbf {\bibinfo {volume} {D62}},\ \bibinfo {pages}
  {023506} (\bibinfo {year} {2000})},\ \Eprint
  {http://arxiv.org/abs/astro-ph/0002127} {arXiv:astro-ph/0002127 [astro-ph]}
  \BibitemShut {NoStop}%
%%CITATION = ASTRO-PH/0002127;%%
\bibitem [{\citenamefont {Brinckmann}\ and\ \citenamefont
  {Lesgourgues}(2018)}]{Brinckmann:2018cvx}%
  \BibitemOpen
  \bibfield  {author} {\bibinfo {author} {\bibfnamefont {T.}~\bibnamefont
  {Brinckmann}}\ and\ \bibinfo {author} {\bibfnamefont {J.}~\bibnamefont
  {Lesgourgues}},\ }\href@noop {} {\  (\bibinfo {year} {2018})},\ \Eprint
  {http://arxiv.org/abs/1804.07261} {arXiv:1804.07261 [astro-ph.CO]}
  \BibitemShut {NoStop}%
%%CITATION = ARXIV:1804.07261;%%
\bibitem [{\citenamefont {Audren}\ \emph {et~al.}(2013)\citenamefont {Audren},
  \citenamefont {Lesgourgues}, \citenamefont {Benabed},\ and\ \citenamefont
  {Prunet}}]{Audren:2012wb}%
  \BibitemOpen
  \bibfield  {author} {\bibinfo {author} {\bibfnamefont {B.}~\bibnamefont
  {Audren}}, \bibinfo {author} {\bibfnamefont {J.}~\bibnamefont {Lesgourgues}},
  \bibinfo {author} {\bibfnamefont {K.}~\bibnamefont {Benabed}}, \ and\
  \bibinfo {author} {\bibfnamefont {S.}~\bibnamefont {Prunet}},\ }\href
  {\doibase 10.1088/1475-7516/2013/02/001} {\bibfield  {journal} {\bibinfo
  {journal} {JCAP}\ }\textbf {\bibinfo {volume} {1302}},\ \bibinfo {pages}
  {001} (\bibinfo {year} {2013})},\ \Eprint {http://arxiv.org/abs/1210.7183}
  {arXiv:1210.7183 [astro-ph.CO]} \BibitemShut {NoStop}%
%%CITATION = ARXIV:1210.7183;%%
\bibitem [{\citenamefont {Gelman}\ and\ \citenamefont
  {Rubin}(1992)}]{Gelman:1992zz}%
  \BibitemOpen
  \bibfield  {author} {\bibinfo {author} {\bibfnamefont {A.}~\bibnamefont
  {Gelman}}\ and\ \bibinfo {author} {\bibfnamefont {D.~B.}\ \bibnamefont
  {Rubin}},\ }\href {\doibase 10.1214/ss/1177011136} {\bibfield  {journal}
  {\bibinfo  {journal} {Statist. Sci.}\ }\textbf {\bibinfo {volume} {7}},\
  \bibinfo {pages} {457} (\bibinfo {year} {1992})}\BibitemShut {NoStop}%
%%CITATION = STSCE,7,457;%%
\bibitem [{\citenamefont {Beutler}\ \emph {et~al.}(2011)\citenamefont
  {Beutler}, \citenamefont {Blake}, \citenamefont {Colless}, \citenamefont
  {Jones}, \citenamefont {Staveley-Smith}, \citenamefont {Campbell},
  \citenamefont {Parker}, \citenamefont {Saunders},\ and\ \citenamefont
  {Watson}}]{Beutler:2011hx}%
  \BibitemOpen
  \bibfield  {author} {\bibinfo {author} {\bibfnamefont {F.}~\bibnamefont
  {Beutler}}, \bibinfo {author} {\bibfnamefont {C.}~\bibnamefont {Blake}},
  \bibinfo {author} {\bibfnamefont {M.}~\bibnamefont {Colless}}, \bibinfo
  {author} {\bibfnamefont {D.~H.}\ \bibnamefont {Jones}}, \bibinfo {author}
  {\bibfnamefont {L.}~\bibnamefont {Staveley-Smith}}, \bibinfo {author}
  {\bibfnamefont {L.}~\bibnamefont {Campbell}}, \bibinfo {author}
  {\bibfnamefont {Q.}~\bibnamefont {Parker}}, \bibinfo {author} {\bibfnamefont
  {W.}~\bibnamefont {Saunders}}, \ and\ \bibinfo {author} {\bibfnamefont
  {F.}~\bibnamefont {Watson}},\ }\href {\doibase
  10.1111/j.1365-2966.2011.19250.x} {\bibfield  {journal} {\bibinfo  {journal}
  {Mon. Not. Roy. Astron. Soc.}\ }\textbf {\bibinfo {volume} {416}},\ \bibinfo
  {pages} {3017} (\bibinfo {year} {2011})},\ \Eprint
  {http://arxiv.org/abs/1106.3366} {arXiv:1106.3366 [astro-ph.CO]} \BibitemShut
  {NoStop}%
%%CITATION = ARXIV:1106.3366;%%
\bibitem [{\citenamefont {Ross}\ \emph {et~al.}(2015)\citenamefont {Ross},
  \citenamefont {Samushia}, \citenamefont {Howlett}, \citenamefont {Percival},
  \citenamefont {Burden},\ and\ \citenamefont {Manera}}]{Ross:2014qpa}%
  \BibitemOpen
  \bibfield  {author} {\bibinfo {author} {\bibfnamefont {A.~J.}\ \bibnamefont
  {Ross}}, \bibinfo {author} {\bibfnamefont {L.}~\bibnamefont {Samushia}},
  \bibinfo {author} {\bibfnamefont {C.}~\bibnamefont {Howlett}}, \bibinfo
  {author} {\bibfnamefont {W.~J.}\ \bibnamefont {Percival}}, \bibinfo {author}
  {\bibfnamefont {A.}~\bibnamefont {Burden}}, \ and\ \bibinfo {author}
  {\bibfnamefont {M.}~\bibnamefont {Manera}},\ }\href {\doibase
  10.1093/mnras/stv154} {\bibfield  {journal} {\bibinfo  {journal} {Mon. Not.
  Roy. Astron. Soc.}\ }\textbf {\bibinfo {volume} {449}},\ \bibinfo {pages}
  {835} (\bibinfo {year} {2015})},\ \Eprint {http://arxiv.org/abs/1409.3242}
  {arXiv:1409.3242 [astro-ph.CO]} \BibitemShut {NoStop}%
%%CITATION = ARXIV:1409.3242;%%
\bibitem [{\citenamefont {Alam}\ \emph {et~al.}(2017)\citenamefont {Alam} \emph
  {et~al.}}]{Alam:2016hwk}%
  \BibitemOpen
  \bibfield  {author} {\bibinfo {author} {\bibfnamefont {S.}~\bibnamefont
  {Alam}} \emph {et~al.} (\bibinfo {collaboration} {BOSS}),\ }\href {\doibase
  10.1093/mnras/stx721} {\bibfield  {journal} {\bibinfo  {journal} {Mon. Not.
  Roy. Astron. Soc.}\ }\textbf {\bibinfo {volume} {470}},\ \bibinfo {pages}
  {2617} (\bibinfo {year} {2017})},\ \Eprint {http://arxiv.org/abs/1607.03155}
  {arXiv:1607.03155 [astro-ph.CO]} \BibitemShut {NoStop}%
%%CITATION = ARXIV:1607.03155;%%
\bibitem [{\citenamefont {Beacom}\ \emph {et~al.}(2003)\citenamefont {Beacom},
  \citenamefont {Bell}, \citenamefont {Hooper}, \citenamefont {Pakvasa},\ and\
  \citenamefont {Weiler}}]{Beacom:2002vi}%
  \BibitemOpen
  \bibfield  {author} {\bibinfo {author} {\bibfnamefont {J.~F.}\ \bibnamefont
  {Beacom}}, \bibinfo {author} {\bibfnamefont {N.~F.}\ \bibnamefont {Bell}},
  \bibinfo {author} {\bibfnamefont {D.}~\bibnamefont {Hooper}}, \bibinfo
  {author} {\bibfnamefont {S.}~\bibnamefont {Pakvasa}}, \ and\ \bibinfo
  {author} {\bibfnamefont {T.~J.}\ \bibnamefont {Weiler}},\ }\href {\doibase
  10.1103/PhysRevLett.90.181301} {\bibfield  {journal} {\bibinfo  {journal}
  {Phys. Rev. Lett.}\ }\textbf {\bibinfo {volume} {90}},\ \bibinfo {pages}
  {181301} (\bibinfo {year} {2003})},\ \Eprint
  {http://arxiv.org/abs/hep-ph/0211305} {arXiv:hep-ph/0211305 [hep-ph]}
  \BibitemShut {NoStop}%
%%CITATION = HEP-PH/0211305;%%
\bibitem [{\citenamefont {Ando}(2004)}]{Ando:2004qe}%
  \BibitemOpen
  \bibfield  {author} {\bibinfo {author} {\bibfnamefont {S.}~\bibnamefont
  {Ando}},\ }\href {\doibase 10.1103/PhysRevD.70.033004} {\bibfield  {journal}
  {\bibinfo  {journal} {Phys. Rev.}\ }\textbf {\bibinfo {volume} {D70}},\
  \bibinfo {pages} {033004} (\bibinfo {year} {2004})},\ \Eprint
  {http://arxiv.org/abs/hep-ph/0405200} {arXiv:hep-ph/0405200 [hep-ph]}
  \BibitemShut {NoStop}%
%%CITATION = HEP-PH/0405200;%%
\bibitem [{\citenamefont {Ando}(2003)}]{Ando:2003ie}%
  \BibitemOpen
  \bibfield  {author} {\bibinfo {author} {\bibfnamefont {S.}~\bibnamefont
  {Ando}},\ }\href {\doibase 10.1016/j.physletb.2003.07.009} {\bibfield
  {journal} {\bibinfo  {journal} {Phys. Lett.}\ }\textbf {\bibinfo {volume}
  {B570}},\ \bibinfo {pages} {11} (\bibinfo {year} {2003})},\ \Eprint
  {http://arxiv.org/abs/hep-ph/0307169} {arXiv:hep-ph/0307169 [hep-ph]}
  \BibitemShut {NoStop}%
%%CITATION = HEP-PH/0307169;%%
\bibitem [{\citenamefont {Abrahão}\ \emph {et~al.}(2015)\citenamefont
  {Abrahão}, \citenamefont {Minakata}, \citenamefont {Nunokawa},\ and\
  \citenamefont {Quiroga}}]{Abrahao:2015rba}%
  \BibitemOpen
  \bibfield  {author} {\bibinfo {author} {\bibfnamefont {T.}~\bibnamefont
  {Abrahão}}, \bibinfo {author} {\bibfnamefont {H.}~\bibnamefont {Minakata}},
  \bibinfo {author} {\bibfnamefont {H.}~\bibnamefont {Nunokawa}}, \ and\
  \bibinfo {author} {\bibfnamefont {A.~A.}\ \bibnamefont {Quiroga}},\ }\href
  {\doibase 10.1007/JHEP11(2015)001} {\bibfield  {journal} {\bibinfo  {journal}
  {JHEP}\ }\textbf {\bibinfo {volume} {11}},\ \bibinfo {pages} {001} (\bibinfo
  {year} {2015})},\ \Eprint {http://arxiv.org/abs/1506.02314} {arXiv:1506.02314
  [hep-ph]} \BibitemShut {NoStop}%
%%CITATION = ARXIV:1506.02314;%%
\bibitem [{\citenamefont {Bustamante}\ \emph {et~al.}(2017)\citenamefont
  {Bustamante}, \citenamefont {Beacom},\ and\ \citenamefont
  {Murase}}]{Bustamante:2016ciw}%
  \BibitemOpen
  \bibfield  {author} {\bibinfo {author} {\bibfnamefont {M.}~\bibnamefont
  {Bustamante}}, \bibinfo {author} {\bibfnamefont {J.~F.}\ \bibnamefont
  {Beacom}}, \ and\ \bibinfo {author} {\bibfnamefont {K.}~\bibnamefont
  {Murase}},\ }\href {\doibase 10.1103/PhysRevD.95.063013} {\bibfield
  {journal} {\bibinfo  {journal} {Phys. Rev.}\ }\textbf {\bibinfo {volume}
  {D95}},\ \bibinfo {pages} {063013} (\bibinfo {year} {2017})},\ \Eprint
  {http://arxiv.org/abs/1610.02096} {arXiv:1610.02096 [astro-ph.HE]}
  \BibitemShut {NoStop}%
%%CITATION = ARXIV:1610.02096;%%
\bibitem [{\citenamefont {Coloma}\ and\ \citenamefont
  {Peres}(2017)}]{Coloma:2017zpg}%
  \BibitemOpen
  \bibfield  {author} {\bibinfo {author} {\bibfnamefont {P.}~\bibnamefont
  {Coloma}}\ and\ \bibinfo {author} {\bibfnamefont {O.~L.~G.}\ \bibnamefont
  {Peres}},\ }\href@noop {} {\  (\bibinfo {year} {2017})},\ \Eprint
  {http://arxiv.org/abs/1705.03599} {arXiv:1705.03599 [hep-ph]} \BibitemShut
  {NoStop}%
%%CITATION = ARXIV:1705.03599;%%
\bibitem [{\citenamefont {Choubey}\ \emph
  {et~al.}(2018{\natexlab{b}})\citenamefont {Choubey}, \citenamefont
  {Goswami},\ and\ \citenamefont {Pramanik}}]{Choubey:2017dyu}%
  \BibitemOpen
  \bibfield  {author} {\bibinfo {author} {\bibfnamefont {S.}~\bibnamefont
  {Choubey}}, \bibinfo {author} {\bibfnamefont {S.}~\bibnamefont {Goswami}}, \
  and\ \bibinfo {author} {\bibfnamefont {D.}~\bibnamefont {Pramanik}},\ }\href
  {\doibase 10.1007/JHEP02(2018)055} {\bibfield  {journal} {\bibinfo  {journal}
  {JHEP}\ }\textbf {\bibinfo {volume} {02}},\ \bibinfo {pages} {055} (\bibinfo
  {year} {2018}{\natexlab{b}})},\ \Eprint {http://arxiv.org/abs/1705.05820}
  {arXiv:1705.05820 [hep-ph]} \BibitemShut {NoStop}%
%%CITATION = ARXIV:1705.05820;%%
\bibitem [{\citenamefont {Choubey}\ \emph
  {et~al.}(2018{\natexlab{c}})\citenamefont {Choubey}, \citenamefont {Goswami},
  \citenamefont {Gupta}, \citenamefont {Lakshmi},\ and\ \citenamefont
  {Thakore}}]{Choubey:2017eyg}%
  \BibitemOpen
  \bibfield  {author} {\bibinfo {author} {\bibfnamefont {S.}~\bibnamefont
  {Choubey}}, \bibinfo {author} {\bibfnamefont {S.}~\bibnamefont {Goswami}},
  \bibinfo {author} {\bibfnamefont {C.}~\bibnamefont {Gupta}}, \bibinfo
  {author} {\bibfnamefont {S.~M.}\ \bibnamefont {Lakshmi}}, \ and\ \bibinfo
  {author} {\bibfnamefont {T.}~\bibnamefont {Thakore}},\ }\href {\doibase
  10.1103/PhysRevD.97.033005} {\bibfield  {journal} {\bibinfo  {journal} {Phys.
  Rev.}\ }\textbf {\bibinfo {volume} {D97}},\ \bibinfo {pages} {033005}
  (\bibinfo {year} {2018}{\natexlab{c}})},\ \Eprint
  {http://arxiv.org/abs/1709.10376} {arXiv:1709.10376 [hep-ph]} \BibitemShut
  {NoStop}%
%%CITATION = ARXIV:1709.10376;%%
\bibitem [{\citenamefont {Ascencio-Sosa}\ \emph {et~al.}(2018)\citenamefont
  {Ascencio-Sosa}, \citenamefont {Calatayud-Cadenillas}, \citenamefont {Gago},\
  and\ \citenamefont {Jones-Pérez}}]{Ascencio-Sosa:2018lbk}%
  \BibitemOpen
  \bibfield  {author} {\bibinfo {author} {\bibfnamefont {M.~V.}\ \bibnamefont
  {Ascencio-Sosa}}, \bibinfo {author} {\bibfnamefont {A.~M.}\ \bibnamefont
  {Calatayud-Cadenillas}}, \bibinfo {author} {\bibfnamefont {A.~M.}\
  \bibnamefont {Gago}}, \ and\ \bibinfo {author} {\bibfnamefont
  {J.}~\bibnamefont {Jones-Pérez}},\ }\href {\doibase
  10.1140/epjc/s10052-018-6276-0} {\bibfield  {journal} {\bibinfo  {journal}
  {Eur. Phys. J.}\ }\textbf {\bibinfo {volume} {C78}},\ \bibinfo {pages} {809}
  (\bibinfo {year} {2018})},\ \Eprint {http://arxiv.org/abs/1805.03279}
  {arXiv:1805.03279 [hep-ph]} \BibitemShut {NoStop}%
%%CITATION = ARXIV:1805.03279;%%
\bibitem [{\citenamefont {Tang}\ \emph {et~al.}(2018)\citenamefont {Tang},
  \citenamefont {Wang},\ and\ \citenamefont {Zhang}}]{Tang:2018rer}%
  \BibitemOpen
  \bibfield  {author} {\bibinfo {author} {\bibfnamefont {J.}~\bibnamefont
  {Tang}}, \bibinfo {author} {\bibfnamefont {T.-C.}\ \bibnamefont {Wang}}, \
  and\ \bibinfo {author} {\bibfnamefont {Y.}~\bibnamefont {Zhang}},\
  }\href@noop {} {\  (\bibinfo {year} {2018})},\ \Eprint
  {http://arxiv.org/abs/1811.05623} {arXiv:1811.05623 [hep-ph]} \BibitemShut
  {NoStop}%
%%CITATION = ARXIV:1811.05623;%%
\bibitem [{\citenamefont {Serpico}(2007)}]{Serpico:2007pt}%
  \BibitemOpen
  \bibfield  {author} {\bibinfo {author} {\bibfnamefont {P.~D.}\ \bibnamefont
  {Serpico}},\ }\href {\doibase 10.1103/PhysRevLett.98.171301} {\bibfield
  {journal} {\bibinfo  {journal} {Phys. Rev. Lett.}\ }\textbf {\bibinfo
  {volume} {98}},\ \bibinfo {pages} {171301} (\bibinfo {year} {2007})},\
  \Eprint {http://arxiv.org/abs/astro-ph/0701699} {arXiv:astro-ph/0701699
  [astro-ph]} \BibitemShut {NoStop}%
%%CITATION = ASTRO-PH/0701699;%%
\bibitem [{\citenamefont {Baumann}\ \emph {et~al.}(2019)\citenamefont
  {Baumann}, \citenamefont {Beutler}, \citenamefont {Flauger}, \citenamefont
  {Green}, \citenamefont {Vargas-Magaña}, \citenamefont {Slosar},
  \citenamefont {Wallisch},\ and\ \citenamefont {Yèche}}]{Baumann:2019tdh}%
  \BibitemOpen
  \bibfield  {author} {\bibinfo {author} {\bibfnamefont {D.}~\bibnamefont
  {Baumann}}, \bibinfo {author} {\bibfnamefont {F.}~\bibnamefont {Beutler}},
  \bibinfo {author} {\bibfnamefont {R.}~\bibnamefont {Flauger}}, \bibinfo
  {author} {\bibfnamefont {D.}~\bibnamefont {Green}}, \bibinfo {author}
  {\bibfnamefont {M.}~\bibnamefont {Vargas-Magaña}}, \bibinfo {author}
  {\bibfnamefont {A.}~\bibnamefont {Slosar}}, \bibinfo {author} {\bibfnamefont
  {B.}~\bibnamefont {Wallisch}}, \ and\ \bibinfo {author} {\bibfnamefont
  {C.}~\bibnamefont {Yèche}},\ }\href {\doibase 10.1038/s41567-019-0435-6}
  {\bibfield  {journal} {\bibinfo  {journal} {Nature Phys.}\ }\textbf {\bibinfo
  {volume} {15}},\ \bibinfo {pages} {465} (\bibinfo {year} {2019})},\ \Eprint
  {http://arxiv.org/abs/1803.10741} {arXiv:1803.10741 [astro-ph.CO]}
  \BibitemShut {NoStop}%
%%CITATION = ARXIV:1803.10741;%%
\bibitem [{\citenamefont {Gondolo}\ and\ \citenamefont
  {Gelmini}(1991)}]{Gondolo:1990dk}%
  \BibitemOpen
  \bibfield  {author} {\bibinfo {author} {\bibfnamefont {P.}~\bibnamefont
  {Gondolo}}\ and\ \bibinfo {author} {\bibfnamefont {G.}~\bibnamefont
  {Gelmini}},\ }\href {\doibase 10.1016/0550-3213(91)90438-4} {\bibfield
  {journal} {\bibinfo  {journal} {Nucl. Phys.}\ }\textbf {\bibinfo {volume}
  {B360}},\ \bibinfo {pages} {145} (\bibinfo {year} {1991})}\BibitemShut
  {NoStop}%
%%CITATION = NUPHA,B360,145;%%
\bibitem [{\citenamefont {Edsjo}\ and\ \citenamefont
  {Gondolo}(1997)}]{Edsjo:1997bg}%
  \BibitemOpen
  \bibfield  {author} {\bibinfo {author} {\bibfnamefont {J.}~\bibnamefont
  {Edsjo}}\ and\ \bibinfo {author} {\bibfnamefont {P.}~\bibnamefont
  {Gondolo}},\ }\href {\doibase 10.1103/PhysRevD.56.1879} {\bibfield  {journal}
  {\bibinfo  {journal} {Phys. Rev.}\ }\textbf {\bibinfo {volume} {D56}},\
  \bibinfo {pages} {1879} (\bibinfo {year} {1997})},\ \Eprint
  {http://arxiv.org/abs/hep-ph/9704361} {arXiv:hep-ph/9704361 [hep-ph]}
  \BibitemShut {NoStop}%
%%CITATION = HEP-PH/9704361;%%
\end{thebibliography}%
%%%%%%%%%%%%%%%%%%%%%%%%%%%%%%%%%%%%%%%%%%%%%%%%%%%%%%

\end{document}